\documentclass[traditabstract]{aa}  
\usepackage{geometry}
\usepackage{xcolor}
\usepackage{colortbl}
\usepackage{tabularx}
\usepackage{graphicx} 
\usepackage{multirow}
\usepackage{placeins}
\definecolor{lightgray}{rgb}{0.83, 0.83, 0.83}
\definecolor{lightblue}{rgb}{0.67, 0.84, 0.90}
\definecolor{lightgreen}{rgb}{0.56, 0.93, 0.56}

\usepackage{natbib}
\bibpunct{(}{)}{;}{a}{}{,} 

\usepackage{ctable}
\usepackage{graphics}   
\usepackage{graphicx}   
\usepackage{epstopdf}
\usepackage{longtable}   
\usepackage{url}      
\usepackage{bm}        
\usepackage{soul}

\usepackage{caption}
\usepackage{subcaption}
\usepackage[T1]{fontenc}

\usepackage{tablefootnote}
\usepackage[varg]{txfonts}
\usepackage{pdflscape}
\usepackage{supertabular}
\usepackage{hyperref}
\usepackage{pifont}

\usepackage{xcolor}
\usepackage[normalem]{ulem}
\usepackage{natbib}
\usepackage{amsmath}
\usepackage{amssymb}
\usepackage{csquotes}

\definecolor{green}{rgb}{0.3,0.7,0.}
\definecolor{purple}{rgb}{0.77, 0.29, 0.55}

\hypersetup{
    bookmarks=true,         
    unicode=false,          
    colorlinks=true,       
    linkcolor=blue,          
    citecolor=blue,        
    filecolor=blue,      
    urlcolor=blue           
}

\usepackage{color}

\begin{document}

\title{Constraining the Physical Properties of Quadruply Lensed Quasars using Optical-to-X-Ray Data}

\titlerunning{Physical properties of lensed quasars using multiwavelength data}
\author{Kriti Kamal Gupta\inst{1,2,3}\thanks{E-mail: kgupta@aip.de}, Dominique Sluse\inst{1}, Maarten Baes\inst{2}, Daniel Gilman\inst{4}, Tommaso Treu\inst{5}, Timo Anguita\inst{6}, Ryan Keeley\inst{7}, and Pritom Mozumdar\inst{5} 
}
\authorrunning{Gupta et al.}

\institute{STAR Institute, Li\`ege Universit\'e, Quartier Agora - All\'ee du six Ao\^ut, 19c B-4000 Li\`ege, Belgium \and
Sterrenkundig Observatorium, Universiteit Gent, Krijgslaan 281 S9, B-9000 Gent, Belgium \and Leibniz-Institut für Astrophysik Potsdam (AIP), An der Sternwarte 16, D-14482 Potsdam, Germany \and Department of Astronomy \& Astrophysics, University of Chicago, Chicago, IL 60637, USA \and Department of Physics and Astronomy, University of California, Los Angeles, CA, 90095, USA \and Instituto de Astrof\'isica, Facultad de Ciencias Exactas, Universidad Andres Bello, Santiago, Chile \and University of California, Merced, 5200 N Lake Road, Merced, CA 95341, USA 
}

\date{}


\abstract{Gravitational lensing of luminous active galactic nuclei (AGN; or quasars) can be used as a natural telescope to zoom in on their inner structures. With more and more lensed AGN being discovered, it is extremely important to have a homogeneous study focused on constraining their key physical properties, especially the bolometric luminosity. The primary limitation for such a study is the availability of observations for a representative sample of lensed AGN in at least the optical, ultraviolet (UV), and X-ray bands, where most of the AGN emission is concentrated. In this paper, we present one of the largest multiwavelength studies of lensed AGN with the aim of accurately measuring some of the fundamental quantities, such as their bolometric luminosities, black hole masses, and Eddington ratios. We compiled photometric and spectroscopic data in optical/UV and X-rays, for 27 quadruply lensed AGN ($0.6<z<3.1$) and calculated their bolometric luminosities by fitting their broadband spectral energy distributions (SEDs) with phenomenological models. We also performed spectral emission line fitting to estimate their black hole masses using the virial method. Additionally, we compared different prescriptions to calculate the bolometric luminosities of AGN from limited data, and our results show that the luminosity-dependent 2--10\,keV X-ray bolometric corrections provide unbiased and reliable predictions of the bolometric luminosity, with a maximum scatter of $\sim 0.5\,$dex. The predictions from optical bolometric corrections are generally overestimated but show a lower scatter once microlensing effects are taken into account. Thanks to the compiled optical/UV and X-ray data for our lensed AGN sample, we also present the well-known UV-X-ray luminosity relation for AGN as a novel way to determine uncertainty in the magnifications of lensed AGN induced by micro- and/or millilensing. 
 
}


\maketitle
\section{Introduction}\label{sect:intro}

Over the years, one of the major challenges in the study of active galactic nuclei (AGN) has been to determine their bolometric luminosity ($L_{\rm bol}$). The total accretion luminosity or the bolometric luminosity is one of the key physical properties of AGN, which is essential to understand their accretion and emission physics, as well as how they interact with their host galaxies. However, since AGN emit radiation at multiple wavelengths (from radio to X-rays) via various physical processes (e.g., \citealp{2013peag.book.....N}), a wealth of multiband data is needed to accurately measure their bolometric luminosity. This primarily requires access to their optical, ultraviolet (UV), and X-ray emission data, as these wavelengths are the major contributors to their bolometric luminosity. The optical/UV photons are emitted by the accretion disk surrounding the central supermassive black hole (SMBH), while the X-rays are produced in a plasma of hot electrons located close to the central black hole via Comptonization (e.g., \citealp{1991ApJ...380L..51H}). Together, these two constitute the bolometric emission of the AGN. Additionally, the entire AGN structure is surrounded by a geometrically and optically thick torus-like structure made up of dust and molecular gas, referred to as the \enquote*{dusty torus} (e.g., \citealp{1995PASP..107..803U}). However, since the torus just reprocesses part of the optical/UV emission from the disk and re-emits in the infrared (IR), it does not contribute to the intrinsic emission of the AGN.

The difficulty of estimating $L_{\rm bol}$ further translates to the calculation of the accretion rate and the Eddington ratio ($\lambda_{\rm Edd}$; mass-normalized accretion rate) of the AGN, which are crucial for black hole growth studies. The lack of reliable estimates of these fundamental properties of AGN is even more prominent for certain classes, such as obscured AGN, low-luminosity AGN, super-Eddington AGN, and lensed AGN. Most of these AGN types are relatively scarce in the general AGN population, due to their low brightness (especially in the optical/UV), which makes it a bit complicated to accurately quantify their bolometric luminosities. But in the case of gravitationally lensed AGN, even though we have large samples of optically selected, extremely bright AGN (quasars\footnote{We use the terms quasars and AGN interchangeably throughout the text as most of the sources in our sample are intrinsically bright.}; \citealp{2018A&A...618A..56D,2019MNRAS.483.4242L,2023MNRAS.520.3305L}), we are still missing a systematic multiwavelength study dedicated to constraining their physical properties. Additionally, since lensed AGN provide a unique way to probe the inner structures of AGN (e.g., \citealp{1988A&A...194...54G}), which are otherwise difficult to resolve, it is extremely important to accurately characterize their key properties, including the bolometric luminosity. However, the available literature on lensed AGN consists only of source-specific studies or those with small samples and limited data (e.g., \citealp{2007ApJ...661...19P,2010ApJ...709..937G,2012A&A...544A..62S,2021A&A...656A.108M,2023A&A...680A..51M}). Through this work, we aim to take a step towards solving this issue, by using available multiband data to characterize the physical properties of lensed AGN.

Generally, the best way to calculate $L_{\rm bol}$ for an AGN is to construct and fit its broadband spectral energy distribution (SED). However, various other methods are employed to calculate the bolometric luminosity of AGN in the absence of multiband data. The most common is to use bolometric corrections. Bolometric corrections ($\kappa_{\lambda}$) are defined as the ratio of the bolometric luminosity of the AGN to the luminosity at a specific wavelength/energy ($L_{\lambda}$; $\kappa_{\lambda} = L_{\rm bol}/L_{\lambda}$). Since they quantify the emission fraction of AGN at a particular wavelength, they can be easily used to estimate $L_{\rm bol}$ from limited data. Optical and X-ray bolometric corrections are widely used by the AGN community to compute $L_{\rm bol}$ for sources without multiwavelength data. Specifically, X-ray bolometric corrections in the 2--10\,keV energy range ($\kappa_{2-10}$) are extremely useful as the X-rays (unlike optical/UV) are not contaminated by the host galaxy light and are less prone to dust extinction.

In the case of gravitationally lensed quasars, X-rays provide an additional advantage because they are also not affected by the light from the foreground lens. While acquiring reliable AGN flux in the optical/UV can be very technologically demanding, as it requires accurate modeling of the lens as well as the host galaxy (in some cases), the X-ray spectral extraction process is quite trivial. Therefore, X-ray bolometric corrections provide a useful tool for estimating the bolometric luminosity of lensed AGN, with far fewer sources of uncertainty compared to the optical. However, contrary to the optical bolometric corrections that are not luminosity dependent, $\kappa_{2-10}$ increases with the AGN luminosity and Eddington ratio (e.g., \citealp{2012MNRAS.425..623L,2020A&A...636A..73D,2024A&A...691A.203G}).  Hence, one needs to be careful when using them to calculate $L_{\rm bol}$, and, in turn, $\lambda_{\rm Edd}$. As a result, most of the studies based on gravitationally lensed quasars commonly use optical bolometric corrections instead of the X-rays (e.g., \citealp{2012A&A...544A..62S,2021A&A...656A.108M,2023A&A...680A..51M}). Additionally, X-rays are more prone to microlensing due to the smaller physical scale of their emitting region, in which case they cannot be a reliable estimator of $L_{\rm bol}$. Furthermore, optical instruments have better spatial resolution compared to X-ray telescopes, which is needed to resolve the multiple images of lensed AGN. In conclusion, both optical and X-rays have specific advantages (and disadvantages) over each other to be used as accurate $L_{\rm bol}$ estimators.

The aim of the work presented in this paper is three-fold. First, we have performed a comprehensive multiwavelength study of a sample of quadruply imaged lensed AGN, with the goal of measuring their important physical properties, such as bolometric luminosity ($L_{\rm bol}$), black hole mass ($M_{\rm BH}$), and Eddington ratio ($\lambda_{\rm Edd}$). This was done by compiling optical, UV, and X-ray photometric and spectroscopic data for a sample of 27 lensed AGN systems ($0.6 < z < 3.1$). To date, this is the largest sample of quadruply lensed AGN for which such an analysis has been carried out in a consistent way. Moreover, for most of the systems in our sample, these measurements have been made for the first time. Second, we have explored the well-known UV-X-ray luminosity relation for AGN as a new method to identify systematic uncertainties in the lensing magnification corrections of lensed AGN, caused by either incorrect lens models or by the presence of micro- or millilensing in one of the lens images. Finally, we have also used the available multiband data to perform specific tests to determine the best way to use the X-ray and optical/UV bolometric corrections as an estimator of the bolometric luminosities in lensed AGN in the absence of multiwavelength data.

The structure of the paper is as follows. In Sect. \ref{sect:sampledata}, we describe the sample of lensed AGN (Sect. \ref{sect:sample}) and the multiwavelength data (Sects. \ref{sect:xray} and \ref{sect:uvo}) used in this work, while in Sect. \ref{sect:data} we illustrate the data analysis process for the X-ray (Sect. \ref{sect:xray_spec}) and optical/UV data (Sects. \ref{sect:sed} and \ref{sect:uvo_spec}). In Sect. \ref{sect:results}, we present our main results, including estimates of $L_{\rm bol}$, $M_{\rm BH}$, and $\lambda_{\rm Edd}$. We present the UV-X-ray luminosity relation in the context of lensed AGN (Sect. \ref{sect:alpha}) and show the most accurate method to use X-ray (Sect. \ref{sect:kx}) and optical/UV (Sect. \ref{sect:kuvo}) bolometric corrections as $L_{\rm bol}$ estimators in Sect. \ref{sect:discuss}. Finally, in Sect. \ref{sect:summary}, we summarize the main takeaways from this paper. Throughout the paper, we assume a cosmological model with $H_{\rm 0}=70\,\rm km\,s^{-1}\,Mpc^{-1}$, $\Omega_{\rm M}=0.3$, and $\Omega_{\Lambda}=0.7$. All magnitudes quoted throughout the text are AB magnitudes. All errors and uncertainties listed in the tables or shown as error bars or shaded regions in the figures are one sigma, unless otherwise stated. The correlations were obtained using various functions from the \texttt{statistics}\footnote{\url{https://docs.scipy.org/doc/scipy/reference/stats.html}} module of the Python library \texttt{scipy} (\citealp{2020SciPy-NMeth}) and the significance of the correlations is determined using the Pearson's correlation test.


\begin{table*}
\centering
\caption{Lens systems in our sample.}
\begin{tabular}{ccccccc}
\hline
\hline
\specialrule{0.1em}{0em}{0.5em}
\vspace{1mm}
Object & $z_{\rm s}$ & $z_{\rm l}$\tablefootmark{a} & R.A. & Dec. &  Distance & Optical/UV Photometry\\
\hline\\
\vspace{1mm}
RXJ1131-1231 & 0.66 & 0.30 & 172.9644 & -12.5329 & 3922.4 & \citealp{Sluse_2006}\\
\vspace{1mm}
DESJ2038-4008 & 0.78 & 0.23 & 309.5113 & -40.1371 & 4840.7 & \citealp{2018MNRAS.479.4345A}\\
\vspace{1mm}
SDSSJ1251+2935 & 0.80 & 0.40 & 192.7815 & 29.5946 & 5032.9 & -\\
\vspace{1mm}
GRAL1131-4419 & 1.09 & 0.47 & 172.7502 & -44.3332 & 7350.1 & -\\
\vspace{1mm}
HE1113-0641 & 1.24 & 0.70 & 169.0981 & -6.9608 & 8578.4 & -\\
\vspace{1mm}
WISE2344-3056 & 1.30 & 0.5 & 356.0706 & -30.9406 & 9123.0 & This work\\
\vspace{1mm}
J0607-2152 & 1.30 & 0.56 & 91.7954 & -21.8716 & 9183.9 & -\\
\vspace{1mm}
SDSSJ0924+0219 & 1.52 & 0.39 & 141.2325 & 2.3236 & 11115.0 & \citealp{2009MNRAS.398..233F}\\
\vspace{1mm}
J2145+6345 & 1.56 & 0.5 & 326.2713 & 63.7614 & 11449.0 & -\\
\vspace{1mm}
WFI2033-4723 & 1.66 & 0.66 & 308.4257 & -47.3956 & 12378.2 & \citealp{2004AJ....127.2617M}\\
\vspace{1mm}
PSJ1606-2333 & 1.70 & 0.92 & 241.5009 & -23.5561 & 12635.4 & \citealp{2018MNRAS.479.5060L}\\
\vspace{1mm}
HE0435-1223 & 1.69 & 0.45 & 69.5619 & -12.2875 & 12663.0 & \citealp{2006ApJ...640...47K}\\
\vspace{1mm}
DESJ0405-3308 & 1.70 & 0.5 & 61.4988 & -33.1475 & 12847.3 & \citealp{2018MNRAS.480.5017A}\\
\vspace{1mm}
J1537-3010 & 1.71 & 0.59 & 234.3556 & -30.1713 & 12911.9 & \citealp{2019MNRAS.483.4242L}\\
\vspace{1mm}
J2017+6204 & 1.72 & 0.5 & 304.4544 & 62.0787 & 12911.9 & -\\
\vspace{1mm}
PG1115+080 & 1.71 & 0.31 & 169.5704 & 7.7663 & 13050.6 & \citealp{2008ApJ...689..755M}\\
\vspace{1mm}
WISEJ0259-1635 & 2.16 & 0.91 & 44.9286 & -16.5953 & 17069.5 & This work\\
\vspace{1mm}
WFI2026-4536 & 2.23 & 0.5 & 306.5435 & -45.6075 & 17746.5 & \citealp{2004AJ....127.2617M}\\
\vspace{1mm}
J0608+4229 & 2.35 & 0.5 & 92.1725 & 42.4935 & 18818.1 & -\\
\vspace{1mm}
PSJ0147+4630 & 2.38 & 0.68 & 26.7923 & 46.5118 & 19180.6 & 
\citealp{2017ApJ...844...90B}\\
\vspace{1mm}
SDSSJ0248+1913 & 2.44 & 0.5 & 42.2031 & 19.2253 & 19800.0 & -\\
\vspace{1mm}
J1042+1641 & 2.50 & 0.59 & 160.5921 & 16.6876 & 20560.8 & This work\\
\vspace{1mm}
H1413+117 & 2.56 & 1.15 & 213.9427 & 11.4954 & 20987.3 & \citealp{2011ApJ...742...67M}\\
\vspace{1mm}
MG0414+0534 & 2.64 & 0.96 & 63.6572 & 5.5786 & 21784.2 & -\\
\vspace{1mm}
2M1134-2103 & 2.77 & 0.66 & 173.6688 & -21.0563 & 23087.4 & \citealp{2019MNRAS.486.4987R}\\
\vspace{1mm}
J0803+3908 & 2.97 & 0.5 & 120.9905 & 39.1398 & 25111.6 & -\\
\vspace{1mm}
J0659+1629 & 3.10 & 0.77 & 104.7670 & 16.4859 & 26265.0 & This work\\
\hline
\end{tabular}
\tablefoot{We have listed the lensed AGN systems along with their coordinates (in degrees, J2000), redshifts of the source and the lens, distances in Mpc, and reference for the optical/UV photometry. For sources with optical/UV photometry extracted in  this work, we used HST observations in the F475X and F814W bands to calculate the AGN fluxes (described in Sect. \ref{sect:uvo}).\\
\tablefoottext{a}{For systems without any estimate for the lens redshift, we assumed $z_{\rm l} = 0.5$ (to be consistent with \citealp{2025arXiv251107765K} and \citealp{2025arXiv251107513G}).}}
\label{tab:src_list}
\end{table*}


\section{Sample and data}\label{sect:sampledata}

\subsection{Sample}\label{sect:sample}

Our parent sample consists of 31 quadruply lensed AGN systems from the JWST GO-2046 proposal (PI Nierenberg). As part of the JWST lensed quasar dark matter survey (\citealp{2024MNRAS.530.2960N}), these sources were observed with JWST/MIRI. For details about the observations, see \cite{2024MNRAS.535.1652K,2025arXiv251107765K}. As mentioned in \cite{2025arXiv251107765K}, the lensed system J0457-7820 is a triplet, and the system J2107-1611 is doubly imaged. Therefore, we do not include these sources in our analysis, and moving forward, we work with 29 sources.

For the 29 quadruply lensed quasars in the original sample, we checked archival data and previous publications to compile all available optical, UV, and X-ray observations. Due to the difference in the availability of multiband photometric and spectroscopic data for the sources in our sample, we used various strategies to optimally utilize the available data and obtain the desired products. For a total of 17/29 (59\%) sources, we were able to combine measurements from the literature (in optical/UV) and archival observations (for X-ray and optical/UV) to construct their broadband SEDs. For the remaining 10/29 (34\%) sources, with limited optical/UV photometry ($< 3$ bands) or lacking X-ray observations, we were unable to model their broadband SEDs accurately (see Sect. \ref{sect:sed} for details). In such cases, we used optical or X-ray bolometric corrections to estimate $L_{\rm bol}$. In Sect. \ref{sect:kx}, we discuss in detail why this is the best approach to calculate $L_{\rm bol}$ for such sources. For the two remaining sources (B2045+265 and J2205-3727), we did not have sufficient data to include them in this multiwavelength study. Hence, our final sample consists of 27 lensed quasars that are listed in Table \ref{tab:src_list}, along with their lens and source redshifts (for details see \citealp{2025arXiv251107513G}), coordinates, and distances. The image naming conventions for these quasars are shown in the appendix (Fig. \ref{fig:lens_image}).

We also compiled rest-frame optical/UV spectra for 19/27 (70\%) systems to estimate their black hole masses using the virial method (Sect. \ref{sect:uvo_spec}) and further calculate their Eddington ratios ($\lambda_{\rm Edd} = L_{\rm bol}/L_{\rm Edd}$; $L_{\rm Edd} = 1.5\times10^{38} \times \frac{M_{\rm BH}}{M_{\odot}}\,\rm erg\,s^{-1}$).


\subsection{X-ray data}\label{sect:xray}

To compile the publicly available X-ray observations for our lensed AGN sample, we used the HEASARC archive\footnote{\url{https://heasarc.gsfc.nasa.gov/cgi-bin/W3Browse/w3browse.pl}}. For this study, we only used observations with the Advanced CCD Imaging Spectrometer (ACIS; \citealp{2003SPIE.4851...28G}) on board the \textit{Chandra} X-ray Observatory\footnote{\url{https://chandra.harvard.edu/}}, due to its high angular resolution ($\sim 0.5"$). This is important because unlike nonlensed AGN, lensed AGN have multiple images that need to be distinguishable on the sky to precisely extract and model their X-ray spectra. We recovered X-ray observations for 26/27 ($\sim$ 96\%) sources (listed in Table \ref{tab:xray_obs}). In case of multiple observations for the same source, we used the one with the largest exposure time to have a significant number of photons in the spectrum.

We used the official software package developed by the Chandra X-Ray Center (CXC) for data analysis: CIAO\footnote{\url{https://cxc.cfa.harvard.edu/ciao/}} (Chandra Interactive Analysis of Observations; \citealp{2006SPIE.6270E..1VF}) v4.15 with CALDB v4.10.4, to reduce the X-ray observations and extract the X-ray spectra. For the quadruply lensed AGN in our sample, the images have separations ranging from 0.3$"$ to 5$"$ (see Fig. \ref{fig:lens_image} for reference). Whenever possible, we tried to get the spectra for all the four lensed images. However, in cases where two or more lensed images were within $\sim 1.5"$ ($\sim 3$ pixels) of each other and could not be resolved, we worked with unresolved images. Additionally, for some sources with low exposure times ($<5\,\rm ks$), it was impossible to distinguish individual lensed images due to low counts, so we could only extract one spectrum for such systems. For 7/26 (27\%) sources, we successfully extracted the X-ray spectra for all images; for 10/26 (38\%) sources, only one spectrum per source was extracted, as the lens system was unresolved; for the remaining 9/26 (35\%) sources, we had a combination of resolved and unresolved images (see Table \ref{tab:xray_obs} for details).

We followed the steps prescribed by the Chandra analysis guide\footnote{\url{https://cxc.cfa.harvard.edu/ciao/guides/index.html}} to extract the X-ray spectra. We first used the \texttt{chandra\_repro} command to reprocess the downloaded observations. The reprocessed event files were then used to create circular source regions around each resolved image (whenever possible) so that there was no contamination from other images. Finally, we used the \texttt{specextract} command to extract the source spectra (.pi) in the specified regions, along with their ancillary (.arf) and response (.rmf) files. A similar procedure was followed in the case of unresolved images for certain lens systems, as well as for completely unresolved systems. All extracted spectra were rebinned to include at least one count per bin.


\begin{table}
\centering
\caption{X-ray observations for the lens systems in our sample.}
\begin{tabular}{cccc}
\hline
\hline
\specialrule{0.1em}{0em}{0.5em}
\vspace{1mm}
Object & Obs. ID & Exposure & Resolved Images\\
\hline\\
\vspace{1mm}
RXJ1131 & 4814 & 10180 & All\\
\vspace{1mm}
DESJ2038 & 21440 & 6050 & All\\
\vspace{1mm}
SDSSJ1251 & 22016 & 1610 & Unresolved\\
\vspace{1mm}
GRAL1131 & 25405 & 29710 & A, B, C+D\\
\vspace{1mm}
HE1113 & 14961 & 29560 & Unresolved\\
\vspace{1mm}
WISE2344 & 22017 & 1610 & Unresolved\\
\vspace{1mm}
J0607 & 25408 & 5380 & A, B+C, D\\
\vspace{1mm}
SDSSJ0924 & 11564 & 22160 & A+C+D, B\\
\vspace{1mm}
J2145 & 28209 & 10080 & A, B, C+D\\
\vspace{1mm}
WFI2033 & 5603 & 15620 & A1+A2, B, C\\
\vspace{1mm}
PSJ1606 & 20490 & 30110 & Unresolved\\
\vspace{1mm}
HE0435 & 14489 & 38060 & All\\
\vspace{1mm}
DESJ0405 & 26343 & 13950 & Unresolved\\
\vspace{1mm}
J1537 & 23828 & 23050 & All\\
\vspace{1mm}
PG1115 & 363 & 26830 & A, B, C+D\\
\vspace{1mm}
WISEJ0259 & 21444 & 25090 & A+D, B, C\\
\vspace{1mm}
WFI2026 & 7758 & 10170 & A1+A2+C, B\\
\vspace{1mm}
J0608 & 25407 & 4960 & Unresolved\\
\vspace{1mm}
PSJ0147 & 21443 & 10660 & All\\
\vspace{1mm}
SDSSJ0248 & 23824 & 4950 & Unresolved\\
\vspace{1mm}
J1042 & 23135 & 24490 & Unresolved\\
\vspace{1mm}
H1413 & 5645 & 90040 & Unresolved\\
\vspace{1mm}
MG0414 & 12800 & 30050 & A+B, C, D\\
\vspace{1mm}
2M1134 & 21442 & 30070 & All\\
\vspace{1mm}
J0803 & 29168 & 11760 & Unresolved\\
\vspace{1mm}
J0659 & 23825 & 15070 & All\\
\hline
\end{tabular}
\tablefoot{We have listed the observation IDs and exposure times (in seconds) for the Chandra X-ray observations for the lens systems in our sample. In the last column, we report the images for each system that were resolved in these observations.}
\label{tab:xray_obs}
\end{table}


\subsection{Optical/UV data} \label{sect:uvo}

For optical/UV photometry, we mainly relied on published data from the literature. Primary preference was given to studies that reported flux estimates in at least three bands for all resolved images for each system. A minimum of three bands were needed to eventually construct statistically reliable SEDs (see Sect. \ref{sect:sed} for more details). This was available for 13/27 (48\%) sources in our sample (see Table \ref{tab:src_list}). For 4/27 (15\%) sources\footnote{HST flux measurements for these sources were published in \cite{2023MNRAS.518.1260S}, where the lens models were optimized for time delay cosmography and the noise of the PSF was boosted to avoid overfitting in the PSF reconstruction. Obtaining accurate point source fluxes was not a goal of that paper, but it is important for this work. Therefore, we did not use those values and re-measured the fluxes ourselves using a procedure optimized for point source flux measurements.}, we measured the AGN fluxes from archival observations\footnote{Cycles 25 and 26 programs HST-GO-15320 and HST-GO-15652} with the Wide Field Camera 3 (WFC3) of the \textit{Hubble} Space Telescope (HST), in the F475X and F814W bands. The third photometric point was estimated from the available spectroscopy for these systems (listed in Table \ref{tab:uvo_spectra}). We calculated the third band flux in the wavelength range $6065\,\AA-7062\,\AA$, to complement the fluxes in HST bands F475X and F814W, and to create a continuous SED.

The AGN fluxes in the two HST bands were calculated by performing lens/source decomposition with four Moffat profiles and a S\'ersic profile, to fit the four lensed images and the lensing galaxy, respectively. All lensed quasars in our sample are optically selected and hence significantly brighter than their host galaxies. For three of the lenses, we did not detect any host light in the fit residuals, and for one system, the host contribution was as low as 10\%. So we fit the four point-like lensed quasar images with only a Moffat profile. The input parameters for the Moffat profile include the coordinates and intensity of the source (lensed image), width and slope of the profile. In all four lensed systems, the input value of the model parameters was estimated by fitting one lensed image at a time, in order of decreasing brightness. For each system, we started by fitting the brightest lensed image with a Moffat profile. After obtaining a good fit for the first image, the corresponding fit parameters were fixed to their best-fit values, and a second Moffat profile was added to the model to fit the second brightest lensed image. This procedure was repeated for every consecutively faint image and finally for the lensing galaxy, which was fitted using a S\'ersic profile. In the final step, the best-fit values obtained so far for the five profiles (four Moffats and one S\'ersic) were used as input values for the model parameters (and left free to vary), and all five sources (four lensed images and one lensing galaxy) were fit simultaneously to obtain the best-fit model for the entire lensed system. All residuals were visually inspected, and the least-squares statistics was employed to obtain the best-fit model. The lensed quasar fluxes thus obtained (listed in Table \ref{tab:hst}) were used to construct the optical/UV SEDs, which were used to complement the X-ray spectra and perform a broadband SED fit for these four systems (Sect. \ref{sect:sed}). For the remaining 10/27 (37\%) sources, there was no useful multiband photometry available.

To estimate the black hole masses of the lensed AGN using the virial method (discussed in Sects. \ref{sect:uvo_spec} and \ref{sect:mbh_er}), we also compiled the rest-frame optical/UV spectra for 19/27 (70\%) systems from various astronomical facilities (see Table \ref{tab:uvo_spectra} for details). For 12/19 (63\%) systems, we acquired spectra from $\sim1500\,\AA$ to $\sim3500\,\AA$ rest-frame, which included the Mg{\scriptsize\,II}$\,\lambda2798$ broad emission line. For 5/19 (26\%) systems, we used the broad C{\scriptsize\,IV}$\,\lambda1549$ from their rest-frame spectra in the wavelength range $\sim1000\,\AA$ to $\sim2500\,\AA$. Finally, for 2/19 (11\%) sources with spectra available at longer wavelengths ($\sim4000\,\AA$ to $\sim7000\,\AA$ rest-frame), we used the H$\alpha$ line to estimate their black hole mass.


\begin{table}
\setlength{\tabcolsep}{0.3\tabcolsep}
\centering
\caption{Rest-frame optical and UV spectroscopy for the lens systems in our sample.}
\begin{tabular}{cccc}
\hline
\hline
\specialrule{0.1em}{0em}{0.5em}
\vspace{1mm}
Object & Telescope/Instrument & Emission Line & Images\\
\hline\\
\vspace{1mm}
DESJ2038 & VLT/MUSE & Mg{\scriptsize\,II} & All\\
\vspace{1mm}
WISE2344 & VLT/XSHOOTER & Mg{\scriptsize\,II} & A+D, B+C\\
\vspace{1mm}
J0607 & Keck/LRIS & Mg{\scriptsize\,II} & A, D\\
\vspace{1mm}
SDSSJ0924 & Keck/ESI & Mg{\scriptsize\,II} & A, B\\
\vspace{1mm}
WFI2033 & VLT/MUSE & Mg{\scriptsize\,II} & All\\
\vspace{1mm}
PSJ1606 & VLT/MUSE & Mg{\scriptsize\,II} & All\\
\vspace{1mm}
HE0435 & VLT/FORS1 & Mg{\scriptsize\,II} & B, D\\
\vspace{1mm}
DESJ0405 & Magellan/IMACS & C{\scriptsize\,IV} & B, C\\
\vspace{1mm}
J1537 & VLT/MUSE & Mg{\scriptsize\,II} & All\\
\vspace{1mm}
J2017 & Keck/LRIS & Mg{\scriptsize\,II} & B\\
\vspace{1mm}
PG1115 & VLT/MUSE & Mg{\scriptsize\,II} & All\\
\vspace{1mm}
WISEJ0259 & VLT/XSHOOTER & Mg{\scriptsize\,II} & A+C, B+D\\
\vspace{1mm}
WFI2026 & VLT/FORS1 & C{\scriptsize\,IV} & A1+A2, B\\
\vspace{1mm}
J0608 & Keck/LRIS & C{\scriptsize\,IV} & Unresolved\\
\vspace{1mm}
PSJ0147 & GTC/EMIR & H$\alpha$ & B, C, D\\
\vspace{1mm}
SDSSJ0248 & Keck/ESI & Mg{\scriptsize\,II} & B, C\\
\vspace{1mm}
H1413 & VLT/SINFONI & H$\,\alpha$ & All\\
\vspace{1mm}
2M1134 & VLT/MUSE & C{\scriptsize\,IV} & All\\
\vspace{1mm}
J0659 & Keck/LRIS & C{\scriptsize\,IV} & All\\
\hline
\end{tabular}
\tablefoot{We have listed the telescope and instrument pair used to obtain the optical/UV spectroscopy for the lens systems in our sample. We also report the emission line used to estimate the black hole mass of the system. The last column indicates the individual lensed images (or pair of images in case of compact systems) for which the spectra were available.}
\label{tab:uvo_spectra}
\end{table}


\section{Data analysis}\label{sect:data}

To calculate the bolometric luminosity of lensed AGN, we first need to obtain their intrinsic X-ray and optical/UV luminosity corresponding to the coronal and disk emission, respectively. In Sect. \ref{sect:xray_spec}, we describe the procedure we followed to fit the X-ray spectra and obtain the X-ray luminosity. Sect. \ref{sect:sed} discusses the method used to convert the optical/UV magnitudes into the total emission from the accretion disk. Finally, in Sect. \ref{sect:uvo_spec} we derive the black hole mass from optical/UV spectroscopy for each system. 


\subsection{X-ray spectral fitting}\label{sect:xray_spec}

To fit the 0.3--10\,keV X-ray spectra for our lensed AGN systems, we used the X-ray spectral fitting package \textsc{XSPEC} (\citealp{1996ASPC..101...17A}). We extracted the X-ray spectra for individual lensed images (whenever possible) and for unresolved images for a total of 26 systems (see Sect.  \ref{sect:xray}). We used a simple powerlaw (\texttt{POWERLAW}) model with Galactic absorption (\texttt{TBABS}) to fit these spectra. The column density ($N_{\rm H}$) for Galactic absorption was fixed at the value obtained at the coordinates of the source from the HI 4-PI Survey (HI4PI; \citealp{2016A&A...594A.116H}), a 21-cm all-sky survey of neutral atomic Hydrogen. The spectra were fit using Cash statistics (C-stat; \citealp{1979ApJ...228..939C}). The best-fit values of all free parameters (photon index: $\Gamma$ and normalization: $K_{\rm x}$) were checked using the \texttt{steppar} command in \textsc{XSPEC} to avoid fits corresponding to local minima. Additionally, the 90\% confidence intervals for all parameters were estimated using the \texttt{error} command in \textsc{XSPEC}. All the fit residuals were visually inspected and their goodness was checked using the reduced chi-square value.

Based on the best fit models, we calculated the total intrinsic (rest-frame and absorption corrected) X-ray flux ($F_{\rm x}$) in the 0.1 to 500\,keV energy range for all our lensed AGN (resolved and unresolved). The uncertainties in the flux were calculated using the upper and lower limits of $K_{\rm x}$. The corresponding luminosities were calculated using the distances listed in Table \ref{tab:src_list}. Finally, all X-ray luminosities were corrected for lensing macro magnifications (listed in Table \ref{tab:params}). This final magnification corrected estimate of the total X-ray luminosity ($L_{\rm x}$) will be later combined with the optical/UV disk luminosity, to estimate the bolometric luminosity of the 17 lensed AGN with multiband optical/UV photometry. For the remaining nine systems, we calculated their intrinsic, magnification corrected luminosities in the 2--10\,keV energy band ($L_{2-10}$) to calculate the bolometric luminosity using the 2--10\,keV bolometric correction ($L_{\rm bol} = \kappa_{2-10} \times L_{2-10}$; see Sect. \ref{sect:lbol}). We provide a detailed discussion and specific tests in Sect. \ref{sect:kx} to justify this approach.


\begin{figure*}
  \begin{subfigure}[t]{0.5\textwidth}
    \centering
    \includegraphics[width=\textwidth]{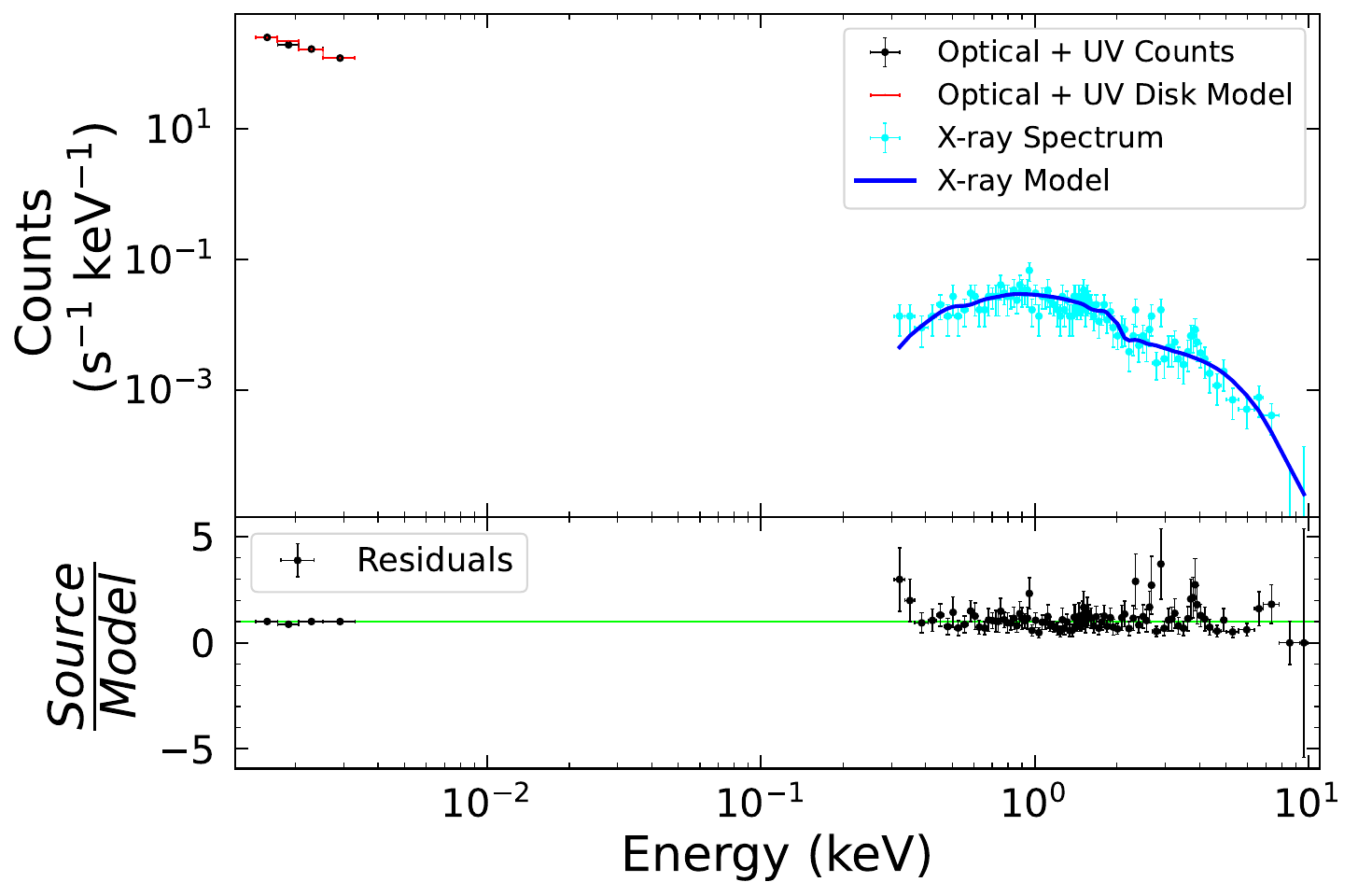} 
    \caption{}        
    \label{fig:1131A}
  \end{subfigure}
  \hspace{-2mm}
  \begin{subfigure}[t]{0.5\textwidth}
    \centering
    \includegraphics[width=\textwidth]{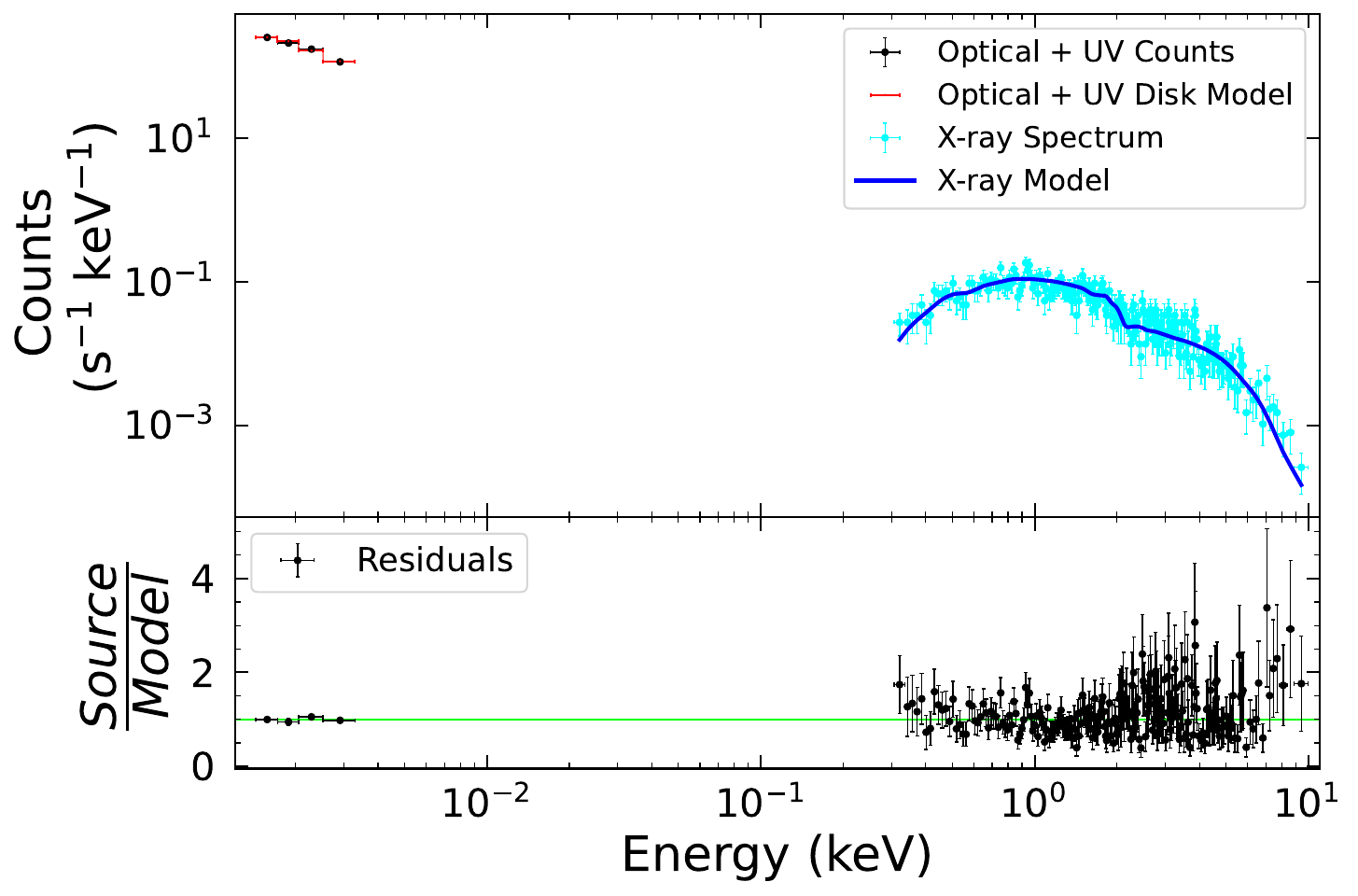}
    \caption{}
    \label{fig:1131B}
  \end{subfigure}

  \begin{subfigure}[t]{0.5\textwidth}
    \centering
    \includegraphics[width=\textwidth]{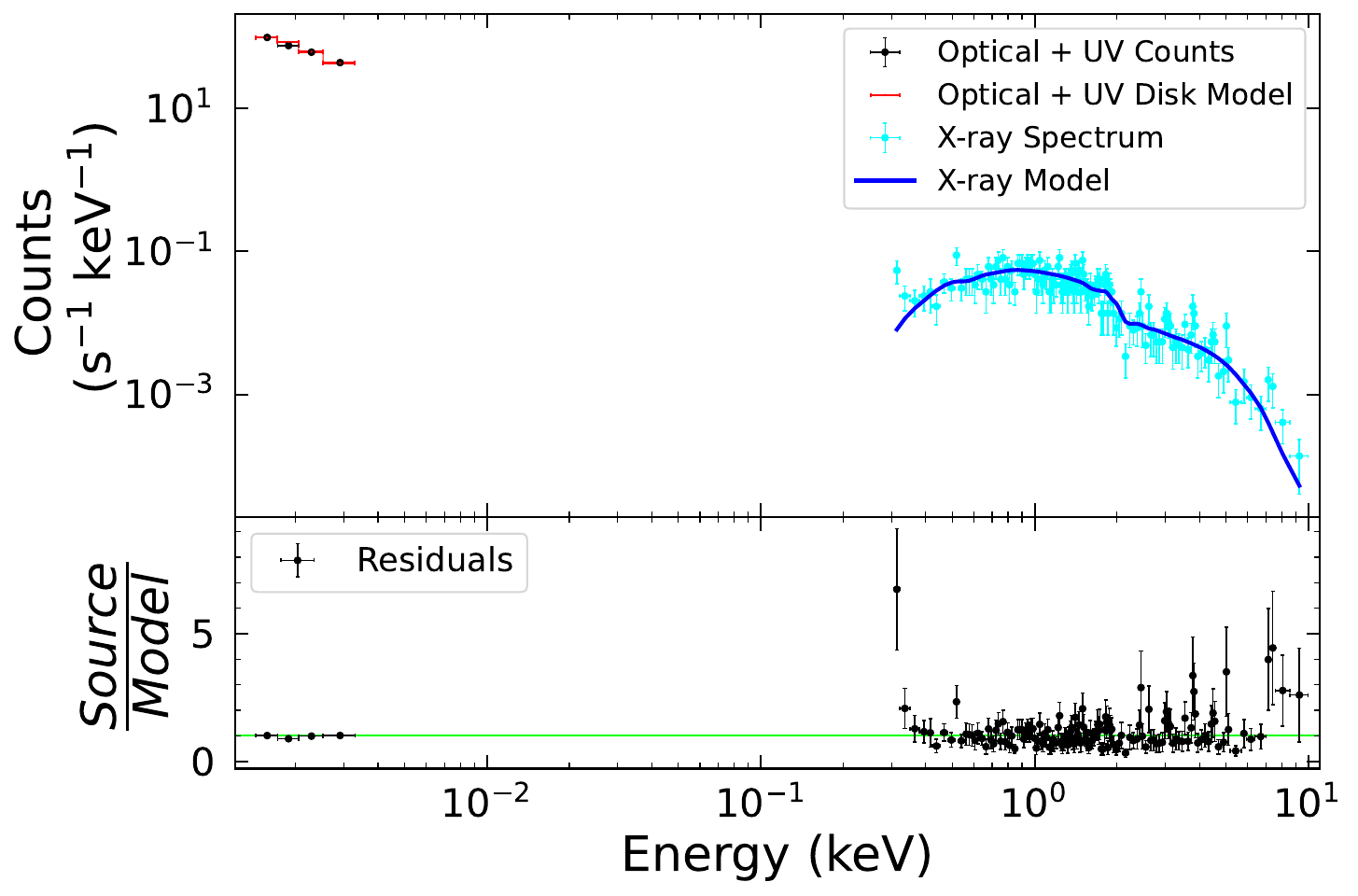}
    \caption{}
    \label{fig:113C}
  \end{subfigure}
  \begin{subfigure}[t]{0.5\textwidth}
    \centering
    \includegraphics[width=\textwidth]{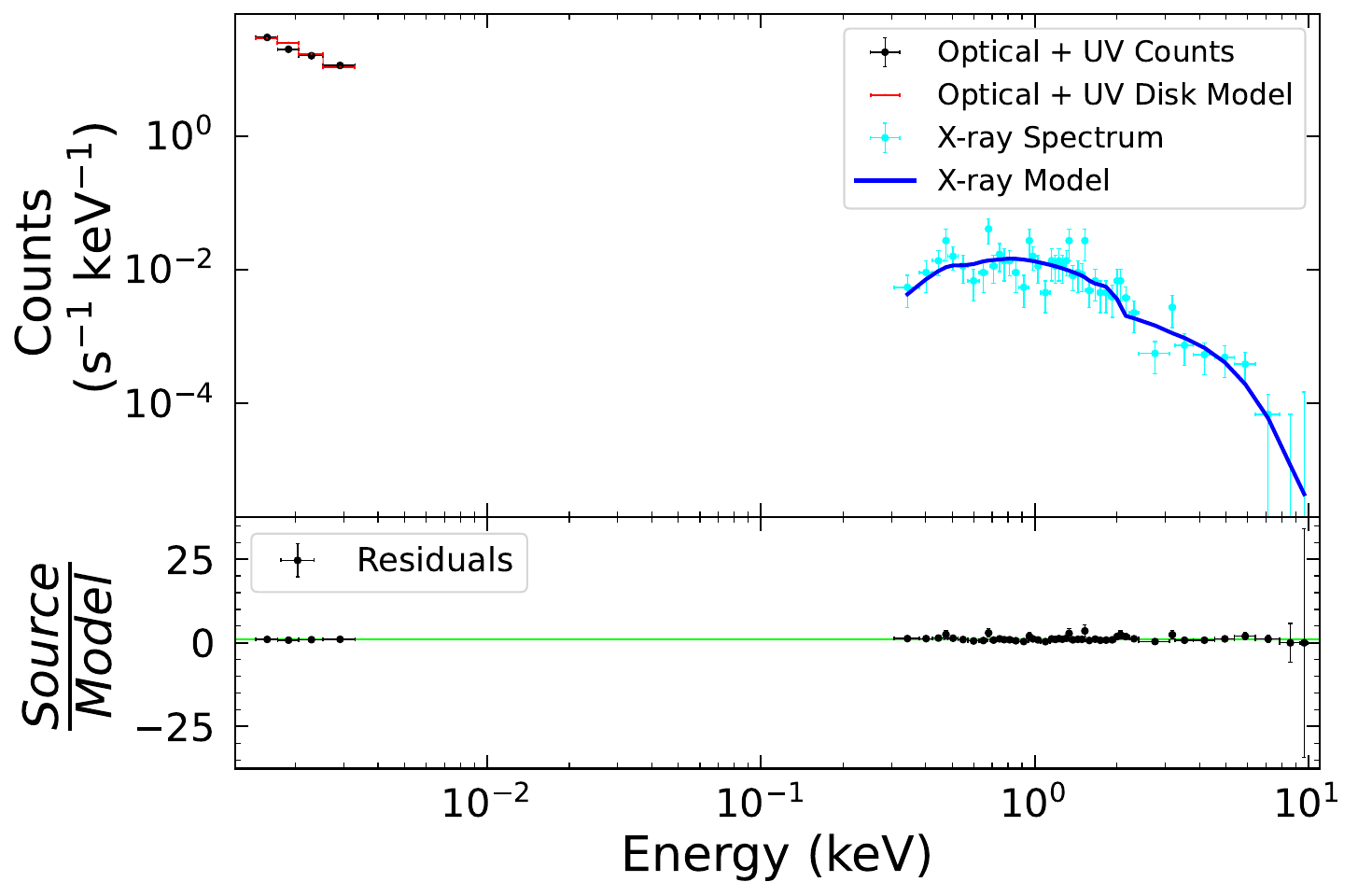}
    \caption{}
    \label{fig:1131D}
  \end{subfigure}

\caption{An example of the optical-to-X-ray SED fitting of the four resolved images of the lensed system RXJ1131. Each of the individual panels show the counts in the optical/UV filters (black points) and the 0.3–10 keV X-ray spectrum (cyan points). They also show the best-fit obtained using the optical/UV disk model including dust extinction (in red) and the X-ray powerlaw model with Galactic absorption (in blue). The fit residuals are shown in the bottom of each panel.}
\label{fig:sed}
\end{figure*}


\subsection{Optical/UV SED fitting}\label{sect:sed}

We used the optical and UV magnitudes compiled for each system in Sect. \ref{sect:uvo} to populate the optical/UV part of their SEDs. The optical/UV SEDs were fitted in \textsc{XSPEC} using the \texttt{DISKPN} model, which is a thermal accretion disk model comprising multiple blackbody components (e.g., \citealp{1984PASJ...36..741M}; \citealp{1986ApJ...308..635M}). This model follows the assumptions of the standard \citet{1973A&A....24..337S} disk around an accreting black hole. Although this model is known to have its own limitations, mainly for sources with higher and lower accretion rates (see \citealp{1999ASPC..161..295B} for a review), it can still successfully reproduce the observed optical/UV emission of most AGN (e.g., \citealp{2015MNRAS.446.3427C,2016MNRAS.460..212C,2024A&A...691A.203G}). Moreover, because of the heterogeneity\footnote{We checked if the heterogeneity in the methods used to calculate the AGN fluxes in the literature affects our SED fits and outputs. We found that our SED fits are not sensitive to errors as low as 0.5\,mag in the AGN fluxes, which is the extent of differences in the AGN photometry estimated using different methods.} in the multiband optical/UV photometry available for our sample, we avoid using more complex disk models to fit the SEDs and instead follow a simple but consistent approach throughout our sample. This approach makes it possible to compare results across our sample in a consistent way. The free parameters for \texttt{DISKPN} include the normalization ($K_{\rm uvo}$) and the maximum temperature of the disk ($kT_{\rm max}$; in units of keV). The inner radius of the disk ($R_{\rm in}$; in units of $R_{\rm g} = GM/c^2$) is fixed to 6.0 gravitational radii\footnote{\url{https://heasarc.gsfc.nasa.gov/xanadu/xspec/manual/node171.html}}. We use the convolution model \texttt{ZASHIFT} to shift the disk spectrum to the redshift of the respective source. In addition to the disk model, we include a component of the \texttt{ZDUST} model to take into account the extinction due to dust grains (\citealp{1992ApJ...395..130P}) in the optical and UV bands, associated with the Milky Way (\texttt{ZDUST}). For this reddening component, we adopt a Milky Way extinction curve with $R_{\rm v} = 3.08$. We fix the $E(B-V)$ value to those estimated by \cite{1998ApJ...500..525S}, who combined the results of IRAS and COBE/DIRBE to create a 100-micron intensity map of the sky.

Based on the number of free parameters in our model, we require optical/UV photometry in at least three bands to obtain a statistically reliable SED fit. These data were available for 17 systems in our sample (see Sect. \ref{sect:uvo}). We fitted the optical/UV SEDs of all four lensed images for these systems with the model described above. The goodness of fit was checked using the reduced chi-square value ($\sim1$) and all residuals were visually inspected. The best-fit values of all free parameters were checked using the \texttt{steppar} command to avoid fits corresponding to local minima. The errors in the free parameters ($kT_{\rm max}$ and $K_{\rm uvo}$) were calculated using the \texttt{error} command to give 90\% confidence intervals. An example of our SED fit, along with the best-fit model and residual, is shown in Fig. \ref{fig:sed}. The final best fits were used to calculate the intrinsic (rest-frame and absorption-corrected) model fluxes and luminosities. The upper and lower limits of these fluxes were calculated considering the upper and lower bounds of $K_{\rm uvo}$. The total intrinsic optical $+$ UV luminosity was estimated in the $1000\mu$m to 0.1\,keV range, since the standard \cite{1973A&A....24..337S} disk typically has very little to no emission in the X-rays. All luminosities were corrected for the respective lens magnifications (listed in Table \ref{tab:params}). The corrected disk luminosity ($L_{\rm uvo}$) was combined with $L_{\rm x}$ (see Sect. \ref{sect:xray_spec}) to estimate the total accretion luminosities ($L_{\rm bol} = L_{\rm uvo}+L_{\rm x}$) of the 17 lensed AGN in our sample with optical-to-X-ray data. 


\subsection{Optical/UV spectral fitting}\label{sect:uvo_spec}

In this section, we describe the spectral emission line fitting method employed to estimate the black hole masses of the 19 lensed AGN systems with good-quality spectra in our sample. To fit and calculate the width of the available broad emission line in the rest-frame optical/UV spectra, we used a Levenberg-Marquardt optimization implemented in the \texttt{mpfit} method \citep{2009ASPC..411..251M} of the Python package \texttt{pyspeckit}\footnote{\url{https://pyspeckit.readthedocs.io/en/latest/}} (\citealp{2011ascl.soft09001G,2022AJ....163..291G}). We followed the step-by-step procedure described by \citet{2016MNRAS.460..187M} to perform the spectral emission line fitting. First, the local continuum surrounding the specific emission line was fitted using a single powerlaw. These continuum windows extends from $2650-2670\,\AA$ and $3030-3070\,\AA$ for Mg{\scriptsize\,II}$\,\lambda2798$, from $1420-1460\,\AA$ and $1680-1720\,\AA$ for C{\scriptsize\,IV}$\,\lambda1549$, and from $6150-6250\,\AA$ and $6950-7150\,\AA$ for H$\alpha$. Next, we modeled the blended iron line emission in the spectra using various iron templates. For the optical region around H$\alpha$ ($4000-7000\,\AA$), we used the \citet{1992ApJS...80..109B} template. For the UV region around C{\scriptsize\,IV} and Mg{\scriptsize\,II}, we used the iron templates from \citet{2001ApJS..134....1V} and \citet{2016MNRAS.460..187M} in the wavelength range 1250--3090$\,\AA$ and 2200--3646$\,\AA$, respectively. For the overlapping wavelength range, preference was given to the \citet{2016MNRAS.460..187M} template for consistency with the fitting method. Finally, after subtracting the continuum and iron emissions, the broad emission line of interest was fitted using a Gaussian profile. In the case of visible doublets (for C{\scriptsize\,IV} and Mg{\scriptsize\,II}), two Gaussian profiles were used. In a few cases, a third Gaussian was used to fit an additional narrow emission arising from the narrow line region.

The full width at half maximum (FWHM) of the broad emission lines extracted from the spectral fitting procedure mentioned above is later used to estimate the black hole mass of the lensed AGN systems using the virial method (Sect. \ref{sect:mbh_er}). Additionally, the intrinsic monochromatic luminosity of the AGN continuum ($L_{\lambda}$; where $\lambda$ = 1450$\,\AA$, 3000$\,\AA$, and 5100$\,\AA$ for C{\scriptsize\,IV}, Mg{\scriptsize\,II}, and H$\alpha$, respectively) is also calculated from the spectral fitting. For 3 systems (J0607, J2017, and J0608) with no photometric data in the optical/UV but available spectroscopy, we also estimated the intrinsic $2500\,\AA$ luminosity from their spectra. All the luminosities are corrected for lensing magnification.


\section{Results}\label{sect:results}

In this section, we present the main results of our multiwavelength analysis of 27 quadruply lensed AGN, including the bolometric luminosities (in Sect. \ref{sect:lbol}), and the black hole masses and Eddington ratios (in Sect. \ref{sect:mbh_er}) of our lensed AGN sample.


\begin{table}
\setlength{\tabcolsep}{0.2\tabcolsep}
\centering
\caption{Summary of available data and $L_{\rm bol}$ estimators for our lensed AGN sample.}
\begin{tabular}{ccccccc}
\hline
\hline
\specialrule{0.1em}{0em}{0.5em}
\vspace{1mm}
Object & X-Ray & Optical/UV & Optical/UV & SED & $L_{\rm bol}$\\
& Spectra & Photometry & Spectra & Fits & Estimator\\
\hline\\
\vspace{1mm}
RXJ1131 & \ding{51}  & \ding{51}  & \ding{55} & \ding{51}  & SED fit\\
\vspace{1mm}
DESJ2038 & \ding{51}  & \ding{51}  & \ding{51}  & \ding{51}  & SED fit\\
\vspace{1mm}
SDSSJ1251 & \ding{51}  & \ding{55} & \ding{55} & \ding{55} & $\kappa_{2-10}$\\
\vspace{1mm}
GRAL1131 & \ding{51} & \ding{55} & \ding{55} & \ding{55} & $\kappa_{2-10}$\\
\vspace{1mm}
HE1113 & \ding{51} & \ding{55} & \ding{55} & \ding{55} & $\kappa_{2-10}$\\
\vspace{1mm}
WISE2344 & \ding{51}  & \ding{51}  & \ding{51}  & \ding{51}  & SED fit\\
\vspace{1mm}
J0607 & \ding{51} & \ding{55} & \ding{51} & \ding{55} & $\kappa_{2-10}$\\
\vspace{1mm}
SDSSJ0924 & \ding{51} & \ding{51}  & \ding{51} & \ding{51} & SED fit\\
\vspace{1mm}
J2145 & \ding{51} & \ding{55} & \ding{55} & \ding{55} & $\kappa_{2-10}$\\
\vspace{1mm}
WFI2033 & \ding{51} & \ding{51}  & \ding{51} & \ding{51} & SED fit\\
\vspace{1mm}
PSJ1606 & \ding{51}  & \ding{51}  & \ding{51}  & \ding{51}  & SED fit\\
\vspace{1mm}
HE0435 & \ding{51}  & \ding{51}  & \ding{51} & \ding{51} & SED fit\\
\vspace{1mm}
DESJ0405 & \ding{51}  & \ding{51}  & \ding{51} & \ding{51}  & SED fit\\
\vspace{1mm}
J1537 & \ding{51}  & \ding{51}  & \ding{51}  & \ding{51}  & SED fit\\
\vspace{1mm}
J2017 & \ding{55} & \ding{55} & \ding{51} & \ding{55} & $\kappa_{3000\AA}$\\
\vspace{1mm}
PG1115 & \ding{51} & \ding{51}  & \ding{51}  & \ding{51} & SED fit\\
\vspace{1mm}
WISEJ0259 & \ding{51}  & \ding{51}  & \ding{51}  & \ding{51}  & SED fit\\
\vspace{1mm}
WFI2026 & \ding{51} & \ding{51}  & \ding{51} & \ding{51} & SED fit\\
\vspace{1mm}
J0608 & \ding{51}  & \ding{55} & \ding{51}  & \ding{55} & $\kappa_{2-10}$\\
\vspace{1mm}
PSJ0147 & \ding{51}  & \ding{51}  & \ding{51}  & \ding{51}  & SED fit\\
\vspace{1mm}
SDSSJ0248 & \ding{51}  & \ding{55} & \ding{51} & \ding{55} & $\kappa_{2-10}$\\
\vspace{1mm}
J1042 & \ding{51}  & \ding{51}  & \ding{55} & \ding{51}  & SED fit\\
\vspace{1mm}
H1413 & \ding{51}  & \ding{51}  & \ding{51}  & \ding{51}  & SED fit\\
\vspace{1mm}
MG0414 & \ding{51} & \ding{55} & \ding{55} & \ding{55} & $\kappa_{2-10}$\\
\vspace{1mm}
2M1134 & \ding{51}  & \ding{51}  & \ding{51}  & \ding{51}  & SED fit\\
\vspace{1mm}
J0803 & \ding{51}  & \ding{55} & \ding{55} & \ding{55} & $\kappa_{2-10}$\\
\vspace{1mm}
J0659 & \ding{51}  & \ding{51}  & \ding{51}  & \ding{51}  & SED fit\\

\hline
\end{tabular}
\tablefoot{We provide a final summary of the multi-wavelength photometric and spectroscopic data available/compiled in this study for our 27 lensed AGN system. We also show whether reliable SED fits were obtained for a system and the method used to calculate the bolometric luminosity for each system, i.e., using the SED fits, X-ray bolometric corrections ($\kappa_{2-10}$), or optical boloemtric corrections ($\kappa_{3000\AA}$).}
\label{tab:summ}
\end{table}


\begin{figure*}
  \begin{subfigure}[t]{0.36\textwidth}
    \centering
    \includegraphics[width=\textwidth]{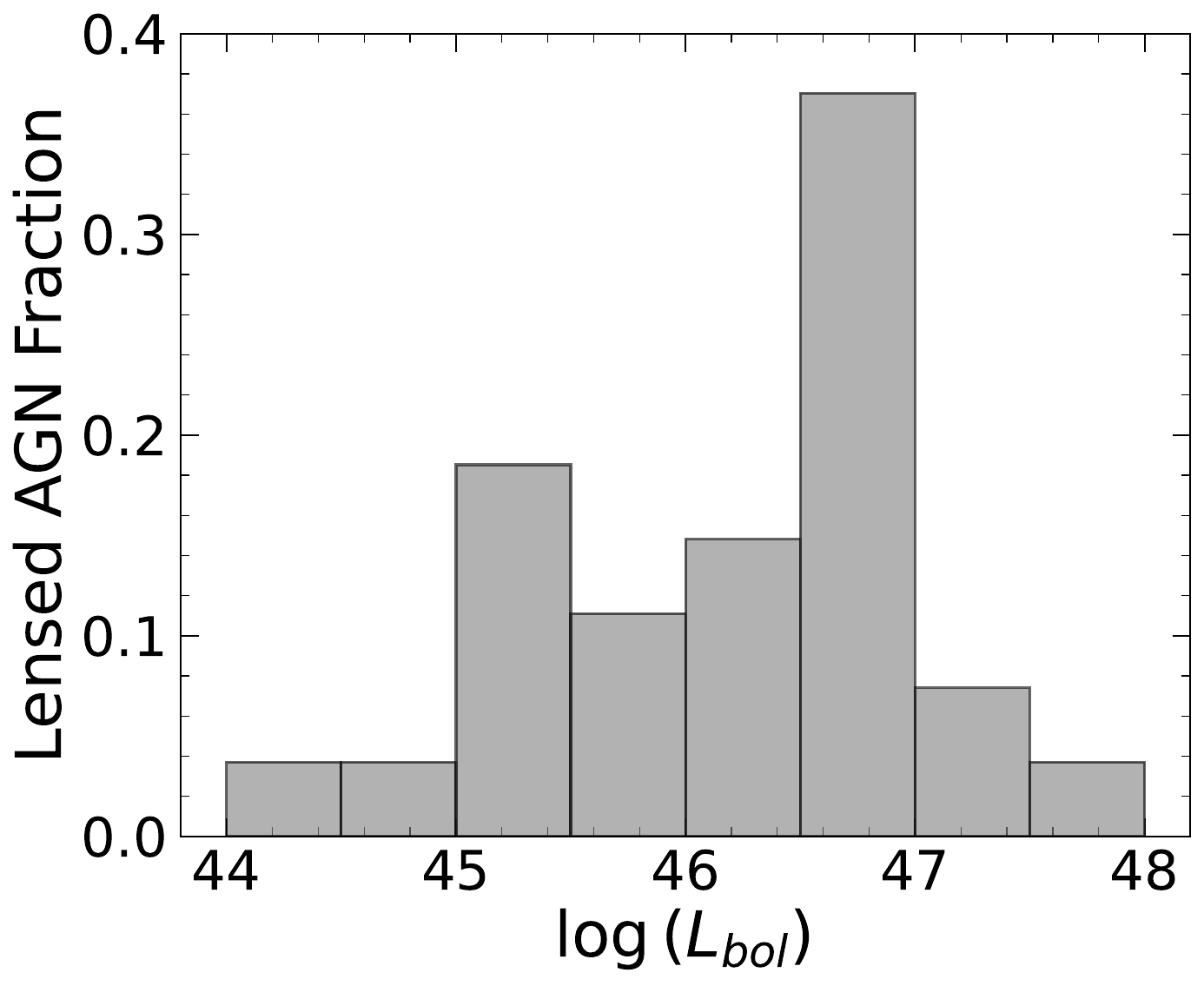} 
    \caption{}        
    \label{fig:lbol}
  \end{subfigure}
  \begin{subfigure}[t]{0.31\textwidth}
    \centering
    \includegraphics[width=\textwidth]{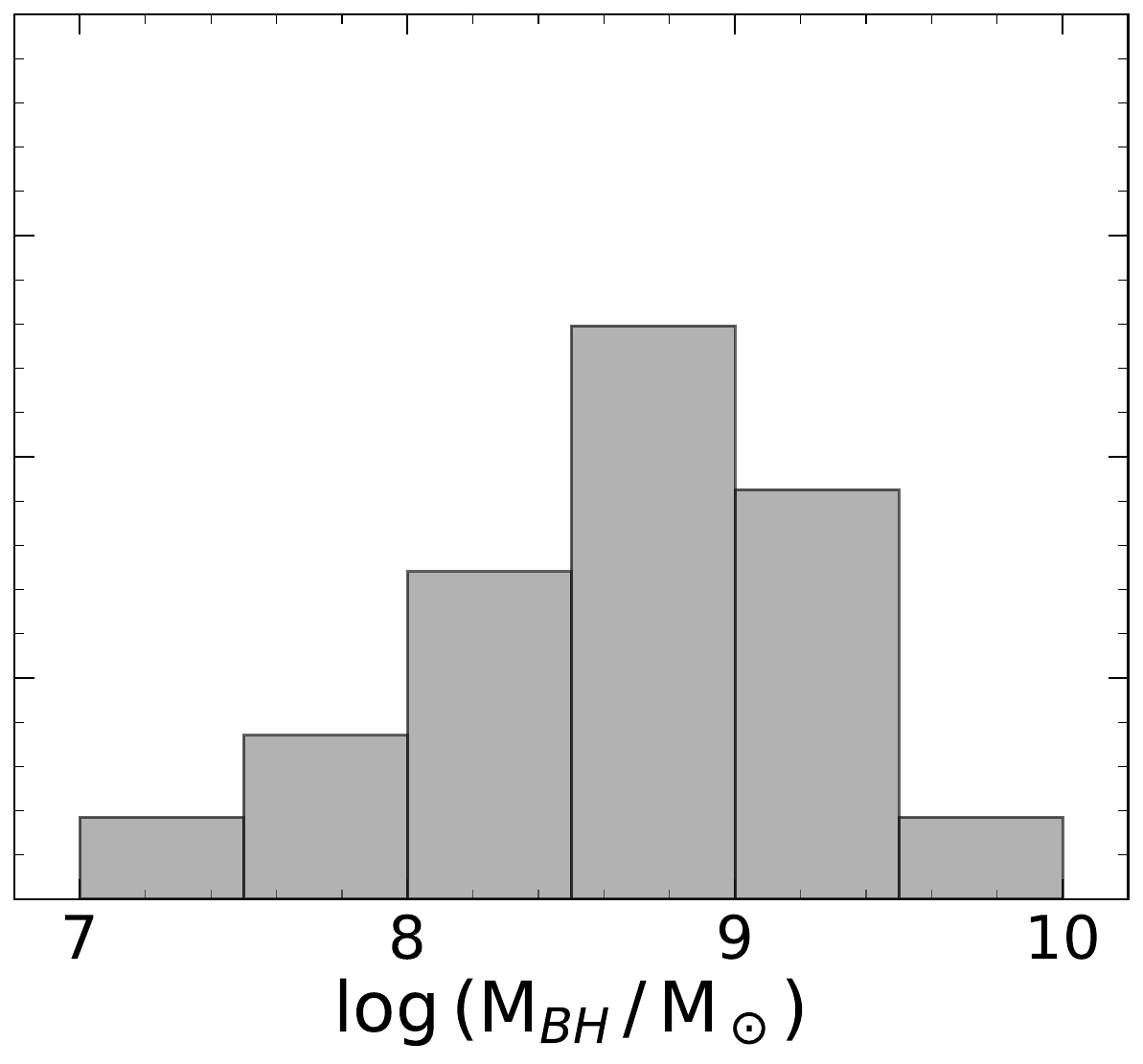}
    \caption{}
    \label{fig:mbh}
  \end{subfigure}
  \hspace{-2mm}
  \begin{subfigure}[t]{0.316\textwidth}
    \centering
    \includegraphics[width=\textwidth]{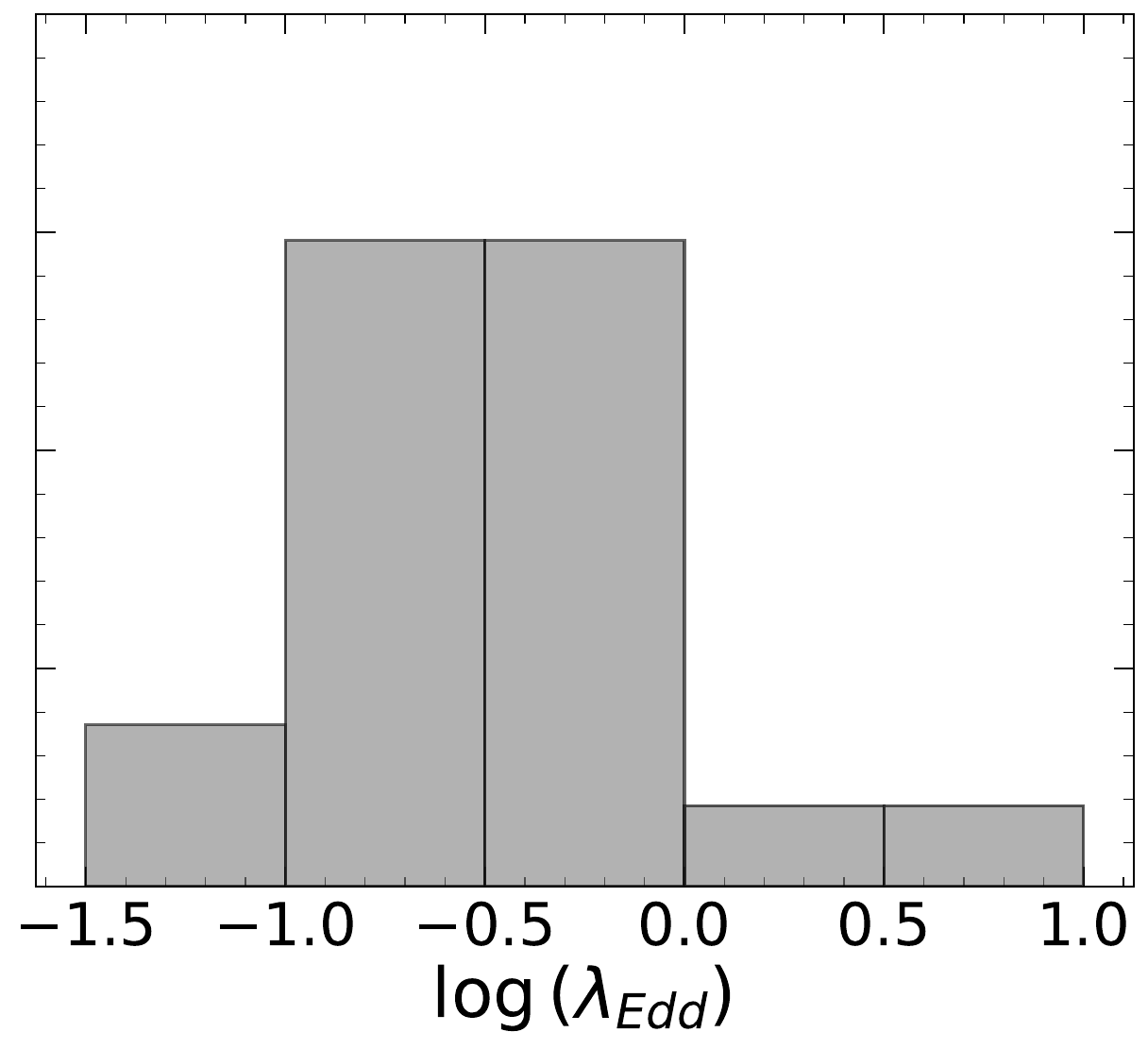}
    \caption{}
    \label{fig:er}
  \end{subfigure}

\caption{Distribution in the (a) bolometric luminosities, (b) black hole masses, and (c) Eddington ratios of our quadruply lensed AGN sample. We have included one measurement per lensed system by averaging over the values for available lensed images for each system.}
\label{fig:prop}
\end{figure*}


\subsection{Bolometric luminosity}\label{sect:lbol}

For 17/27 systems with multiband optical/UV photometry and X-ray spectroscopy, we used their optical-to-X-ray SED fits to calculate the bolometric luminosity. This was done by adding the X-ray luminosity in the 0.1 to 500\,keV energy range ($L_{\rm x}$) and the optical/UV disk luminosity in the $1000\,\mu$m to 0.1\,keV range ($L_{\rm uvo}$). A possible caveat of this method arises from the fact that AGN are known to show variability at different wavelengths with different timescales (e.g., \citealp{1997ARA&A..35..445U}). Since the multiwavelength data used in this study have been compiled from different instruments, they are spread over a few years; specifically, the time difference between the X-ray and optical/UV observations in a few cases is as large as six years. To quantify the impact of using non-simultaneous observations for SED fitting on our final results, we selected the five most variable sources in our sample, based on their optical light curves obtained from the Zwicky Transient Facility\footnote{ZTF: \url{https://irsa.ipac.caltech.edu/Missions/ztf.html}} and redid the SED fitting for their brightest and faintest state. We then calculated the intrinsic luminosities ($L_{\rm x}, L_{\rm uvo}, L_{\rm bol}$) at the two states to determine how much they deviate from our original estimate made using non-simultaneous SEDs. We found that the maximum scatter in $L_{\rm bol}$ due to variability is 0.2\,dex, which is similar to the typical uncertainties we obtained in $L_{\rm bol}$ from the SED modeling. Considering that the scatter of 0.2\,dex is measured for the most variable source of our sample, we can safely assume that for all other sources, any scatter due to variability is much smaller than the existing uncertainties in $L_{\rm bol}$ from the SED fitting.

For 9/27 systems lacking optical/UV photometry, we used X-ray bolometric corrections in the 2--10\,keV ($\kappa_{2-10}$) to estimate their bolometric luminosities from the 2--10\,keV X-ray luminosity ($L_{\rm bol} = \kappa_{2-10}\times L_{2-10}$). Studies have shown that $\kappa_{2-10}$ does not show any significant dependence on redshift (e.g., \citealp{2020A&A...636A..73D,2024A&A...691A.203G}). However, they strongly correlate with the bolometric luminosity, especially for $L_{\rm bol} > 10^{45}\,\rm erg\,s^{-1}$ (e.g., \citealp{2020A&A...636A..73D,2024A&A...691A.203G}). Hence, we employ luminosity-dependent $\kappa_{2-10}$ (using Eq. 1 from \citealp{2025ApJ...990...86G}) to calculate $L_{\rm bol}$ for these nine lensed systems (see Sect. \ref{sect:kx} for more details). For one system, we used the optical bolometric correction (at $3000\AA$) to estimate $L_{\rm bol}$ (more details in Sect. for \ref{sect:kuvo}). The individual methods used to estimate $L_{\rm bol}$ for each system are summarised in Table \ref{tab:summ}.


\subsection{Black hole mass and Eddington ratio}\label{sect:mbh_er}

Next, to calculate the black hole mass of our lensed AGN, we used the virial method. This method basically assumes virialized motion of the gas in the broad line region (BLR) and by employing the empirical relations between the continuum luminosity ($L_{\lambda}$) and the size of the BLR (from reverberation mapping), the mass of the black hole can be calculated using the width (FWHM) of certain broad emission lines. In this study, based on the wavelength ranges of the available rest-frame optical/UV spectra, we used the broad C{\scriptsize\,IV}, Mg{\scriptsize\,II}, and H$\alpha$ lines (see Table \ref{tab:uvo_spectra} for details). The following relation was used to estimate $M_{\rm BH}$:

\begin{equation}\label{eq:mbh}
    M_{\rm BH} = fG^{-1} \times R_{\rm BLR} \times V_{\rm BLR}^2 = K(\lambda L_{\lambda})^\alpha\,{\rm FWHM}^2
\end{equation}\\
where, $R_{\rm BLR}$ and $V_{\rm BLR}$ are the radius and velocity of the BLR, respectively, and $f$ is a general geometric function (assumed to be 1; \citealp{2016MNRAS.460..187M}). In agreement with our fitting procedure, we used values of  $K$ and $\alpha$ corresponding to a local fit in \citeauthor{2016MNRAS.460..187M} (\citeyear{2016MNRAS.460..187M}; listed in columns 1--3 of Table 7) to calculate $M_{\rm BH}$. The typical uncertainties on our $M_{\rm BH}$ estimates are dominated by those introduced due to the relation used (Eq. \ref{eq:mbh}). This corresponds to errors of 0.25\,dex (for Mg{\scriptsize\,II}), 0.33\,dex (for C{\scriptsize\,IV}), and 0.16\,dex (for H$\alpha$), as reported by \cite{2016MNRAS.460..187M}. Finally, we also estimated the Eddington ratios for the 19 systems with $M_{\rm{BH}}$ and $L_{\rm{bol}}$ measurements. We have reported the bolometric luminosities, black hole masses, and Eddington ratios for the resolved lensed images (whenever possible) for all 27 quasar systems in Table \ref{tab:params}. We show the distribution in these physical properties of lensed quasars (averaged over available lensed images) in Fig. \ref{fig:prop}.


\begin{figure*}
    \centering
    \includegraphics[width=15.9cm]{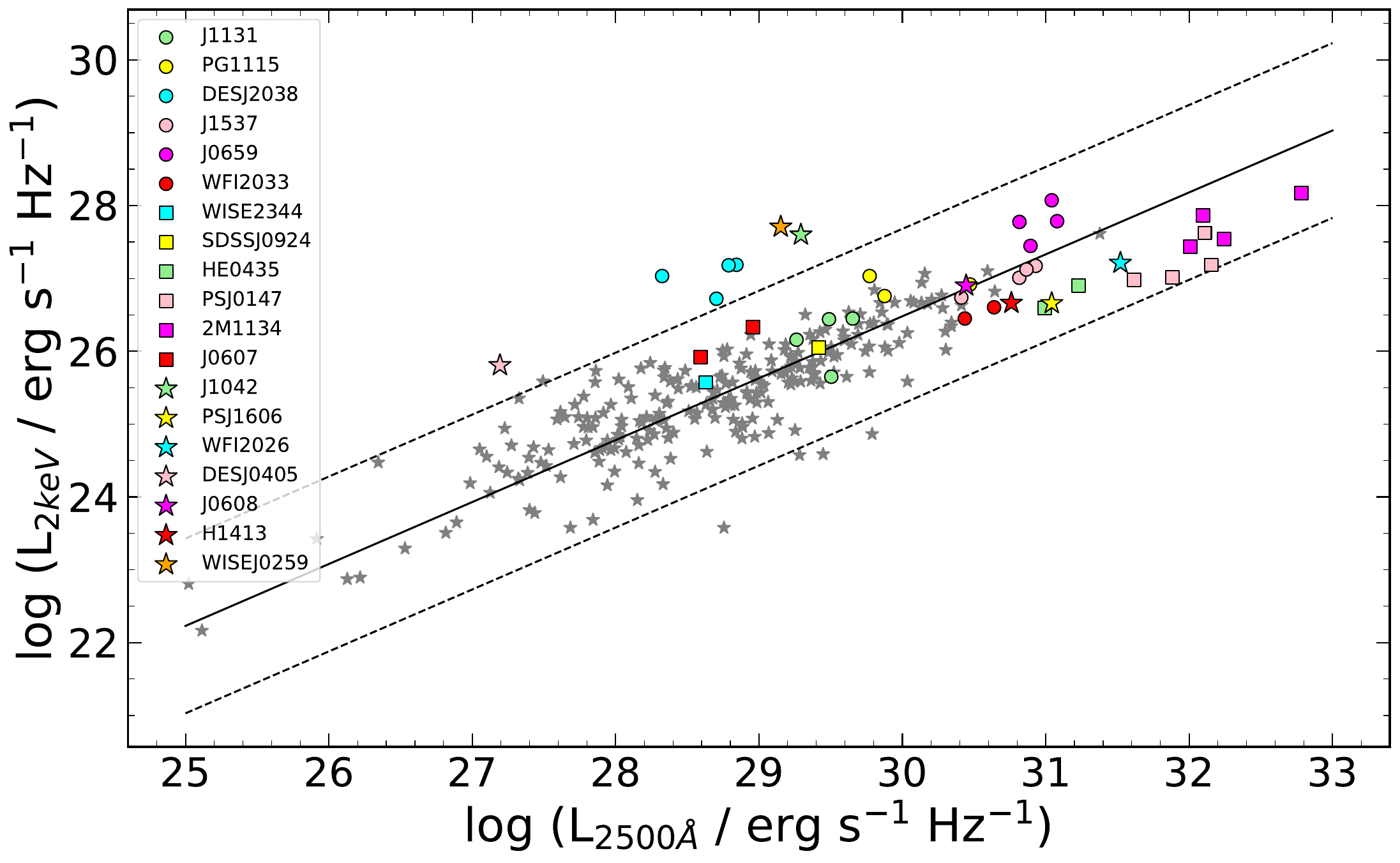}
    \caption{Monochromatic luminosity at 2500\,$\rm\AA$ versus the monochromatic luminosity at 2\,keV. Both luminosity estimates are intrinsic and have been corrected for macro-magnification. The best-fit relation (Eq. 6 from \citealp{2024A&A...691A.203G}) with 3$\sigma$ uncertainty  are shown as solid and dashed black lines, respectively. Individual systems are shown as different markers as listed in the legend. Multiple measurements for the same system correspond to each resolved image for that system.}
    \label{fig:l2500_vs_l2keV}
\end{figure*}


\section{Discussion}\label{sect:discuss}

In this section, we look at some of the established relations for AGN in the context of lensed AGN and their implications. We first investigate the well-known UV-X-ray luminosity relation to see if lensed AGN follow the same trends as nonlensed AGN (in Sect. \ref{sect:alpha}). Next, we study the evolution of X-ray bolometric corrections from lensed AGN with their physical properties (specifically luminosity) to determine the best and most accurate method to use them as $L_{\rm bol}$ estimators (Sect. \ref{sect:kx}). We also briefly explore the scatter in $L_{\rm bol}$ when predicted from optical/UV bolometric corrections (Sect. \ref{sect:kuvo}). 


\subsection{UV-X-ray luminosity relation}\label{sect:alpha}

It is well known that the monochromatic X-ray and UV luminosities at 2\,keV and 2500\,$\rm \AA$, respectively show a strong nonlinear correlation for AGN (e.g., \citealp{1986ApJ...305...83A,1994ApJS...92...53W,2006AJ....131.2826S,2010A&A...512A..34L,2024A&A...691A.203G}). Although the exact physics driving this relation is still under debate (e.g., \citealp{2017A&A...602A..79L}), it is useful to better understand the link between the optical/UV disk and X-ray coronal emission. Moreover, the absence of any significant redshift evolution of this relation (e.g., \citealp{2010A&A...512A..34L,2015ApJ...815...33R,2019NatAs...3..272R}) further confirms its universality for all AGN populations. In the case of gravitationally lensed AGN, this relation requires a prior magnification correction for the lensed image luminosities, which is usually calculated from the macro-model (corresponding to the foreground lens). However, since this relation is non-linear, the same, but incorrect, multiplicative factor (i.e., magnification) applied to the luminosities can move an object away from the expected relation. There are two main reasons that could lead to wrong magnification corrections: (a) the macro-model being biased due to incorrect modeling assumptions or (b) one or more of the lensed images being subject to a strong micro- or millilensing event. In the first case, it is more likely that all the lensed images will be similarly shifted in the same direction, whereas the effect of micro-/millilensing is expected to dominate for only one image. However, it is important to realize that to shift an object off the relation based on just the macro-model bias, relatively large changes of the magnification would be needed. One can easily show that a bias of a factor 10 in magnification (corresponding to a change of 0.1\,dex in $L_{\rm 2500\,\AA}$) translates into a change of 0.15\,dex in $L_{\rm 2\,keV}$. On the contrary, because the optical and X-ray emitting regions in AGN have different physical scales and could potentially not be co-spatial (e.g., \citealp{2022ApJ...931...68S,2025ApJ...987...75R}), microlensing (and its time variability) could produce larger shifts of the object luminosities in the $L_{\rm 2\,keV}-L_{\rm 2500\,\AA}$ plot. In this section, we investigate whether the lensed AGN systems with multiple images show a similar trend as nonlensed AGN between their X-ray and UV luminosity and use their position on this relation as a diagnostic for macro-model bias and/or possibility of micro-/millilensing in the lensed AGN.

In Fig. \ref{fig:l2500_vs_l2keV}, we plot the rest-frame monochromatic luminosity at 2\,keV ($L_{\rm 2\,keV}$) as a function of the rest-frame monochromatic luminosity at 2500\,$\rm \AA$ ($L_{2500\AA}$) for all the lensed images (if resolved) for our lensed AGN sample (values are listed in Table \ref{tab:params}). We also show local, nonlensed AGN in the background (as gray stars), along with the linear fit from \citeauthor{2024A&A...691A.203G} (\citeyear{2024A&A...691A.203G}; Eq. 6). The dashed lines mark the 3$\sigma$ scatter around this relation. Most of the lensed AGN images occupy the region within 3$\sigma$ of the UV-X-ray luminosity trend observed for nonlensed AGN. This is not surprising, as we do not expect lensing to change the intrinsic accretion and emission physics of AGN. However, we do find some outliers. Three of these are from the unresolved systems DESJ0405 (pink star), J1042 (green star), and WISEJ0259 (orange star). Very few observations of these systems exist in the literature, which limits our ability to identify the origin of the deviation. However, it is worth noting that both DESJ0405 and J1042 show evidence of flux anomalies due to microlensing in at least one of their images, which could possibly shift the points away from the expected relation. In the case of DESJ0405, the chromatic study of the flux ratios by \citealp{2024MNRAS.530.2960N} suggests some microlensing in the optical/UV, but it is of too small amplitude to shift the object out of the relation. A plausible scenario to explain the observed deviation would be a large differential microlensing between the UV and X-ray domains. For instance, microlensing demagnification of an image in UV and microlensing magnification in X-ray would move a system towards lower $L_{\rm 2500\,\AA}$ but higher $L_{\rm 2\,keV}$, as observed for DESJ0405. A shift of 0.5\,dex in $L_{\rm 2\,keV}$ (but negligible microlensing in UV) is roughly needed to reconcile DESJ0405 with the relation. This corresponds to a micro-magnification factor $\sim 3$ (1.25\,mag), that has been previously observed for a single lensed image in other systems (e.g., see Fig. 3 of \citealp{2007ApJ...661...19P}). A similar amplitude of differential microlensing between X-ray and UV is also possible if the two epochs of observations are not exactly simultaneous, as is the case for DESJ0405 (e.g., see Fig. 4 and 5 of \citealp{2010ApJ...709..278D}). For J1042, the HST data from \citet{2023MNRAS.518.1260S} indicates some flux ratio anomaly in one of the images, further supporting the important role of microlensing in this system. In short, the common occurrence of microlensing in quadruply lensed quasars suggests that strong-amplitude microlensing may be at work in the three unresolved systems, pushing them away from the observed UV-X-ray relation. 


\begin{figure*}
  \begin{subfigure}[t]{0.5\textwidth}
    \centering
    \includegraphics[width=\textwidth]{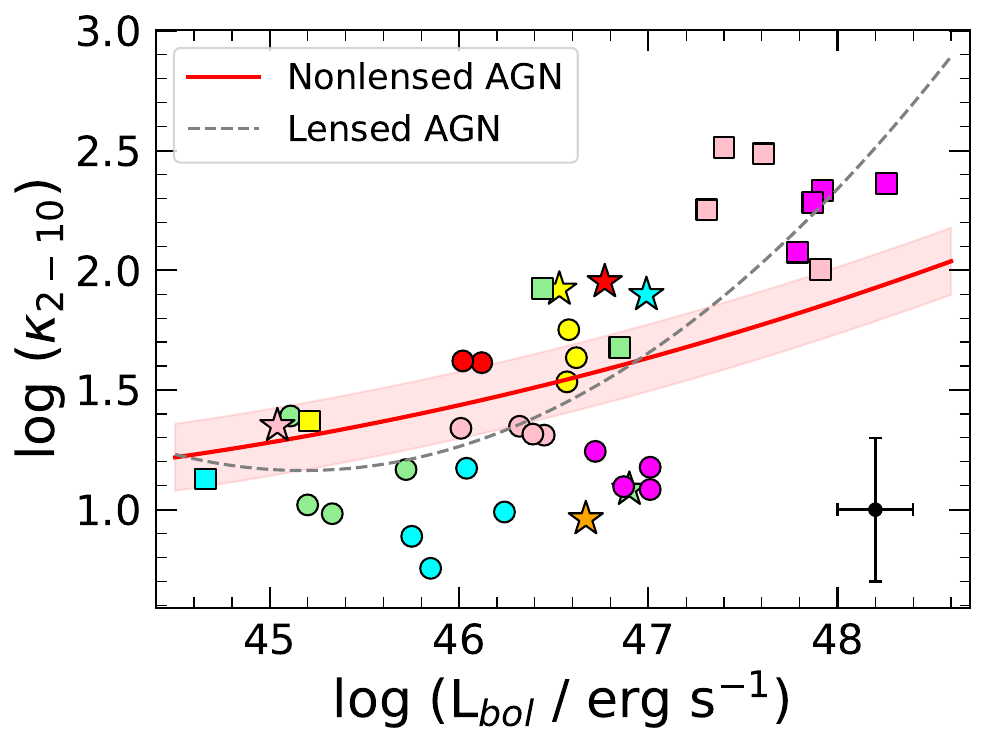}
    \caption{}
    \label{fig:kx_lb}
  \end{subfigure}
  \begin{subfigure}[t]{0.5\textwidth}
    \centering
    \includegraphics[width=\textwidth]{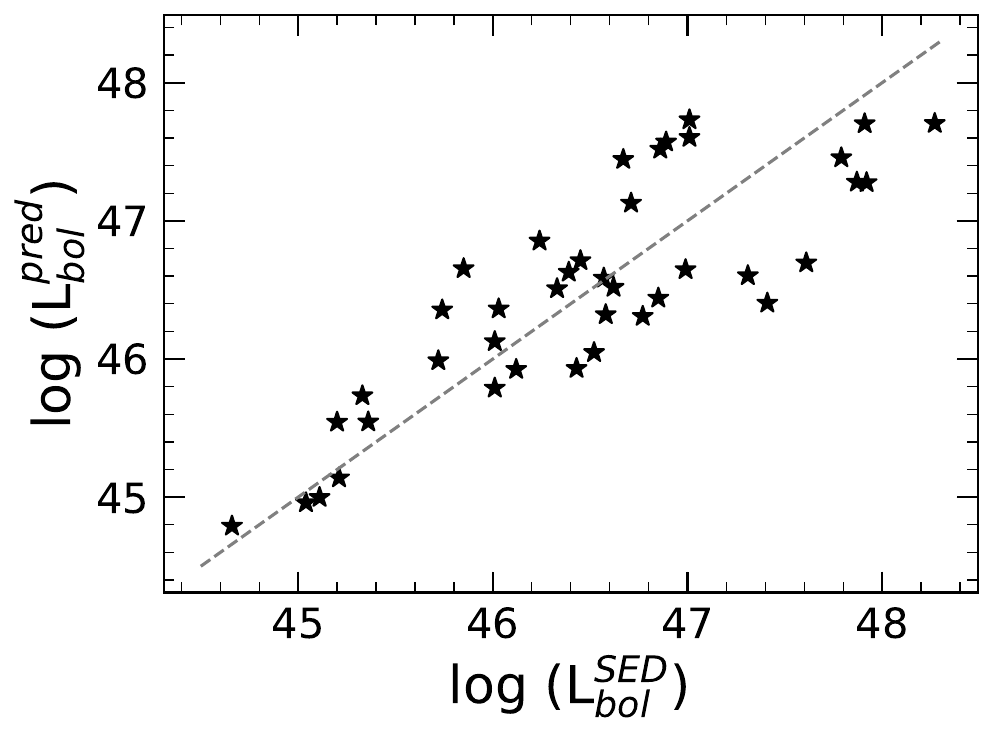}
    \caption{}
    \label{fig:lb_test}
  \end{subfigure}

\caption{(a) The 2--10\,keV X-ray bolometric correction ($\kappa_{2-10}$) as a function of the bolometric luminosity. The best-fit relation for nonlensed AGN from \cite{2025ApJ...990...86G} is shown as a solid red line. The shaded region (in light red) marks the $1\sigma$ scatter on the best-fit relation. For visual aid, we also show the linear fit for our lensed AGN sample (as dashed gray line). Individual systems are shown as different markers as listed in the legend of Fig. \ref{fig:l2500_vs_l2keV} and there is one point per resolved image. The marker at the bottom right corner shows the typical uncertainties on the quantities. (b) Comparison between the predicted bolometric luminosity (from luminosity-dependent $\kappa_{2-10}$) and that obtained from the broadband SED analysis. The dashed gray line shows the 1:1 relation.}
\label{fig:kx}
\end{figure*}


In contrast, for the lensed system of DESJ2038 (blue circles), microlensing does not seem to be a likely explanation because all four images occupy a similar region outside of the 3$\sigma$ scatter of the expected trend, such that they either show a higher X-ray luminosity or a lower UV luminosity.

Finally, in the case of systems PSJ0147 (pink squares) and 2M1134 (magenta squares), although both agree with the expected UV-X-ray relation, they are clearly more luminous compared to the rest of the sample. The four images of PSJ0417 have roughly the same luminosity within uncertainties, suggesting that microlensing is not a major source of the visible systematic offset. In fact, PSJ0147 is one of the brightest lensed quasars known, suggesting that it is genuinely an intrinsically bright source (e.g., \citealp{2017ApJ...844...90B,2023ApJ...955..140S}). For 2M1134, there might be two effects in place. Firstly, one of the images seems to be much brighter than the others, which could possibly be due to micro- or millilensing. Secondly, this system is suspected to have a galaxy cluster nearby ($\sim 30"$), whose mass is not explicitly accounted for in the lens model and would affect its accuracy (e.g., \citealp{2019MNRAS.486.4987R}). Therefore, the high brightness of the whole system may be caused by a large underestimation of the resultant magnifications.


\begin{figure*}
  \begin{subfigure}[t]{0.5\textwidth}
    \centering
    \includegraphics[width=\textwidth]{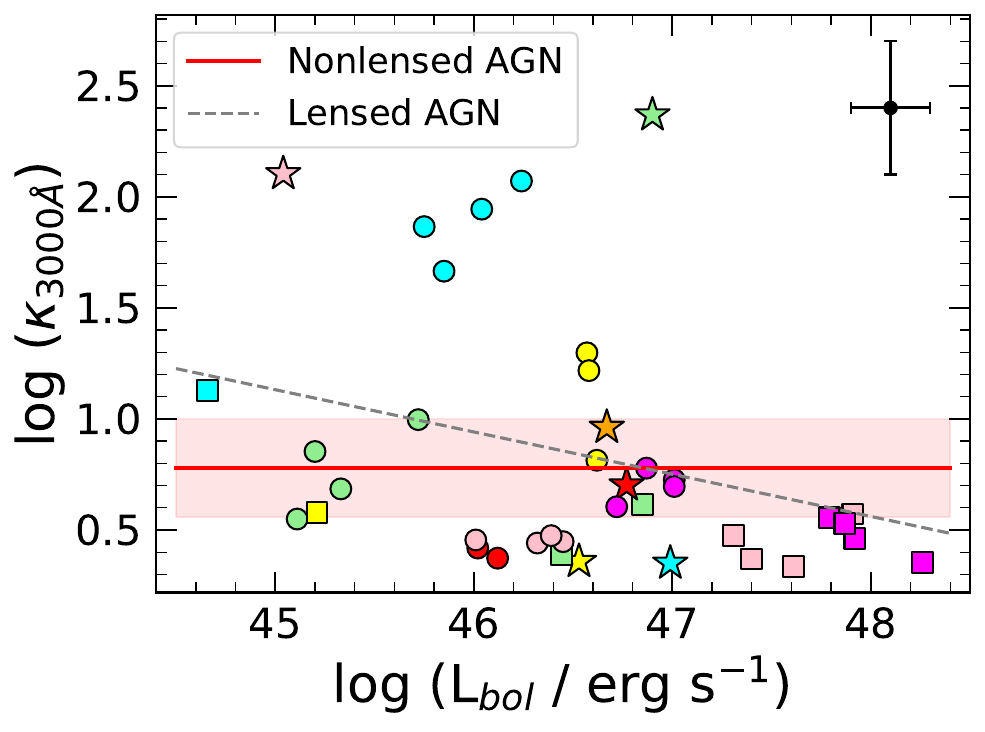}
    \caption{}
    \label{fig:kuvo_lb}
  \end{subfigure}
  \begin{subfigure}[t]{0.5\textwidth}
    \centering
    \includegraphics[width=\textwidth]{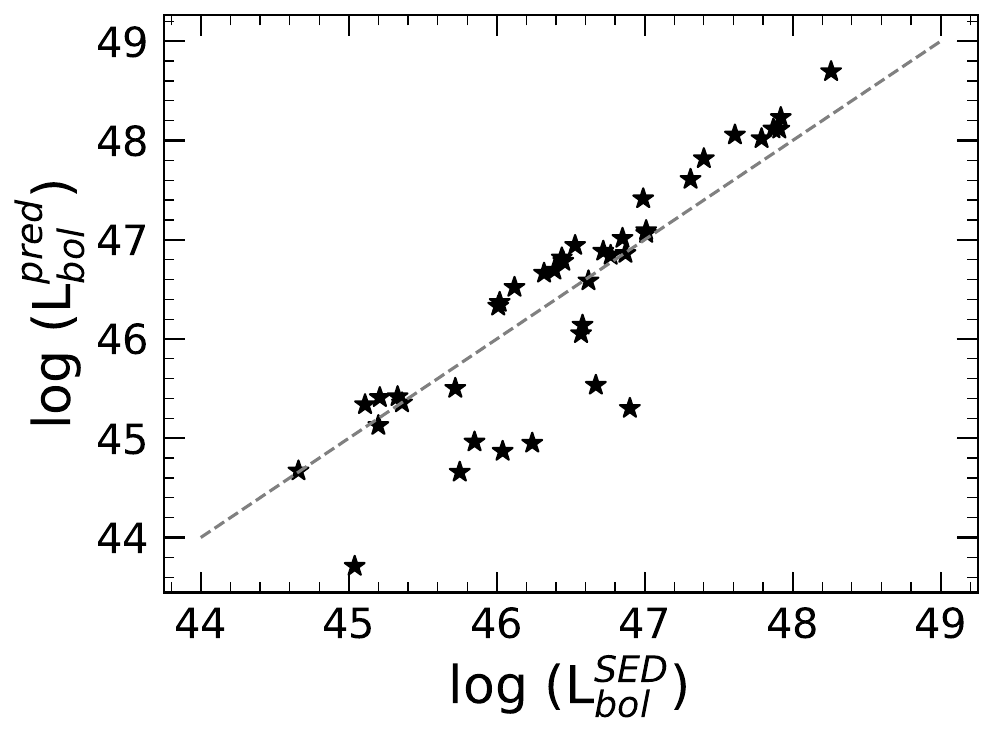}
    \caption{}
    \label{fig:lb_test2}
  \end{subfigure}
   
\caption{(a) The $3000\,\AA$ bolometric correction ($\kappa_{3000\AA}$) as a function of the bolometric luminosity. The best-fit relation for nonlensed AGN from \cite{2024A&A...691A.203G} is shown as a solid red line. The shaded region (in light red) marks the $1\sigma$ scatter on the best-fit relation. For visual aid, we also show the linear fit for our lensed AGN sample (as dashed gray line). Individual systems are shown as different markers as listed in the legend of Fig. \ref{fig:l2500_vs_l2keV} and there is one point per resolved image. The marker at the top right corner shows the typical uncertainties on the quantities. (b) Comparison between the predicted bolometric luminosity (from constant $\kappa_{3000\AA}$) and that obtained from the broadband SED analysis. The dashed gray line shows the 1:1 relation.}
\label{fig:kuvo}
\end{figure*}


\subsection{X-ray bolometric correction}\label{sect:kx}

In this section, we check the reliability of X-ray bolometric corrections to predict the bolometric luminosity of lensed AGN. To do so, we first confirmed that the 2--10\,keV X-ray bolometric corrections for our lensed AGN sample show a similar evolution with luminosity to the one expected from nonlensed AGN. Thanks to the broadband SED analysis of the 17 lensed systems in our sample, we could calculate their 2--10\,keV X-ray bolometric corrections. Here, we have included measurements from both resolved and unresolved systems to maximize our sample size. We illustrate the evolution of $\kappa_{2-10}$ with the bolometric luminosity in Fig. \ref{fig:kx_lb}. Individual lensed AGN are shown by different markers and colors (similar to Fig. \ref{fig:l2500_vs_l2keV}). We clearly see an increase in $\kappa_{2-10}$ with $L_{\rm bol}$, as expected from previous studies using nonlensed AGN. For comparison, we show the best-fit relation reported by \citet{2025ApJ...990...86G} based on a large study of hard-X-ray-selected AGN in the local Universe (solid red line). The visible scatter in Fig. \ref{fig:kx_lb} is consistent with the expected 3$\sigma$ dispersion in $\kappa_{2-10}$ \citep{2025ApJ...990...86G}. The primary outliers to the relation are systems with $L_{\rm bol} > 10^{47}\,\rm erg\,s^{-1}$ (2M1134 and PSJ0147), which systematically show high optical luminosities, leading to high $\kappa_{2-10}$ values. We also show the fit obtained for our lensed AGN sample (dashed gray line), but purely for visual aid, as this relation is heavily biased (discussed in detail in Appendix \ref{sect:appendixa}).

Next, we used the best-fit relation from \citet{2025ApJ...990...86G} to estimate $\kappa_{2-10}$ based on the already calculated $L_{\rm bol}$ (from SED analysis) to check if the resultant value of $L_{\rm bol}$ agrees with the expected one from the broadband SEDs. We solved the following equation to obtain the bolometric luminosity:
\begin{equation}
    L_{\rm bol} = \kappa_{2-10} \times L_{2-10}\,\,\,;\,\,\,{\rm where,}\,\,\kappa_{2-10} =  f(L_{\rm bol,G25})
\end{equation}
Here, $f(L_{\rm bol,G25})$ is the equation that describes $\kappa_{2-10}$ as a function of $L_{\rm bol}$ from \citeauthor{2025ApJ...990...86G} (\citeyear{2025ApJ...990...86G}; Eq. 1). This method only requires, as input, the 2--10\,keV X-ray luminosity of the source to finally obtain the bolometric luminosity for that source. In Fig. \ref{fig:lb_test}, we compare the values of $L_{\rm bol}$ obtained from the SED fitting ($L_{\rm bol}^{\rm SED}$) with the predicted value of $L_{\rm bol}$ ($L_{\rm bol}^{\rm pred}$) calculated from the luminosity-dependent X-ray bolometric corrections. As visible in Fig. \ref{fig:lb_test}, the predicted value of $L_{\rm bol}$ matches quite well with the one from the SED fitting, with a scatter of only $\sim 0.5$\,dex. Moreover, the predictions are uniformly spread around the 1:1 line (gray dashed), without significant evidence for bias towards over- or under-estimation. For points at $L_{\rm bol} > 10^{47}\,\rm erg\,s^{-1}$, the predictions might appear to be underestimated. However, these points correspond to the four images of the two lensed sources, 2M1134 and PSJ0147, with higher than usual optical luminosities contributing significantly to their total bolometric luminosity estimated from the SED (see Sect. \ref{sect:alpha}). Therefore, their $L_{\rm bol}$ predictions from $\kappa_{2-10}$ are generally lower.

We also checked if other estimates of $\kappa_{2-10}$ give a better prediction of the bolometric luminosity compared to the one discussed above. We first tested constant bolometric corrections with $\kappa_{2-10}=15$ \citep{2024A&A...691A.203G} and $\kappa_{2-10}=20$ \citep{2009MNRAS.392.1124V}. For both cases the bolometric luminosity predictions agreed well with the real values for low luminosity sources ($L_{\rm bol} < 10^{46}\,\rm erg\,s^{-1}$). However, as we go to higher luminosities, $L_{\rm bol}$ was underestimated by up to $\sim 1$\,dex. This is expected as the constant bolometric corrections only apply for a small range of bolometric luminosities and do not include the obvious effects of increasing luminosity on the X-ray bolometric corrections. Finally, we tested another $\kappa_{2-10}-L_{\rm bol}$ relation from \citet{2020A&A...636A..73D} to see if it could decrease the scatter in the values $L_{\rm bol}$ predicted by the \citet{2025ApJ...990...86G} relation. We used the general relation reported in \citeauthor{2020A&A...636A..73D} (\citeyear{2020A&A...636A..73D}; Table 1) to estimate $L_{\rm bol}$ from $L_{2-10}$ using the same iterative method mentioned previously. On comparing the resultant $L_{\rm bol}^{\rm pred}$ with $L_{\rm bol}^{\rm SED}$, we found that the value of the bolometric luminosity was severely overestimated with a deviation up to $\sim2$\,dex.

Based on the tests discussed in this section, we strongly recommend the use of luminosity-dependent X-ray bolometric corrections for the estimation of $L_{\rm bol}$ in the case when only X-ray data is available. Additionally, our analysis shows that using the $\kappa_{2-10}-L_{\rm bol}$ function reported by \citet{2025ApJ...990...86G} to estimate the bolometric luminosity results in the least scatter ($\sim0.5$\,dex) in the value of $L_{\rm bol}$. Therefore, we used the same method for the nine lensed AGN in our sample (without optical/UV data) to estimate their bolometric luminosities (listed in Table \ref{tab:params}).


\subsection{Optical/UV bolometric correction}\label{sect:kuvo}

For sources lacking X-ray data, constant bolometric corrections at $1250\,\AA$, $3000\,\AA$, and/or $5100\,\AA$ are commonly used to estimate the bolometric luminosity. In this section, we investigate the goodness of the $L_{\rm bol}$ estimates from these corrections. Going forward, we present all results using the $3000\,\AA$ bolometric correction ($\kappa_{3000\AA}$). However, they are also applicable for other wavelengths in the optical/UV. We first checked the dependence of $\kappa_{3000\AA}$ on $L_{\rm bol}$ for our sample of 17 lensed AGN (with both resolved and unresolved images) in Fig. \ref{fig:kuvo_lb}. In agreement with previous studies, we did not obtain any significant trend in $\kappa_{3000\AA}$ with luminosity. For reference, we plot the constant $\kappa_{3000\AA} = 6$ line from \cite{2024A&A...691A.203G}, who reported a median value of $\kappa_{3000\AA}$ based on the broadband SED analysis of nearby AGN. Comparing to the expected 3$\sigma$ scatter in $\kappa_{3000\AA}$ \citep{2024A&A...691A.203G}, we obtain three systems as outliers. Two of these systems (DESJ0405 and J1042) show signatures of microlensing (see Sect. 5.1), which could explain their large deviation from the expected trend.  The last outlier (DESJ2038) is already identified as an unusual system based on Fig. 3 (discussed in Sect. 5.1) and requires a dedicated study to determine why it behaves differently.

Next, we used $\kappa_{3000\AA} = 6$ to calculate $L_{\rm bol}$ from $L_{3000\AA}$ ($L_{\rm bol}^{\rm pred} = 6
\times L_{3000\AA}$) and checked these predictions against the more accurate values of $L_{\rm bol}$ measured from the SED fitting ($L_{\rm bol}^{\rm SED}$). Fig. \ref{fig:lb_test2} shows that the use of a constant $\kappa_{3000\AA}$ leads to a slight overestimation of the predicted $L_{\rm bol}$ in most cases. But the scatter in $L_{\rm bol}^{\rm pred}$ is generally very small, except for certain sources where the scatter is as high as $\sim 1$\,dex. On closer inspection, we find that these sources are specifically the ones that are outliers in the UV-X-ray relation shown in Fig. \ref{fig:l2500_vs_l2keV} (discussed in Sect. \ref{sect:alpha}). Therefore, we can attribute this large deviation in the predicted $L_{\rm bol}$ to a microlensing event or biases in the lens model. If luminosities are corrected for these effects in advance, optical bolometric corrections can provide extremely reliable estimates of $L_{\rm bol}$, although biased toward overprediction. Similar results were also obtained when using $\kappa_{3000\AA} = 5.15$, as reported by \citealp{2008ApJ...680..169S}.


\section{Summary}\label{sect:summary}

In this paper, we present one of the largest multiwavelength studies of 27 quadruply lensed AGN ($0.6 < z < 3.1$), with the aim to calculate their bolometric luminosities, black hole masses, and Eddington ratios. We compiled optical, UV, and X-ray photometric and spectroscopic observations for all sources. We constructed and fit the optical-to-X-ray SEDs (from $1000\,\mu$m to 500\,keV) for 17 lensed AGN using a dust-reddened multi-temperature blackbody to model the optical/UV accretion disk and a simple X-ray powerlaw model with Galactic absorption to model the coronal emission. We then used the best-fit SED models to calculate the total accretion luminosity of our sources. We also performed spectral emission line fitting in rest-frame optical/UV to measure the black hole masses and eventually the Eddington ratios of 19 systems. We list all these important quantities in Table \ref{tab:params}.

For ten sources with limited data, we used bolometric corrections to estimate their bolometric luminosities. We performed multiple tests to determine the best method to use X-ray bolometric corrections as $L_{\rm bol}$ estimators. Our analysis showed that using luminosity-dependent 2--10\,keV bolometric corrections (from Eq. 1 in \citealp{2025ApJ...990...86G}) results in $L_{\rm bol}$ predictions within $\sim0.5$\,dex of the value calculated from SED fitting. Using other methods could bias the predictions towards over- or under-estimation with deviations up to $\sim1-2$\,dex. We also showed that using constant optical/UV bolometric corrections (at $1250\,\AA$, $3000\,\AA$, or $5100\,\AA$) to calculate $L_{\rm bol}$ results in a lower scatter in the predictions, if one has already corrected the luminosities for microlensing events. However, these predictions are generally biased towards overestimation.

We also investigated the well-known UV-X-ray luminosity relation for AGN in the context of lensed systems. As expected, we found that lensed AGN majorly occupy a similar parameter space in terms of the monochromatic 2\,keV and $2500\,\AA$ luminosity. However, we also find some sources that deviate from the expected trend. We suspect that the main reason for this offset is large differential microlensing between the UV and X-ray domains, which could also be enhanced by the lack of simultaneous UV and X-ray data. As a result, we propose that the UV-X-ray relation for AGN can be a useful tool for identifying microlensing events in lensed AGN.

Through this multiwavelength study, we have performed a comprehensive broadband SED fitting analysis of a sample of lensed AGN, to provide the community estimates of some useful quantities, such as bolometric luminosities, black hole masses, and Eddington ratios. We have explored possible sources of errors that might affect our analysis, including AGN variability, dust extinction, and host galaxy contamination. We have also presented and compared prescriptions to derive bolometric luminosities for lensed AGN from limited data, which we believe would be extremely useful for the community.


\section{Data availability}

All the optical-to-X-ray SEDs and optical/UV spectral fits generated in this work are available upon reasonable request to the corresponding author.\\


\begin{acknowledgements}

We would like to thank the anonymous referee for their useful suggestions that helped improve the clarity of this manuscript. This work made use of data from the NASA/IPAC Infrared Science Archive and NASA/IPAC Extragalactic Database (NED), which are operated by the Jet Propulsion Laboratory, California Institute of Technology, under contract with the National Aeronautics and Space Administration. This research has made use of data and/or software provided by the High Energy Astrophysics Science Archive Research Center (HEASARC), which is a service of the Astrophysics Science Division at NASA/GSFC and the High Energy Astrophysics Division of the Smithsonian Astrophysical Observatory. KKG and DS acknowledge financial support from the Belgian Federal Science Policy Office (BELSPO) in the framework of the PRODEX Programme of the European Space Agency under contract number 4000142531. TA acknowledges support from ANID-FONDECYT Regular Project 1240105 and the ANID BASAL project FB210003.

\end{acknowledgements}


\bibliographystyle{aa}
\bibliography{References}

@article{Sluse_2006,
	author = {Sluse, D. and Claeskens, J.-F. and Altieri, B. and Cabanac, R. A. and Garcet, O. and Hutsem{\'e}kers, D. and Jean, C. and Smette, A. and Surdej, J.},
	doi = {10.1051/0004-6361:20053148},
	issn = {1432-0746},
	journal = {\aap},
	month = mar,
	number = {2},
	pages = {539--550},
	publisher = {EDP Sciences},
	title = {Multi-wavelength study of the gravitational lens system RXS J113155.4-123155: I. Multi-epoch optical and near infrared imaging},
	url = {http://dx.doi.org/10.1051/0004-6361:20053148},
	volume = {449},
	year = {2006}}

@article{2025arXiv251107513G,
	adsurl = {https://ui.adsabs.harvard.edu/abs/2025arXiv251107513G},
	archiveprefix = {arXiv},
	author = {{Gilman}, Daniel and {Nierenberg}, A.~M. and {Treu}, T. and {Gannon}, C. and {Du}, X. and {Paugnat}, H. and {Birrer}, S. and {Benson}, A.~J. and {Mozumdar}, P. and {Wong}, K.~C. and {Williams}, D. and {Keeley}, R.~E. and {Abazajian}, K.~N. and {Anguita}, T. and {Bennert}, V.~N. and {Djorgovski}, S.~G. and {Kusenko}, A. and {Malkan}, M. and {Morishita}, T. and {Motta}, V. and {Moustakas}, L.~A. and {Sheu}, W. and {Sluse}, D. and {Stiavelli}, M.},
	doi = {10.48550/arXiv.2511.07513},
	eid = {arXiv:2511.07513},
	eprint = {2511.07513},
	journal = {arXiv e-prints},
	month = nov,
	pages = {arXiv:2511.07513},
	primaryclass = {astro-ph.CO},
	title = {{JWST lensed quasar dark matter survey IV: Stringent warm dark matter constraints from the joint reconstruction of extended lensed arcs and quasar flux ratios}},
	year = 2025}

@article{2025arXiv251107765K,
	adsurl = {https://ui.adsabs.harvard.edu/abs/2025arXiv251107765K},
	archiveprefix = {arXiv},
	author = {{Keeley}, R.~E. and {Nierenberg}, A.~M. and {Gilman}, D. and {Treu}, T. and {Du}, X. and {Gannon}, C. and {Mozumdar}, P. and {Wong}, K.~C. and {Paugnat}, H. and {Birrer}, S. and {Malkan}, M. and {Benson}, A.~J. and {Abazajian}, K.~N. and {Anguita}, T. and {Bennert}, V.~N. and {Djorgovski}, S.~G. and {Hoenig}, S.~F. and {Kusenko}, A. and {Morishita}, T. and {Motta}, V. and {Moustakas}, L.~A. and {Sheu}, W. and {Sluse}, D. and {Stern}, D. and {Stiavelli}, M. and {Williams}, D.},
	doi = {10.48550/arXiv.2511.07765},
	eid = {arXiv:2511.07765},
	eprint = {2511.07765},
	journal = {arXiv e-prints},
	month = nov,
	pages = {arXiv:2511.07765},
	primaryclass = {astro-ph.CO},
	title = {{JWST Lensed Quasar Dark Matter Survey III: Dark Matter Sensitive Flux Ratios and Warm Dark Matter Constraint from the Full Sample}},
	year = 2025}

@article{2019NatAs...3..272R,
	adsurl = {https://ui.adsabs.harvard.edu/abs/2019NatAs...3..272R},
	archiveprefix = {arXiv},
	author = {{Risaliti}, G. and {Lusso}, E.},
	doi = {10.1038/s41550-018-0657-z},
	eprint = {1811.02590},
	journal = {Nature Astronomy},
	month = jan,
	pages = {272-277},
	primaryclass = {astro-ph.CO},
	title = {{Cosmological Constraints from the Hubble Diagram of Quasars at High Redshifts}},
	volume = {3},
	year = 2019}

@article{2025ApJ...987...75R,
	adsurl = {https://ui.adsabs.harvard.edu/abs/2025ApJ...987...75R},
	archiveprefix = {arXiv},
	author = {{Rogers}, Alysa and {Schwartz}, Daniel and {Spingola}, Cristiana and {Barnacka}, Anna},
	doi = {10.3847/1538-4357/add71a},
	eid = {75},
	eprint = {2505.08077},
	journal = {\apj},
	month = jul,
	number = {1},
	pages = {75},
	primaryclass = {astro-ph.HE},
	title = {{Milliarcsecond X-Ray Positions and X-Ray Varstrometry for the Strongly Lensed Active Galactic Nucleus HE 0435-1223}},
	volume = {987},
	year = 2025}

@article{2022ApJ...931...68S,
	adsurl = {https://ui.adsabs.harvard.edu/abs/2022ApJ...931...68S},
	archiveprefix = {arXiv},
	author = {{Spingola}, Cristiana and {Schwartz}, Daniel and {Barnacka}, Anna},
	doi = {10.3847/1538-4357/ac68eb},
	eid = {68},
	eprint = {2203.04245},
	journal = {\apj},
	month = may,
	number = {1},
	pages = {68},
	primaryclass = {astro-ph.HE},
	title = {{Milliarcsecond X-Ray Astrometry to Resolve Inner Regions of AGN at z > 1 Using Gravitational Lensing}},
	volume = {931},
	year = 2022}

@article{2023ApJ...955..140S,
	adsurl = {https://ui.adsabs.harvard.edu/abs/2023ApJ...955..140S},
	archiveprefix = {arXiv},
	author = {{Shalyapin}, Vyacheslav N. and {Goicoechea}, Luis J. and {Dyrland}, Karianne and {Dahle}, H{\r{a}}kon},
	doi = {10.3847/1538-4357/acee7e},
	eid = {140},
	eprint = {2309.04285},
	journal = {\apj},
	month = oct,
	number = {2},
	pages = {140},
	primaryclass = {astro-ph.GA},
	title = {{Andromeda's Parachute: Time Delays and Hubble Constant}},
	volume = {955},
	year = 2023}

@inproceedings{2009ASPC..411..251M,
	adsurl = {https://ui.adsabs.harvard.edu/abs/2009ASPC..411..251M},
	archiveprefix = {arXiv},
	author = {{Markwardt}, C.~B.},
	booktitle = {Astronomical Data Analysis Software and Systems XVIII},
	doi = {10.48550/arXiv.0902.2850},
	editor = {{Bohlender}, D.~A. and {Durand}, D. and {Dowler}, P.},
	eprint = {0902.2850},
	month = sep,
	pages = {251},
	primaryclass = {astro-ph.IM},
	series = {Astronomical Society of the Pacific Conference Series},
	title = {{Non-linear Least-squares Fitting in IDL with MPFIT}},
	volume = {411},
	year = 2009}

@article{2010ApJ...709..937G,
	adsurl = {https://ui.adsabs.harvard.edu/abs/2010ApJ...709..937G},
	archiveprefix = {arXiv},
	author = {{Greene}, Jenny E. and {Peng}, Chien Y. and {Ludwig}, Randi R.},
	doi = {10.1088/0004-637X/709/2/937},
	eprint = {0911.0685},
	journal = {\apj},
	month = feb,
	number = {2},
	pages = {937-949},
	primaryclass = {astro-ph.CO},
	title = {{Redshift Evolution in Black Hole-Bulge Relations: Testing C IV-Based Black Hole Masses}},
	volume = {709},
	year = 2010}

@article{2018A&A...618A..56D,
	adsurl = {https://ui.adsabs.harvard.edu/abs/2018A&A...618A..56D},
	author = {{Ducourant}, C. and {Wertz}, O. and {Krone-Martins}, A. and {Teixeira}, R. and {Le Campion}, J. -F. and {Galluccio}, L. and {Kl{\"u}ter}, J. and {Delchambre}, L. and {Surdej}, J. and {Mignard}, F. and {Wambsganss}, J. and {Bastian}, U. and {Graham}, M.~J. and {Djorgovski}, S.~G. and {Slezak}, E.},
	doi = {10.1051/0004-6361/201833480},
	eid = {A56},
	journal = {\aap},
	month = oct,
	pages = {A56},
	title = {{Gaia GraL: Gaia DR2 gravitational lens systems. II. The known multiply imaged quasars}},
	volume = {618},
	year = 2018}

@article{2023MNRAS.520.3305L,
	adsurl = {https://ui.adsabs.harvard.edu/abs/2023MNRAS.520.3305L},
	archiveprefix = {arXiv},
	author = {{Lemon}, C. and {Anguita}, T. and {Auger-Williams}, M.~W. and {Courbin}, F. and {Galan}, A. and {McMahon}, R. and {Neira}, F. and {Oguri}, M. and {Schechter}, P. and {Shajib}, A. and {Treu}, T. and {Agnello}, A. and {Spiniello}, C.},
	doi = {10.1093/mnras/stac3721},
	eprint = {2206.07714},
	journal = {\mnras},
	month = apr,
	number = {3},
	pages = {3305-3328},
	primaryclass = {astro-ph.GA},
	title = {{Gravitationally lensed quasars in Gaia - IV. 150 new lenses, quasar pairs, and projected quasars}},
	volume = {520},
	year = 2023}

@article{2008ApJ...680..169S,
	adsurl = {https://ui.adsabs.harvard.edu/abs/2008ApJ...680..169S},
	archiveprefix = {arXiv},
	author = {{Shen}, Yue and {Greene}, Jenny E. and {Strauss}, Michael A. and {Richards}, Gordon T. and {Schneider}, Donald P.},
	doi = {10.1086/587475},
	eprint = {0709.3098},
	journal = {\apj},
	month = jun,
	number = {1},
	pages = {169-190},
	primaryclass = {astro-ph},
	title = {{Biases in Virial Black Hole Masses: An SDSS Perspective}},
	volume = {680},
	year = 2008}

@article{2025ApJ...990...86G,
	adsurl = {https://ui.adsabs.harvard.edu/abs/2025ApJ...990...86G},
	author = {{Gupta}, Kriti Kamal and {Ricci}, Claudio and {Tortosa}, Alessia and {Temple}, Matthew J. and {Koss}, Michael J. and {Trakhtenbrot}, Benny and {Bauer}, Franz E. and {Treister}, Ezequiel and {Mushotzky}, Richard and {Kammoun}, Elias and {Papadakis}, Iossif and {Oh}, Kyuseok and {Rojas}, Alejandra and {Chang}, Chin-Shin and {Diaz}, Yaherlyn and {Jana}, Arghajit and {Kakkad}, Darshan and {del Moral-Castro}, Ignacio and {Peca}, Alessandro and {Powell}, Meredith C. and {Stern}, Daniel and {Urry}, C. Megan and {Harrison}, Fiona},
	doi = {10.3847/1538-4357/adf0f8},
	eid = {86},
	journal = {\apj},
	month = sep,
	number = {1},
	pages = {86},
	title = {{BASS. LIII. The Eddington Ratio as the Primary Regulator of the Fraction of X-Ray Emission in Active Galactic Nuclei}},
	volume = {990},
	year = 2025}

@article{2023A&A...680A..51M,
	adsurl = {https://ui.adsabs.harvard.edu/abs/2023A&A...680A..51M},
	archiveprefix = {arXiv},
	author = {{Melo}, A. and {Motta}, V. and {Mej{\'\i}a-Restrepo}, J. and {Assef}, R.~J. and {Godoy}, N. and {Mediavilla}, E. and {Falco}, E. and {Kochanek}, C.~S. and {{\'A}vila-Vera}, F. and {Jerez}, R.},
	doi = {10.1051/0004-6361/202347078},
	eid = {A51},
	eprint = {2306.03472},
	journal = {\aap},
	month = dec,
	pages = {A51},
	primaryclass = {astro-ph.GA},
	title = {{Black hole masses for 14 gravitationally lensed quasars}},
	volume = {680},
	year = 2023}

@article{2021A&A...656A.108M,
	adsurl = {https://ui.adsabs.harvard.edu/abs/2021A&A...656A.108M},
	archiveprefix = {arXiv},
	author = {{Melo}, A. and {Motta}, V. and {Godoy}, N. and {Mejia-Restrepo}, J. and {Assef}, R.~J. and {Mediavilla}, E. and {Falco}, E. and {{\'A}vila-Vera}, F. and {Jerez}, R.},
	doi = {10.1051/0004-6361/202141869},
	eid = {A108},
	eprint = {2110.01575},
	journal = {\aap},
	month = dec,
	pages = {A108},
	primaryclass = {astro-ph.GA},
	title = {{First black hole mass estimation for the quadruple lensed system WGD2038-4008}},
	volume = {656},
	year = 2021}

@software{2011ascl.soft09001G,
	adsurl = {https://ui.adsabs.harvard.edu/abs/2011ascl.soft09001G},
	author = {{Ginsburg}, Adam and {Mirocha}, Jordan},
	eid = {ascl:1109.001},
	howpublished = {Astrophysics Source Code Library, record ascl:1109.001},
	month = sep,
	title = {{PySpecKit: Python Spectroscopic Toolkit}},
	year = 2011}

@article{2022AJ....163..291G,
	adsurl = {https://ui.adsabs.harvard.edu/abs/2022AJ....163..291G},
	archiveprefix = {arXiv},
	author = {{Ginsburg}, Adam and {Sokolov}, Vlas and {de Val-Borro}, Miguel and {Rosolowsky}, Erik and {Pineda}, Jaime E. and {Sip{\H{o}}cz}, Brigitta M. and {Henshaw}, Jonathan D.},
	doi = {10.3847/1538-3881/ac695a},
	eid = {291},
	eprint = {2205.04987},
	journal = {\aj},
	month = jun,
	number = {6},
	pages = {291},
	primaryclass = {astro-ph.IM},
	title = {{Pyspeckit: A Spectroscopic Analysis and Plotting Package}},
	volume = {163},
	year = 2022}

@article{2012A&A...544A..62S,
	adsurl = {https://ui.adsabs.harvard.edu/abs/2012A&A...544A..62S},
	archiveprefix = {arXiv},
	author = {{Sluse}, D. and {Hutsem{\'e}kers}, D. and {Courbin}, F. and {Meylan}, G. and {Wambsganss}, J.},
	doi = {10.1051/0004-6361/201219125},
	eid = {A62},
	eprint = {1206.0731},
	journal = {\aap},
	month = aug,
	pages = {A62},
	primaryclass = {astro-ph.CO},
	title = {{Microlensing of the broad line region in 17 lensed quasars}},
	volume = {544},
	year = 2012}

@inproceedings{2003SPIE.4851...28G,
	adsurl = {https://ui.adsabs.harvard.edu/abs/2003SPIE.4851...28G},
	author = {{Garmire}, Gordon P. and {Bautz}, Mark W. and {Ford}, Peter G. and {Nousek}, John A. and {Ricker}, Jr., George R.},
	booktitle = {X-Ray and Gamma-Ray Telescopes and Instruments for Astronomy.},
	doi = {10.1117/12.461599},
	editor = {{Truemper}, Joachim E. and {Tananbaum}, Harvey D.},
	month = mar,
	pages = {28-44},
	series = {Society of Photo-Optical Instrumentation Engineers (SPIE) Conference Series},
	title = {{Advanced CCD imaging spectrometer (ACIS) instrument on the Chandra X-ray Observatory}},
	volume = {4851},
	year = 2003}

@article{2010ApJ...709..278D,
	adsurl = {https://ui.adsabs.harvard.edu/abs/2010ApJ...709..278D},
	archiveprefix = {arXiv},
	author = {{Dai}, X. and {Kochanek}, C.~S. and {Chartas}, G. and {Koz{\l}owski}, S. and {Morgan}, C.~W. and {Garmire}, G. and {Agol}, E.},
	doi = {10.1088/0004-637X/709/1/278},
	eprint = {0906.4342},
	journal = {\apj},
	month = jan,
	number = {1},
	pages = {278-285},
	primaryclass = {astro-ph.HE},
	title = {{The Sizes of the X-ray and Optical Emission Regions of RXJ 1131-1231}},
	volume = {709},
	year = 2010}

@article{1988A&A...194...54G,
	adsurl = {https://ui.adsabs.harvard.edu/abs/1988A&A...194...54G},
	author = {{Grieger}, B. and {Kayser}, R. and {Refsdal}, S.},
	journal = {\aap},
	month = apr,
	pages = {54-64},
	title = {{Gravitational micro-lensing as a clue to quasar structure.}},
	volume = {194},
	year = 1988}

@article{2001ApJS..134....1V,
	adsurl = {https://ui.adsabs.harvard.edu/abs/2001ApJS..134....1V},
	archiveprefix = {arXiv},
	author = {{Vestergaard}, M. and {Wilkes}, B.~J.},
	doi = {10.1086/320357},
	eprint = {astro-ph/0104320},
	journal = {\apjs},
	month = may,
	number = {1},
	pages = {1-33},
	primaryclass = {astro-ph},
	title = {{An Empirical Ultraviolet Template for Iron Emission in Quasars as Derived from I Zwicky 1}},
	volume = {134},
	year = 2001}

@article{1992ApJS...80..109B,
	adsurl = {https://ui.adsabs.harvard.edu/abs/1992ApJS...80..109B},
	author = {{Boroson}, Todd A. and {Green}, Richard F.},
	doi = {10.1086/191661},
	journal = {\apjs},
	month = may,
	pages = {109},
	title = {{The Emission-Line Properties of Low-Redshift Quasi-stellar Objects}},
	volume = {80},
	year = 1992}

@article{2024A&A...691A.203G,
	adsurl = {https://ui.adsabs.harvard.edu/abs/2024A&A...691A.203G},
	archiveprefix = {arXiv},
	author = {{Gupta}, Kriti K. and {Ricci}, Claudio and {Temple}, Matthew J. and {Tortosa}, Alessia and {Koss}, Michael J. and {Assef}, Roberto J. and {Bauer}, Franz E. and {Mushotzy}, Richard and {Ricci}, Federica and {Ueda}, Yoshihiro and {Rojas}, Alejandra F. and {Trakhtenbrot}, Benny and {Chang}, Chin-Shin and {Oh}, Kyuseok and {Li}, Ruancun and {Kawamuro}, Taiki and {Diaz}, Yaherlyn and {Powell}, Meredith C. and {Stern}, Daniel and {Megan Urry}, C. and {Harrison}, Fiona and {Cenko}, Brad},
	doi = {10.1051/0004-6361/202450567},
	eid = {A203},
	eprint = {2409.12239},
	journal = {\aap},
	month = nov,
	pages = {A203},
	primaryclass = {astro-ph.GA},
	title = {{BASS: XLIII. Optical, UV, and X-ray emission properties of unobscured Swift/BAT active galactic nuclei}},
	volume = {691},
	year = 2024}

@article{2024MNRAS.530.2960N,
	adsurl = {https://ui.adsabs.harvard.edu/abs/2024MNRAS.530.2960N},
	archiveprefix = {arXiv},
	author = {{Nierenberg}, A.~M. and {Keeley}, R.~E. and {Sluse}, D. and {Gilman}, D. and {Birrer}, S. and {Treu}, T. and {Abazajian}, K.~N. and {Anguita}, T. and {Benson}, A.~J. and {Bennert}, V.~N. and {Djorgovski}, S.~G. and {Du}, X. and {Fassnacht}, C.~D. and {Hoenig}, S.~F. and {Kusenko}, A. and {Lemon}, C. and {Malkan}, M. and {Motta}, V. and {Moustakas}, L.~A. and {Stern}, D. and {Wechsler}, R.~H.},
	doi = {10.1093/mnras/stae499},
	eprint = {2309.10101},
	journal = {\mnras},
	month = may,
	number = {3},
	pages = {2960-2971},
	primaryclass = {astro-ph.CO},
	title = {{JWST lensed quasar dark matter survey - I. Description and first results}},
	volume = {530},
	year = 2024}

@inproceedings{2006SPIE.6270E..1VF,
	adsurl = {https://ui.adsabs.harvard.edu/abs/2006SPIE.6270E..1VF},
	author = {{Fruscione}, Antonella and {McDowell}, Jonathan C. and {Allen}, Glenn E. and {Brickhouse}, Nancy S. and {Burke}, Douglas J. and {Davis}, John E. and {Durham}, Nick and {Elvis}, Martin and {Galle}, Elizabeth C. and {Harris}, Daniel E. and {Huenemoerder}, David P. and {Houck}, John C. and {Ishibashi}, Bish and {Karovska}, Margarita and {Nicastro}, Fabrizio and {Noble}, Michael S. and {Nowak}, Michael A. and {Primini}, Frank A. and {Siemiginowska}, Aneta and {Smith}, Randall K. and {Wise}, Michael},
	booktitle = {Observatory Operations: Strategies, Processes, and Systems},
	doi = {10.1117/12.671760},
	editor = {{Silva}, David R. and {Doxsey}, Rodger E.},
	eid = {62701V},
	month = jun,
	pages = {62701V},
	series = {Society of Photo-Optical Instrumentation Engineers (SPIE) Conference Series},
	title = {{CIAO: Chandra's data analysis system}},
	volume = {6270},
	year = 2006}

@article{2024MNRAS.535.1652K,
	adsurl = {https://ui.adsabs.harvard.edu/abs/2024MNRAS.535.1652K},
	archiveprefix = {arXiv},
	author = {{Keeley}, Ryan E. and {Nierenberg}, A.~M. and {Gilman}, D. and {Gannon}, C. and {Birrer}, S. and {Treu}, T. and {Benson}, A.~J. and {Du}, X. and {Abazajian}, K.~N. and {Anguita}, T. and {Bennert}, V.~N. and {Djorgovski}, S.~G. and {Gupta}, K.~K. and {Hoenig}, S.~F. and {Kusenko}, A. and {Lemon}, C. and {Malkan}, M. and {Motta}, V. and {Moustakas}, L.~A. and {Oh}, Maverick S.~H. and {Sluse}, D. and {Stern}, D. and {Wechsler}, R.~H.},
	doi = {10.1093/mnras/stae2458},
	eprint = {2405.01620},
	journal = {\mnras},
	month = dec,
	number = {2},
	pages = {1652-1671},
	primaryclass = {astro-ph.CO},
	title = {{JWST lensed quasar dark matter survey - II. Strongest gravitational lensing limit on the dark matter free streaming length to date}},
	volume = {535},
	year = 2024}

@article{2007ApJ...661...19P,
	adsurl = {https://ui.adsabs.harvard.edu/abs/2007ApJ...661...19P},
	archiveprefix = {arXiv},
	author = {{Pooley}, David and {Blackburne}, Jeffrey A. and {Rappaport}, Saul and {Schechter}, Paul L.},
	doi = {10.1086/512115},
	eprint = {astro-ph/0607655},
	journal = {\apj},
	month = may,
	number = {1},
	pages = {19-29},
	primaryclass = {astro-ph},
	title = {{X-Ray and Optical Flux Ratio Anomalies in Quadruply Lensed Quasars. I. Zooming in on Quasar Emission Regions}},
	volume = {661},
	year = 2007}

@article{2023MNRAS.518.1260S,
	adsurl = {https://ui.adsabs.harvard.edu/abs/2023MNRAS.518.1260S},
	archiveprefix = {arXiv},
	author = {{Schmidt}, T. and {Treu}, T. and {Birrer}, S. and {Shajib}, A.~J. and {Lemon}, C. and {Millon}, M. and {Sluse}, D. and {Agnello}, A. and {Anguita}, T. and {Auger-Williams}, M.~W. and {McMahon}, R.~G. and {Motta}, V. and {Schechter}, P. and {Spiniello}, C. and {Kayo}, I. and {Courbin}, F. and {Ertl}, S. and {Fassnacht}, C.~D. and {Frieman}, J.~A. and {More}, A. and {Schuldt}, S. and {Suyu}, S.~H. and {Aguena}, M. and {Andrade-Oliveira}, F. and {Annis}, J. and {Bacon}, D. and {Bertin}, E. and {Brooks}, D. and {Burke}, D.~L. and {Carnero Rosell}, A. and {Carrasco Kind}, M. and {Carretero}, J. and {Conselice}, C. and {Costanzi}, M. and {da Costa}, L.~N. and {Pereira}, M.~E.~S. and {De Vicente}, J. and {Desai}, S. and {Doel}, P. and {Everett}, S. and {Ferrero}, I. and {Friedel}, D. and {Garc{\'\i}a-Bellido}, J. and {Gaztanaga}, E. and {Gruen}, D. and {Gruendl}, R.~A. and {Gschwend}, J. and {Gutierrez}, G. and {Hinton}, S.~R. and {Hollowood}, D.~L. and {Honscheid}, K. and {James}, D.~J. and {Kuehn}, K. and {Lahav}, O. and {Menanteau}, F. and {Miquel}, R. and {Palmese}, A. and {Paz-Chinch{\'o}n}, F. and {Pieres}, A. and {Plazas Malag{\'o}n}, A.~A. and {Prat}, J. and {Rodriguez-Monroy}, M. and {Romer}, A.~K. and {Sanchez}, E. and {Scarpine}, V. and {Sevilla-Noarbe}, I. and {Smith}, M. and {Suchyta}, E. and {Tarle}, G. and {To}, C. and {Varga}, T.~N. and {DES Collaboration}},
	doi = {10.1093/mnras/stac2235},
	eprint = {2206.04696},
	journal = {\mnras},
	month = jan,
	number = {1},
	pages = {1260-1300},
	primaryclass = {astro-ph.CO},
	title = {{STRIDES: automated uniform models for 30 quadruply imaged quasars}},
	volume = {518},
	year = 2023}

@article{2008ApJ...689..755M,
	adsurl = {https://ui.adsabs.harvard.edu/abs/2008ApJ...689..755M},
	archiveprefix = {arXiv},
	author = {{Morgan}, Christopher W. and {Kochanek}, Christopher. S. and {Dai}, Xinyu and {Morgan}, Nicholas D. and {Falco}, Emilio E.},
	doi = {10.1086/592767},
	eprint = {0802.1210},
	journal = {\apj},
	month = dec,
	number = {2},
	pages = {755-761},
	primaryclass = {astro-ph},
	title = {{X-Ray and Optical Microlensing in the Lensed Quasar PG 1115+080}},
	volume = {689},
	year = 2008}

@article{2009MNRAS.398..233F,
	adsurl = {https://ui.adsabs.harvard.edu/abs/2009MNRAS.398..233F},
	archiveprefix = {arXiv},
	author = {{Floyd}, David J.~E. and {Bate}, N.~F. and {Webster}, R.~L.},
	doi = {10.1111/j.1365-2966.2009.15045.x},
	eprint = {0905.2651},
	journal = {\mnras},
	month = sep,
	number = {1},
	pages = {233-239},
	primaryclass = {astro-ph.HE},
	title = {{The accretion disc in the quasar SDSS J0924+0219}},
	volume = {398},
	year = 2009}

@article{2006ApJ...640...47K,
	adsurl = {https://ui.adsabs.harvard.edu/abs/2006ApJ...640...47K},
	archiveprefix = {arXiv},
	author = {{Kochanek}, C.~S. and {Morgan}, N.~D. and {Falco}, E.~E. and {McLeod}, B.~A. and {Winn}, J.~N. and {Dembicky}, J. and {Ketzeback}, B.},
	doi = {10.1086/499766},
	eprint = {astro-ph/0508070},
	journal = {\apj},
	month = mar,
	number = {1},
	pages = {47-61},
	primaryclass = {astro-ph},
	title = {{The Time Delays of Gravitational Lens HE 0435-1223: An Early-Type Galaxy with a Rising Rotation Curve}},
	volume = {640},
	year = 2006}

@article{2004AJ....127.2617M,
	adsurl = {https://ui.adsabs.harvard.edu/abs/2004AJ....127.2617M},
	archiveprefix = {arXiv},
	author = {{Morgan}, Nicholas D. and {Caldwell}, John A.~R. and {Schechter}, Paul L. and {Dressler}, Alan and {Egami}, Eiichi and {Rix}, Hans-Walter},
	doi = {10.1086/383295},
	eprint = {astro-ph/0312478},
	journal = {\aj},
	month = may,
	number = {5},
	pages = {2617-2630},
	primaryclass = {astro-ph},
	title = {{WFI J2026-4536 and WFI J2033-4723: Two New Quadruple Gravitational Lenses}},
	volume = {127},
	year = 2004}

@article{2011ApJ...742...67M,
	adsurl = {https://ui.adsabs.harvard.edu/abs/2011ApJ...742...67M},
	archiveprefix = {arXiv},
	author = {{Mu{\~n}oz}, J.~A. and {Mediavilla}, E. and {Kochanek}, C.~S. and {Falco}, E.~E. and {Mosquera}, A.~M.},
	doi = {10.1088/0004-637X/742/2/67},
	eid = {67},
	eprint = {1107.5932},
	journal = {\apj},
	month = dec,
	number = {2},
	pages = {67},
	primaryclass = {astro-ph.CO},
	title = {{A Study of Gravitational Lens Chromaticity with the Hubble Space Telescope}},
	volume = {742},
	year = 2011}

@article{2017ApJ...844...90B,
	adsurl = {https://ui.adsabs.harvard.edu/abs/2017ApJ...844...90B},
	archiveprefix = {arXiv},
	author = {{Berghea}, C.~T. and {Nelson}, George J. and {Rusu}, C.~E. and {Keeton}, C.~R. and {Dudik}, R.~P.},
	doi = {10.3847/1538-4357/aa7aa6},
	eid = {90},
	eprint = {1705.08359},
	journal = {\apj},
	month = aug,
	number = {2},
	pages = {90},
	primaryclass = {astro-ph.GA},
	title = {{Discovery of the First Quadruple Gravitationally Lensed Quasar Candidate with Pan-STARRS}},
	volume = {844},
	year = 2017}

@article{2019MNRAS.486.4987R,
	adsurl = {https://ui.adsabs.harvard.edu/abs/2019MNRAS.486.4987R},
	archiveprefix = {arXiv},
	author = {{Rusu}, Cristian E. and {Berghea}, Ciprian T. and {Fassnacht}, Christopher D. and {More}, Anupreeta and {Seman}, Erica and {Nelson}, George J. and {Chen}, Geoff C. -F.},
	doi = {10.1093/mnras/stz1142},
	eprint = {1803.07175},
	journal = {\mnras},
	month = jul,
	number = {4},
	pages = {4987-5007},
	primaryclass = {astro-ph.GA},
	title = {{A search for gravitationally lensed quasars and quasar pairs in Pan-STARRS1: spectroscopy and sources of shear in the diamond 2M1134-2103}},
	volume = {486},
	year = 2019}

@article{2018MNRAS.480.5017A,
	adsurl = {https://ui.adsabs.harvard.edu/abs/2018MNRAS.480.5017A},
	archiveprefix = {arXiv},
	author = {{Anguita}, T. and {Schechter}, P.~L. and {Kuropatkin}, N. and {Morgan}, N.~D. and {Ostrovski}, F. and {Abramson}, L.~E. and {Agnello}, A. and {Apostolovski}, Y. and {Fassnacht}, C.~D. and {Hsueh}, J.~W. and {Motta}, V. and {Rojas}, K. and {Rusu}, C.~E. and {Treu}, T. and {Williams}, P. and {Auger}, M. and {Buckley-Geer}, E. and {Lin}, H. and {McMahon}, R. and {Abbott}, T.~M.~C. and {Allam}, S. and {Annis}, J. and {Bernstein}, R.~A. and {Bertin}, E. and {Brooks}, D. and {Burke}, D.~L. and {Carnero Rosell}, A. and {Carrasco-Kind}, M. and {Carretero}, J. and {Cunha}, C.~E. and {D'Andrea}, C.~B. and {De Vicente}, J. and {DePoy}, D.~L. and {Desai}, S. and {Diehl}, H.~T. and {Doel}, P. and {Flaugher}, B. and {Garc{\'\i}a-Bellido}, J. and {Gerdes}, D.~W. and {Gruen}, D. and {Gruendl}, R.~A. and {Gschwend}, J. and {Hartley}, W.~G. and {Hollowood}, D.~L. and {Honscheid}, K. and {James}, D.~J. and {Kuehn}, K. and {Lima}, M. and {Maia}, M.~A.~G. and {Miquel}, R. and {Plazas}, A.~A. and {Sanchez}, E. and {Scarpine}, V. and {Smith}, M. and {Soares-Santos}, M. and {Sobreira}, F. and {Suchyta}, E. and {Tarle}, G. and {Walker}, A.~R.},
	doi = {10.1093/mnras/sty2172},
	eprint = {1805.12151},
	journal = {\mnras},
	month = nov,
	number = {4},
	pages = {5017-5028},
	primaryclass = {astro-ph.GA},
	title = {{The STRong lensing Insights into the Dark Energy Survey (STRIDES) 2016 follow-up campaign - II. New quasar lenses from double component fitting}},
	volume = {480},
	year = 2018}

@article{2019MNRAS.483.4242L,
	adsurl = {https://ui.adsabs.harvard.edu/abs/2019MNRAS.483.4242L},
	archiveprefix = {arXiv},
	author = {{Lemon}, Cameron A. and {Auger}, Matthew W. and {McMahon}, Richard G.},
	doi = {10.1093/mnras/sty3366},
	eprint = {1810.04480},
	journal = {\mnras},
	month = mar,
	number = {3},
	pages = {4242-4258},
	primaryclass = {astro-ph.GA},
	title = {{Gravitationally lensed quasars in Gaia - III. 22 new lensed quasars from Gaia data release 2}},
	volume = {483},
	year = 2019}

@article{2018MNRAS.479.5060L,
	adsurl = {https://ui.adsabs.harvard.edu/abs/2018MNRAS.479.5060L},
	archiveprefix = {arXiv},
	author = {{Lemon}, Cameron A. and {Auger}, Matthew W. and {McMahon}, Richard G. and {Ostrovski}, Fernanda},
	doi = {10.1093/mnras/sty911},
	eprint = {1803.07601},
	journal = {\mnras},
	month = oct,
	number = {4},
	pages = {5060-5074},
	primaryclass = {astro-ph.GA},
	title = {{Gravitationally lensed quasars in Gaia - II. Discovery of 24 lensed quasars}},
	volume = {479},
	year = 2018}

@article{2018MNRAS.479.4345A,
	adsurl = {https://ui.adsabs.harvard.edu/abs/2018MNRAS.479.4345A},
	archiveprefix = {arXiv},
	author = {{Agnello}, A. and {Lin}, H. and {Kuropatkin}, N. and {Buckley-Geer}, E. and {Anguita}, T. and {Schechter}, P.~L. and {Morishita}, T. and {Motta}, V. and {Rojas}, K. and {Treu}, T. and {Amara}, A. and {Auger}, M.~W. and {Courbin}, F. and {Fassnacht}, C.~D. and {Frieman}, J. and {More}, A. and {Marshall}, P.~J. and {McMahon}, R.~G. and {Meylan}, G. and {Suyu}, S.~H. and {Glazebrook}, K. and {Morgan}, N. and {Nord}, B. and {Abbott}, T.~M.~C. and {Abdalla}, F.~B. and {Annis}, J. and {Bechtol}, K. and {Benoit-L{\'e}vy}, A. and {Bertin}, E. and {Bernstein}, R.~A. and {Brooks}, D. and {Burke}, D.~L. and {Rosell}, A. Carnero and {Carretero}, J. and {Cunha}, C.~E. and {D'Andrea}, C.~B. and {da Costa}, L.~N. and {Desai}, S. and {Drlica-Wagner}, A. and {Eifler}, T.~F. and {Flaugher}, B. and {Garc{\'\i}a-Bellido}, J. and {Gaztanaga}, E. and {Gerdes}, D.~W. and {Gruen}, D. and {Gruendl}, R.~A. and {Gschwend}, J. and {Gutierrez}, G. and {Honscheid}, K. and {James}, D.~J. and {Kuehn}, K. and {Lahav}, O. and {Lima}, M. and {Maia}, M.~A.~G. and {March}, M. and {Menanteau}, F. and {Miquel}, R. and {Ogando}, R.~L.~C. and {Plazas}, A.~A. and {Sanchez}, E. and {Scarpine}, V. and {Schindler}, R. and {Schubnell}, M. and {Sevilla-Noarbe}, I. and {Smith}, M. and {Soares-Santos}, M. and {Sobreira}, F. and {Suchyta}, E. and {Swanson}, M.~E.~C. and {Tarle}, G. and {Tucker}, D. and {Wechsler}, R.},
	doi = {10.1093/mnras/sty1419},
	eprint = {1711.03971},
	journal = {\mnras},
	month = oct,
	number = {4},
	pages = {4345-4354},
	primaryclass = {astro-ph.CO},
	title = {{DES meets Gaia: discovery of strongly lensed quasars from a multiplet search}},
	volume = {479},
	year = 2018}

@article{2020SciPy-NMeth,
	adsurl = {https://rdcu.be/b08Wh},
	author = {Virtanen, Pauli and Gommers, Ralf and Oliphant, Travis E. and Haberland, Matt and Reddy, Tyler and Cournapeau, David and Burovski, Evgeni and Peterson, Pearu and Weckesser, Warren and Bright, Jonathan and {van der Walt}, St{\'e}fan J. and Brett, Matthew and Wilson, Joshua and Millman, K. Jarrod and Mayorov, Nikolay and Nelson, Andrew R. J. and Jones, Eric and Kern, Robert and Larson, Eric and Carey, C J and Polat, {\.I}lhan and Feng, Yu and Moore, Eric W. and {VanderPlas}, Jake and Laxalde, Denis and Perktold, Josef and Cimrman, Robert and Henriksen, Ian and Quintero, E. A. and Harris, Charles R. and Archibald, Anne M. and Ribeiro, Ant{\^o}nio H. and Pedregosa, Fabian and {van Mulbregt}, Paul and {SciPy 1.0 Contributors}},
	doi = {10.1038/s41592-019-0686-2},
	journal = {Nature Methods},
	pages = {261--272},
	title = {{{SciPy} 1.0: Fundamental Algorithms for Scientific Computing in Python}},
	volume = {17},
	year = {2020}}

@article{1997ARA&A..35..445U,
	adsurl = {https://ui.adsabs.harvard.edu/abs/1997ARA&A..35..445U},
	author = {{Ulrich}, Marie-Helene and {Maraschi}, Laura and {Urry}, C. Megan},
	doi = {10.1146/annurev.astro.35.1.445},
	journal = {\araa},
	month = jan,
	pages = {445-502},
	title = {{Variability of Active Galactic Nuclei}},
	volume = {35},
	year = 1997}

@article{2015ApJ...815...33R,
	adsurl = {https://ui.adsabs.harvard.edu/abs/2015ApJ...815...33R},
	archiveprefix = {arXiv},
	author = {{Risaliti}, G. and {Lusso}, E.},
	doi = {10.1088/0004-637X/815/1/33},
	eid = {33},
	eprint = {1505.07118},
	journal = {\apj},
	month = dec,
	number = {1},
	pages = {33},
	primaryclass = {astro-ph.CO},
	title = {{A Hubble Diagram for Quasars}},
	volume = {815},
	year = 2015}

@article{2017A&A...602A..79L,
	adsurl = {https://ui.adsabs.harvard.edu/abs/2017A&A...602A..79L},
	archiveprefix = {arXiv},
	author = {{Lusso}, E. and {Risaliti}, G.},
	doi = {10.1051/0004-6361/201630079},
	eid = {A79},
	eprint = {1703.05299},
	journal = {\aap},
	month = jun,
	pages = {A79},
	primaryclass = {astro-ph.HE},
	title = {{Quasars as standard candles. I. The physical relation between disc and coronal emission}},
	volume = {602},
	year = 2017}

@article{1986ApJ...305...83A,
	adsurl = {https://ui.adsabs.harvard.edu/abs/1986ApJ...305...83A},
	author = {{Avni}, Y. and {Tananbaum}, H.},
	doi = {10.1086/164230},
	journal = {\apj},
	month = jun,
	pages = {83},
	title = {{X-Ray Properties of Optically Selected QSOs}},
	volume = {305},
	year = 1986}

@article{1994ApJS...92...53W,
	adsurl = {https://ui.adsabs.harvard.edu/abs/1994ApJS...92...53W},
	author = {{Wilkes}, Belinda J. and {Tananbaum}, Harvey and {Worrall}, D.~M. and {Avni}, Yoram and {Oey}, M.~S. and {Flanagan}, Joan},
	doi = {10.1086/191959},
	journal = {\apjs},
	month = may,
	pages = {53},
	title = {{The Einstein Database of IPC X-Ray Observations of Optically Selected and Radio-selected Quasars. I.}},
	volume = {92},
	year = 1994}

@article{2016MNRAS.460..187M,
	adsurl = {https://ui.adsabs.harvard.edu/abs/2016MNRAS.460..187M},
	archiveprefix = {arXiv},
	author = {{Mej{\'\i}a-Restrepo}, J.~E. and {Trakhtenbrot}, B. and {Lira}, P. and {Netzer}, H. and {Capellupo}, D.~M.},
	doi = {10.1093/mnras/stw568},
	eprint = {1603.03437},
	journal = {\mnras},
	month = jul,
	number = {1},
	pages = {187-211},
	primaryclass = {astro-ph.GA},
	title = {{Active galactic nuclei at z {\ensuremath{\sim}} 1.5 - II. Black hole mass estimation by means of broad emission lines}},
	volume = {460},
	year = 2016}

@inproceedings{1999ASPC..161..295B,
	adsurl = {https://ui.adsabs.harvard.edu/abs/1999ASPC..161..295B},
	archiveprefix = {arXiv},
	author = {{Beloborodov}, A.~M.},
	booktitle = {High Energy Processes in Accreting Black Holes},
	doi = {10.48550/arXiv.astro-ph/9901108},
	editor = {{Poutanen}, Juri and {Svensson}, Roland},
	eprint = {astro-ph/9901108},
	month = jan,
	pages = {295},
	primaryclass = {astro-ph},
	series = {Astronomical Society of the Pacific Conference Series},
	title = {{Accretion Disk Models}},
	volume = {161},
	year = 1999}

@article{1998ApJ...500..525S,
	adsurl = {https://ui.adsabs.harvard.edu/abs/1998ApJ...500..525S},
	archiveprefix = {arXiv},
	author = {{Schlegel}, David J. and {Finkbeiner}, Douglas P. and {Davis}, Marc},
	doi = {10.1086/305772},
	eprint = {astro-ph/9710327},
	journal = {\apj},
	month = jun,
	number = {2},
	pages = {525-553},
	primaryclass = {astro-ph},
	title = {{Maps of Dust Infrared Emission for Use in Estimation of Reddening and Cosmic Microwave Background Radiation Foregrounds}},
	volume = {500},
	year = 1998}

@article{1984PASJ...36..741M,
	adsurl = {https://ui.adsabs.harvard.edu/abs/1984PASJ...36..741M},
	author = {{Mitsuda}, K. and {Inoue}, H. and {Koyama}, K. and {Makishima}, K. and {Matsuoka}, M. and {Ogawara}, Y. and {Shibazaki}, N. and {Suzuki}, K. and {Tanaka}, Y. and {Hirano}, T.},
	journal = {\pasj},
	month = jan,
	pages = {741-759},
	title = {{Energy spectra of low-mass binary X-ray sources observed from Tenma.}},
	volume = {36},
	year = 1984}

@article{1986ApJ...308..635M,
	adsurl = {https://ui.adsabs.harvard.edu/abs/1986ApJ...308..635M},
	author = {{Makishima}, K. and {Maejima}, Y. and {Mitsuda}, K. and {Bradt}, H.~V. and {Remillard}, R.~A. and {Tuohy}, I.~R. and {Hoshi}, R. and {Nakagawa}, M.},
	doi = {10.1086/164534},
	journal = {\apj},
	month = sep,
	pages = {635},
	title = {{Simultaneous X-Ray and Optical Observations of GX 339-4 in an X-Ray High State}},
	volume = {308},
	year = 1986}

@article{1992ApJ...395..130P,
	adsurl = {https://ui.adsabs.harvard.edu/abs/1992ApJ...395..130P},
	author = {{Pei}, Yichuan C.},
	doi = {10.1086/171637},
	journal = {\apj},
	month = aug,
	pages = {130},
	title = {{Interstellar Dust from the Milky Way to the Magellanic Clouds}},
	volume = {395},
	year = 1992}

@article{2016A&A...594A.116H,
	adsurl = {https://ui.adsabs.harvard.edu/abs/2016A&A...594A.116H},
	archiveprefix = {arXiv},
	author = {{HI4PI Collaboration} and {Ben Bekhti}, N. and {Fl{\"o}er}, L. and {Keller}, R. and {Kerp}, J. and {Lenz}, D. and {Winkel}, B. and {Bailin}, J. and {Calabretta}, M.~R. and {Dedes}, L. and {Ford}, H.~A. and {Gibson}, B.~K. and {Haud}, U. and {Janowiecki}, S. and {Kalberla}, P.~M.~W. and {Lockman}, F.~J. and {McClure-Griffiths}, N.~M. and {Murphy}, T. and {Nakanishi}, H. and {Pisano}, D.~J. and {Staveley-Smith}, L.},
	doi = {10.1051/0004-6361/201629178},
	eid = {A116},
	eprint = {1610.06175},
	journal = {\aap},
	month = oct,
	pages = {A116},
	primaryclass = {astro-ph.GA},
	title = {{HI4PI: A full-sky H I survey based on EBHIS and GASS}},
	volume = {594},
	year = 2016}

@article{1979ApJ...228..939C,
	adsurl = {https://ui.adsabs.harvard.edu/abs/1979ApJ...228..939C},
	author = {{Cash}, W.},
	doi = {10.1086/156922},
	journal = {\apj},
	month = mar,
	pages = {939-947},
	title = {{Parameter estimation in astronomy through application of the likelihood ratio.}},
	volume = {228},
	year = 1979}

@article{2016MNRAS.460..212C,
	adsurl = {https://ui.adsabs.harvard.edu/abs/2016MNRAS.460..212C},
	archiveprefix = {arXiv},
	author = {{Capellupo}, D.~M. and {Netzer}, H. and {Lira}, P. and {Trakhtenbrot}, B. and {Mej{\'\i}a-Restrepo}, J.},
	doi = {10.1093/mnras/stw937},
	eprint = {1604.05310},
	journal = {\mnras},
	month = jul,
	number = {1},
	pages = {212-226},
	primaryclass = {astro-ph.GA},
	title = {{Active galactic nuclei at z {\ensuremath{\sim}} 1.5 - III. Accretion discs and black hole spin}},
	volume = {460},
	year = 2016}

@article{2015MNRAS.446.3427C,
	adsurl = {https://ui.adsabs.harvard.edu/abs/2015MNRAS.446.3427C},
	archiveprefix = {arXiv},
	author = {{Capellupo}, D.~M. and {Netzer}, H. and {Lira}, P. and {Trakhtenbrot}, B. and {Mej{\'\i}a-Restrepo}, Juli{\'a}n},
	doi = {10.1093/mnras/stu2266},
	eprint = {1410.8137},
	journal = {\mnras},
	month = feb,
	number = {4},
	pages = {3427-3446},
	primaryclass = {astro-ph.GA},
	title = {{Active galactic nuclei at z {\ensuremath{\sim}} 1.5 - I. Spectral energy distribution and accretion discs}},
	volume = {446},
	year = 2015}

@article{1973A&A....24..337S,
	adsurl = {https://ui.adsabs.harvard.edu/abs/1973A&A....24..337S},
	author = {{Shakura}, N.~I. and {Sunyaev}, R.~A.},
	journal = {\aap},
	month = jun,
	pages = {33-51},
	title = {{Reprint of 1973A\&A....24..337S. Black holes in binary systems. Observational appearance.}},
	volume = {500},
	year = 1973}

@article{2006AJ....131.2826S,
	adsurl = {https://ui.adsabs.harvard.edu/abs/2006AJ....131.2826S},
	archiveprefix = {arXiv},
	author = {{Steffen}, A.~T. and {Strateva}, I. and {Brandt}, W.~N. and {Alexand er}, D.~M. and {Koekemoer}, A.~M. and {Lehmer}, B.~D. and {Schneider}, D.~P. and {Vignali}, C.},
	doi = {10.1086/503627},
	eprint = {astro-ph/0602407},
	journal = {\aj},
	month = jun,
	number = {6},
	pages = {2826-2842},
	primaryclass = {astro-ph},
	title = {{The X-Ray-to-Optical Properties of Optically Selected Active Galaxies over Wide Luminosity and Redshift Ranges}},
	volume = {131},
	year = 2006}

@article{2020A&A...636A..73D,
	adsurl = {https://ui.adsabs.harvard.edu/abs/2020A&A...636A..73D},
	archiveprefix = {arXiv},
	author = {{Duras}, F. and {Bongiorno}, A. and {Ricci}, F. and {Piconcelli}, E. and {Shankar}, F. and {Lusso}, E. and {Bianchi}, S. and {Fiore}, F. and {Maiolino}, R. and {Marconi}, A. and {Onori}, F. and {Sani}, E. and {Schneider}, R. and {Vignali}, C. and {La Franca}, F.},
	doi = {10.1051/0004-6361/201936817},
	eid = {A73},
	eprint = {2001.09984},
	journal = {\aap},
	month = apr,
	pages = {A73},
	primaryclass = {astro-ph.GA},
	title = {{Universal bolometric corrections for active galactic nuclei over seven luminosity decades}},
	volume = {636},
	year = 2020}

@article{2012MNRAS.425..623L,
	adsurl = {https://ui.adsabs.harvard.edu/abs/2012MNRAS.425..623L},
	archiveprefix = {arXiv},
	author = {{Lusso}, E. and {Comastri}, A. and {Simmons}, B.~D. and {Mignoli}, M. and {Zamorani}, G. and {Vignali}, C. and {Brusa}, M. and {Shankar}, F. and {Lutz}, D. and {Trump}, J.~R. and {Maiolino}, R. and {Gilli}, R. and {Bolzonella}, M. and {Puccetti}, S. and {Salvato}, M. and {Impey}, C.~D. and {Civano}, F. and {Elvis}, M. and {Mainieri}, V. and {Silverman}, J.~D. and {Koekemoer}, A.~M. and {Bongiorno}, A. and {Merloni}, A. and {Berta}, S. and {Le Floc'h}, E. and {Magnelli}, B. and {Pozzi}, F. and {Riguccini}, L.},
	doi = {10.1111/j.1365-2966.2012.21513.x},
	eprint = {1206.2642},
	journal = {\mnras},
	month = sep,
	number = {1},
	pages = {623-640},
	primaryclass = {astro-ph.CO},
	title = {{Bolometric luminosities and Eddington ratios of X-ray selected active galactic nuclei in the XMM-COSMOS survey}},
	volume = {425},
	year = 2012}

@article{2010A&A...512A..34L,
	adsurl = {https://ui.adsabs.harvard.edu/abs/2010A&A...512A..34L},
	archiveprefix = {arXiv},
	author = {{Lusso}, E. and {Comastri}, A. and {Vignali}, C. and {Zamorani}, G. and {Brusa}, M. and {Gilli}, R. and {Iwasawa}, K. and {Salvato}, M. and {Civano}, F. and {Elvis}, M. and {Merloni}, A. and {Bongiorno}, A. and {Trump}, J.~R. and {Koekemoer}, A.~M. and {Schinnerer}, E. and {Le Floc'h}, E. and {Cappelluti}, N. and {Jahnke}, K. and {Sargent}, M. and {Silverman}, J. and {Mainieri}, V. and {Fiore}, F. and {Bolzonella}, M. and {Le F{\`e}vre}, O. and {Garilli}, B. and {Iovino}, A. and {Kneib}, J.~P. and {Lamareille}, F. and {Lilly}, S. and {Mignoli}, M. and {Scodeggio}, M. and {Vergani}, D.},
	doi = {10.1051/0004-6361/200913298},
	eid = {A34},
	eprint = {0912.4166},
	journal = {\aap},
	month = mar,
	pages = {A34},
	primaryclass = {astro-ph.CO},
	title = {{The X-ray to optical-UV luminosity ratio of X-ray selected type 1 AGN in XMM-COSMOS}},
	volume = {512},
	year = 2010}

@article{2009MNRAS.392.1124V,
	adsurl = {https://ui.adsabs.harvard.edu/abs/2009MNRAS.392.1124V},
	archiveprefix = {arXiv},
	author = {{Vasudevan}, R.~V. and {Fabian}, A.~C.},
	doi = {10.1111/j.1365-2966.2008.14108.x},
	eprint = {0810.3777},
	journal = {\mnras},
	month = jan,
	number = {3},
	pages = {1124-1140},
	primaryclass = {astro-ph},
	title = {{Simultaneous X-ray/optical/UV snapshots of active galactic nuclei from XMM-Newton: spectral energy distributions for the reverberation mapped sample}},
	volume = {392},
	year = 2009}

@book{2013peag.book.....N,
	adsurl = {https://ui.adsabs.harvard.edu/abs/2013peag.book.....N},
	author = {{Netzer}, Hagai},
	title = {{The Physics and Evolution of Active Galactic Nuclei}},
	year = 2013}

@article{1995PASP..107..803U,
	adsurl = {https://ui.adsabs.harvard.edu/abs/1995PASP..107..803U},
	archiveprefix = {arXiv},
	author = {{Urry}, C. Megan and {Padovani}, Paolo},
	doi = {10.1086/133630},
	eprint = {astro-ph/9506063},
	journal = {\pasp},
	month = sep,
	pages = {803},
	primaryclass = {astro-ph},
	title = {{Unified Schemes for Radio-Loud Active Galactic Nuclei}},
	volume = {107},
	year = 1995}

@inbook{1996ASPC..101...17A,
	adsurl = {https://ui.adsabs.harvard.edu/abs/1996ASPC..101...17A},
	author = {{Arnaud}, K.~A.},
	booktitle = {Astronomical Data Analysis Software and Systems V},
	editor = {{Jacoby}, George H. and {Barnes}, Jeannette},
	pages = {17},
	series = {Astronomical Society of the Pacific Conference Series},
	title = {{XSPEC: The First Ten Years}},
	volume = {101},
	year = 1996}

@article{1991ApJ...380L..51H,
	adsurl = {https://ui.adsabs.harvard.edu/abs/1991ApJ...380L..51H},
	author = {{Haardt}, F. and {Maraschi}, L.},
	doi = {10.1086/186171},
	journal = {\apjl},
	month = oct,
	pages = {L51},
	title = {{A Two-Phase Model for the X-Ray Emission from Seyfert Galaxies}},
	volume = {380},
	year = 1991}


\begin{appendix} 


\section{$\kappa_{2-10}$ vs $L_{\rm bol}$ for lensed AGN}\label{sect:appendixa}

As expected, the 2--10\,keV X-ray bolometric corrections for the lensed AGN show an increasing trend with the bolometric luminosity (Fig. \ref{fig:kx_lb}). The best-fit relation corresponding to the lensed images (dashed gray line) agrees within uncertainties with the fit (solid red line) reported for nonlensed AGN by \cite{2025ApJ...990...86G}. However, the relation obtained from the current study is biased due to many factors and therefore cannot be used to estimate $L_{\rm bol}$ for other lensed AGN. Firstly, it is based on a statistically small sample. Secondly, as discussed in Sect. \ref{sect:alpha}, certain resolved or unresolved lensed images do not represent the intrinsic properties of the entire system. In some cases, they can be affected by micro-/millilensing, while sometimes the lens model might be poorly constrained. As a result, based on the study and sample presented in this paper, we cannot provide a one-to-one relation between $\kappa_{2-10}$ and $L_{\rm bol}$ specific to lensed AGN. However, lensing should not affect the intrinsic accretion mechanisms of AGN and therefore, we can use the established $\kappa_{2-10}-L_{\rm bol}$ relations from the literature to estimate $L_{\rm bol}$ for lensed AGN from limited data.


\newpage
\section{Physical properties of lensed AGN}\label{sect:appendixb}


\begin{table*}
\centering
\caption{Outputs from the optical-to-X-ray SED and optical/UV spectral fitting.}
\begin{tabular}{cccccccc}
\hline
\hline
\specialrule{0.1em}{0em}{0.5em}
\vspace{1mm}
Object & Image & Macro-Magnification\tablefootmark{a} & $L_{2500\,{\rm \AA}}$ & $L_{\rm 2\,keV}$ & $L_{\rm bol}$ & $M_{\rm BH}$ & $\lambda_{\rm Edd}$\\
\hline\\
RXJ1131 & A & 18.86 & 44.58 & 43.33 & 45.11 & \,\,\,8.11\tablefootmark{b} & $-1.17$\\
        & B & 12.05 & 44.73 & 44.13 & 45.72 & \,\,\,8.11\tablefootmark{b} & $-0.57$\\
        & C & 9.95 & 44.34 & 43.84 & 45.20 & \,\,\,8.11\tablefootmark{b} & $-1.08$\\
\vspace{2mm}
        & D & 1.05 & 44.57 & 44.12 & 45.33 & \,\,\,8.11\tablefootmark{b} & $-0.96$\\
DESJ2038 & A & 3.04 & 43.92 & 44.87 & 45.85 & 8.49 & $-0.82$\\
         & B & 3.65 & 43.78 & 44.40 & 46.04 & 8.53 & $-0.67$\\
         & C & 2.85 & 43.87 & 44.86 & 46.24 & 8.58 & $-0.51$\\
\vspace{2mm}
         & D & 1.33 & 43.40 & 44.72 & 45.75 & 8.68 & $-1.11$\\
\vspace{2mm}
SDSSJ1251 & A+B+C+D & 23.00 & - & 42.66 & 44.45 & - & -\\
GRAL1131 & A & 2.62 & - & 43.95 & 45.60 & - & -\\
         & B & 4.17 & - & 43.54 & 45.33 & - & -\\
\vspace{2mm}
         & C+D & 18.63 & - & 44.02 & 45.71 & - & -\\
\vspace{2mm}
HE1113 & A+B+C+D & 54.60 & - & 43.64 & 45.16 & - & -\\
\vspace{2mm}
WISE2344 & A+B+C+D & 33.00 & 43.71 & 43.26 & 44.66 & 7.66 & $-1.17$\\
J0607 & A & 7.84 & 44.04 & 44.02 & 45.48 & 8.17 & $-0.87$\\
\vspace{2mm}
      & D & 17.55 & 43.67 & 43.61 & 45.22 & 7.95 & $-0.91$\\
\vspace{2mm}
SDSSJ0924 & B & 4.08 & 44.50 & 43.74 & 45.21 & 8.48 & $-1.45$\\
J2145 & A & 16.23 & - & 44.35 & 46.07 & - & -\\
      & B & 18.03 & - & 43.55 & 45.38 & - & -\\
\vspace{2mm}
      & C+D & 6.3 & - & 44.96 & 46.87 & - & -\\
WFI2033 & B & 3.74 & 45.72 & 44.29 & 46.12 & 8.54 & $-0.60$\\
\vspace{2mm}
        & C & 3.66 & 45.51 & 44.13 & 46.02 & 8.47 & $-0.63$\\
\vspace{2mm}
PSJ1606 & A+B+C+D & 8.00 & 46.12 & 44.34 & 46.53 & 8.86 & $-0.51$\\
HE0435 & B & 5.18 & 46.07 & 44.28 & 46.44 & 8.74 & $-0.48$\\
\vspace{2mm}
       & D & 3.42 & 46.31 & 44.59 & 46.85 & 8.72 & $-0.05$\\
\vspace{2mm}
DESJ0405 & A+B+C+D & 93.30 & 42.27 & 43.49 & 45.04 & 7.02 & $-0.16$\\
J1537 & A & 2.93 & 45.89 & 44.69 & 46.32 & 9.08 & $-0.93$\\
      & B & 1.65 & 46.01 & 44.86 & 46.45 & 8.94 & $-0.66$\\
      & C & 2.70 & 45.94 & 44.81 & 46.39 & 8.94 & $-0.73$\\
\vspace{2mm}
      & D & 1.92 & 45.49 & 44.42 & 46.01 & 8.79 & $-0.96$\\
\vspace{2mm}
J2017 & B & 5.4 & 45.53 & - & 45.66 & 7.94 & $-0.46$\\
PG1115 & A & 2.49 & 44.85 & 44.72 & 46.57 & 9.20 & $-0.80$\\
       & B & 3.31 & 45.55 & 44.60 & 46.62 & 9.27 & $-0.83$\\
\vspace{2mm}
       & C+D & 22.63 & 44.95 & 44.44 & 46.58 & 9.25 & $-0.85$\\  
\vspace{2mm}
WISEJ0259 & A+B+C+D & 12.64 & 44.23 & 45.39 & 46.67 & 8.07 & \,\,\,\,0.42\\
\vspace{2mm}
WFI2026 & B & 2.78 & 46.60 & 44.90 & 46.99 & 9.11 & $-0.30$\\
\vspace{2mm}
J0608 & A+B+C+D & 24.60 & 45.52 & 44.59 & 46.54 & 8.51 & $-0.15$\\
PSJ0147 & B & 9.36 & 46.69 & 44.66 & 47.31 & 9.52 & $-0.39$\\
        & C & 16.44 & 46.96 & 44.70 & 47.40 & 9.52 & $-0.29$\\
\vspace{2mm}
        & D & 9.76 & 47.23 & 44.87 & 47.61 & 9.60 & $-0.17$\\
\vspace{2mm}
SDSSJ0248 & A+B+C+D & 27.60 & - & 44.02 & 46.46 & 8.50 & $-0.21$\\
\vspace{2mm}
J1042 & A+B+C+D & 47.31 & 44.37 & 45.28 & 46.90 & - & -\\
\hline
\end{tabular}
\tablefoot{We have reported the magnification-corrected intrinsic 2500\,${\rm \AA}$, 2\,keV, and bolometric luminosities, in ${\rm log}\,(L/\rm erg\,s^{-1})$. We also present the black hole masses ($M_{\rm BH}$) of our sources in units of ${\rm log}\,(M_{\rm BH}/{\rm M_{\odot}}$). Finally, we list the estimated value of the Eddington ratios ($\lambda_{\rm Edd}$ in log).\\
\tablefoottext{a}{\citealp{2025arXiv251107513G}}\\
\tablefoottext{b}{Black hole mass estimate from \citet{2012A&A...544A..62S}}}
\label{tab:params}
\end{table*}


\begin{table*}
\centering
\caption*{Table \ref{tab:params} continued.}
\begin{tabular}{cccccccc}
\hline
\hline
\specialrule{0.1em}{0em}{0.5em}
\vspace{1mm}
Object & Image & Macro-Magnification\tablefootmark{a} & $L_{2500\,{\rm \AA}}$ & $L_{\rm 2\,keV}$ & $L_{\rm bol}$ & $M_{\rm BH}$ & $\lambda_{\rm Edd}$\\
\hline\\
\vspace{2mm}
H1413 & A+B+C+D & 31.54 & 45.84 & 44.35 & 46.77 & 9.48 & $-0.88$\\
MG0414 & A+B & 29.91 & - & 44.97 & 46.85 & - & -\\
       & C & 4.29 & - & 45.15 & 47.23 & - & -\\
\vspace{2mm}
       & D & 1.82 & - & 45.11 & 47.23 & - & -\\
2M1134 & A & 1.78 & 47.40 & 45.30 & 47.92 & 9.28 & \,\,\,0.46\\
       & B & 0.57 & 47.35 & 45.39 & 47.79 & 9.22 & \,\,\,0.40\\
       & C & 1.85 & 47.25 & 45.62 & 47.87 & 9.11 & \,\,\,0.59\\
\vspace{2mm}
       & D & 1.01 & 48.04 & 46.04 & 48.26 & 9.01 & \,\,\,1.08\\
 \vspace{2mm}
J0803 & A+B+C+D & 23.00 & - & 44.75 & 46.77 & - & -\\
J0659 & A & 4.84 & 45.97 & 45.13 & 46.72 & 9.15 & $-0.61$\\
      & B & 2.94 & 46.12 & 45.76 & 47.01 & 9.26 & $-0.42$\\
        & C & 3.59 & 45.90 & 45.46 & 46.87 & 9.22 & $-0.53$\\
\vspace{2mm}
      & D & 10.17 & 46.16 & 45.47 & 47.01 & 9.28 & $-0.44$\\
\hline
\end{tabular}
\end{table*}


\newpage
\section{Additional material}\label{sect:appendixc}

\begin{figure*}
  \begin{subfigure}[t]{0.15\textwidth}
    \centering
    \includegraphics[width=\textwidth]{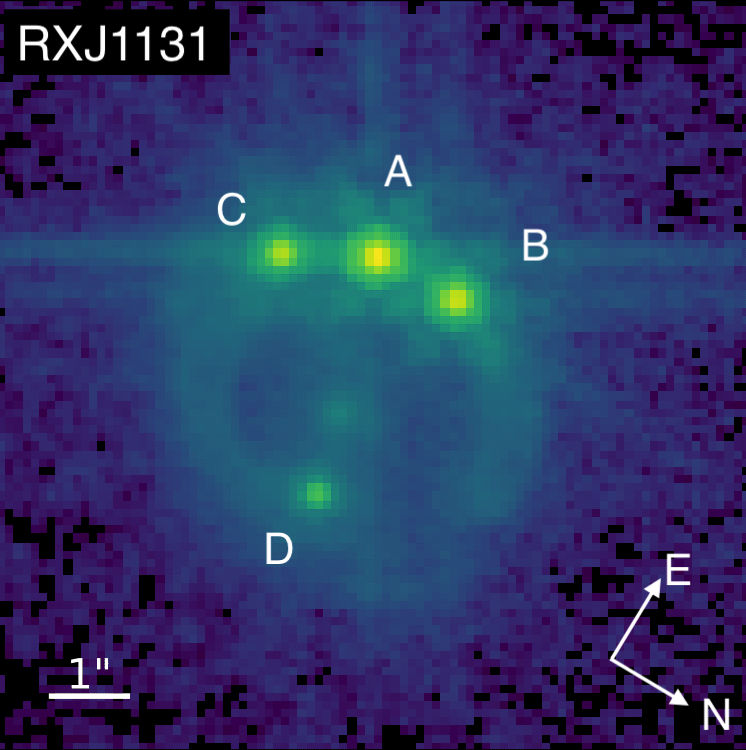}
  \end{subfigure}
  \begin{subfigure}[t]{0.15\textwidth}
    \centering
    \includegraphics[width=\textwidth]{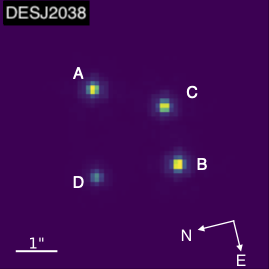}
  \end{subfigure}
  \begin{subfigure}[t]{0.15\textwidth}
    \centering
    \includegraphics[width=\textwidth]{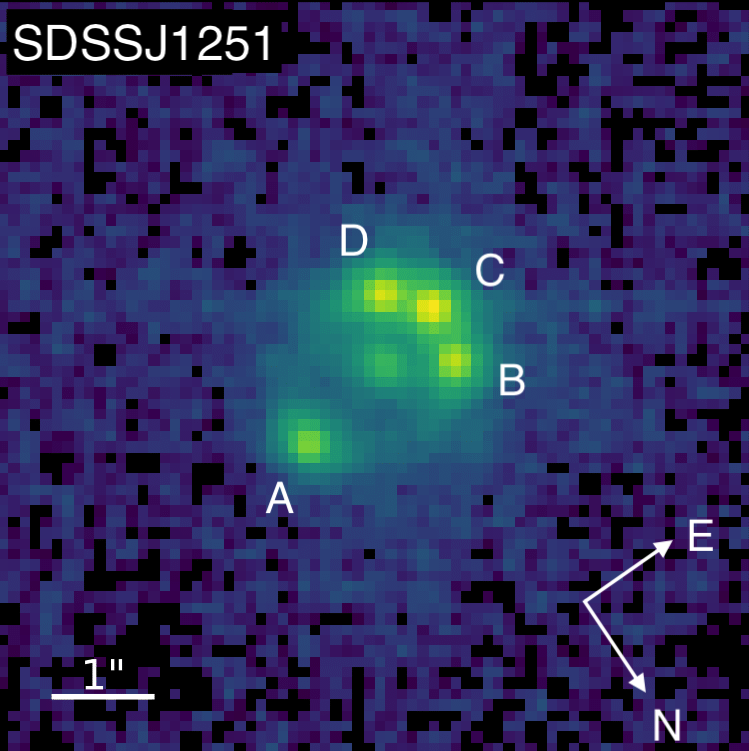}
  \end{subfigure}
  \begin{subfigure}[t]{0.15\textwidth}
    \centering
    \includegraphics[width=\textwidth]{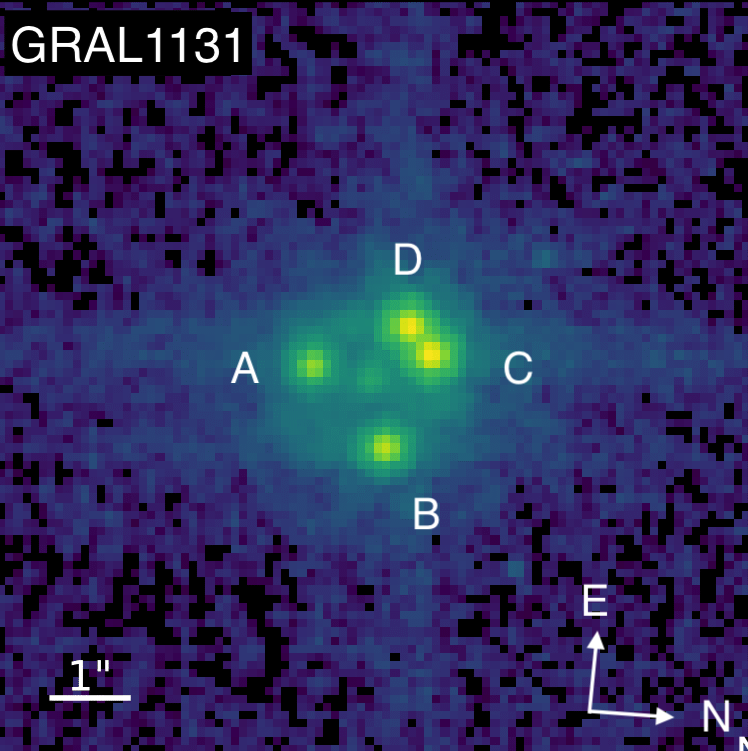}
  \end{subfigure}
  \begin{subfigure}[t]{0.15\textwidth}
    \centering
    \includegraphics[width=\textwidth]{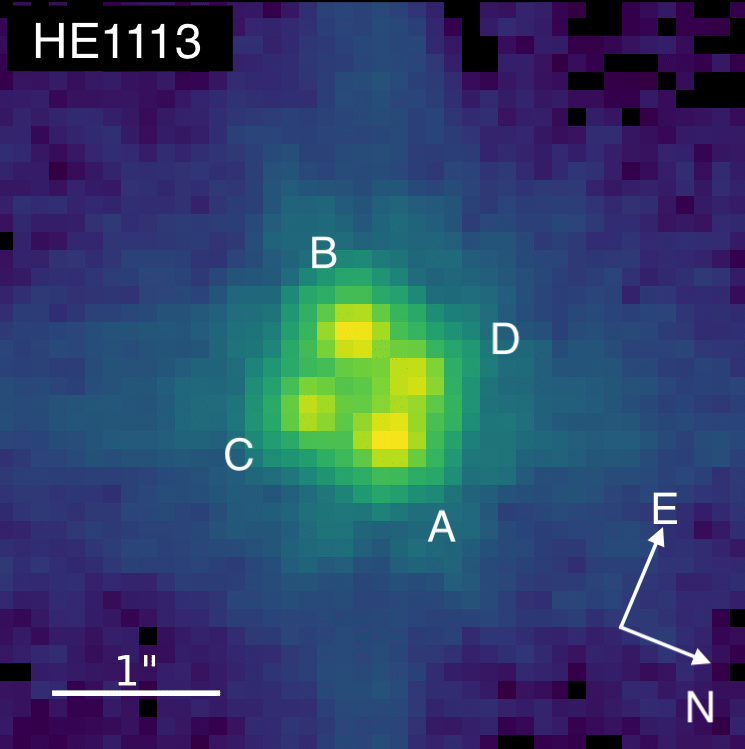}
  \end{subfigure}
  \begin{subfigure}[t]{0.15\textwidth}
    \centering
    \includegraphics[width=\textwidth]{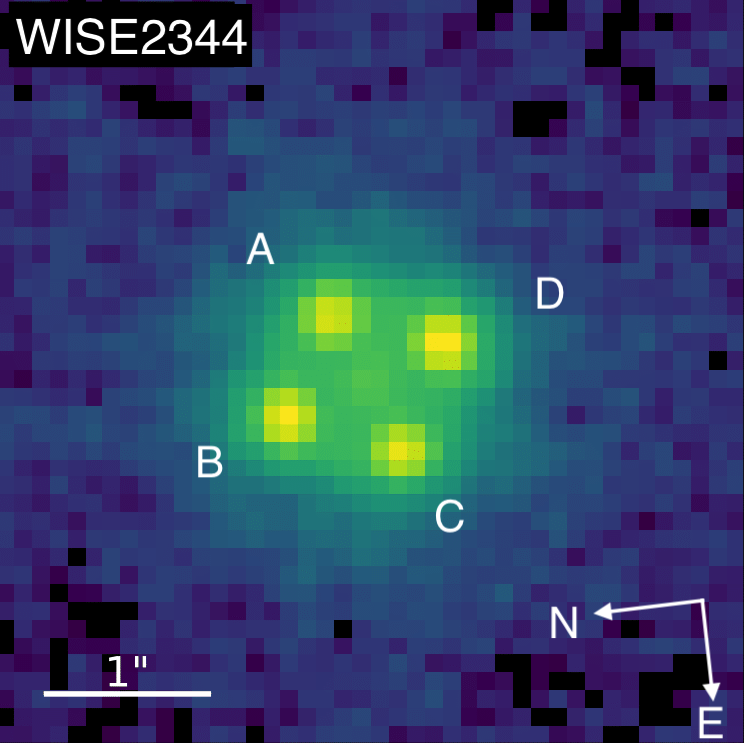}
  \end{subfigure}

  \begin{subfigure}[t]{0.15\textwidth}
    \centering
    \includegraphics[width=\textwidth]{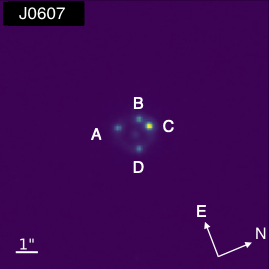}
  \end{subfigure}
  \begin{subfigure}[t]{0.15\textwidth}
    \centering
    \includegraphics[width=\textwidth]{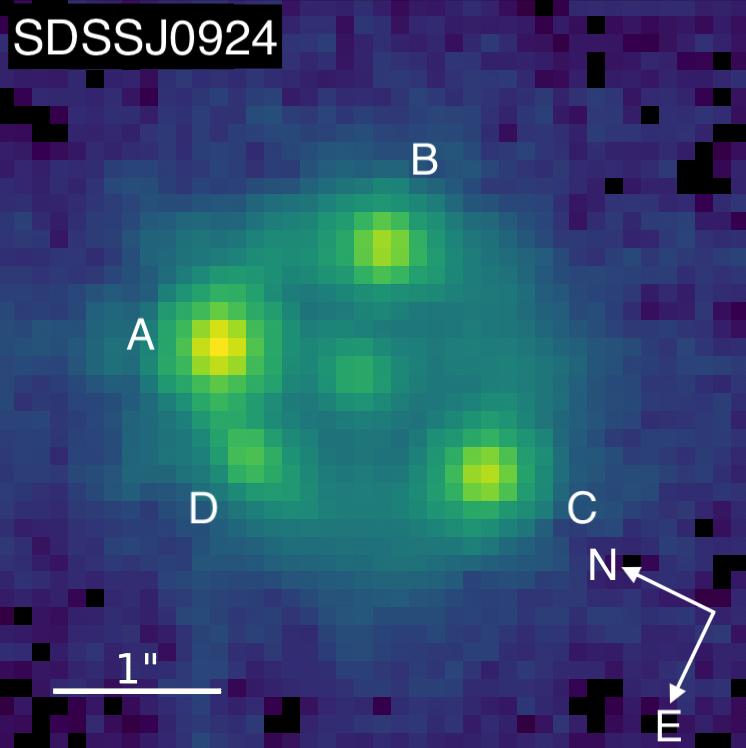}
  \end{subfigure}
  \begin{subfigure}[t]{0.15\textwidth}
    \centering
    \includegraphics[width=\textwidth]{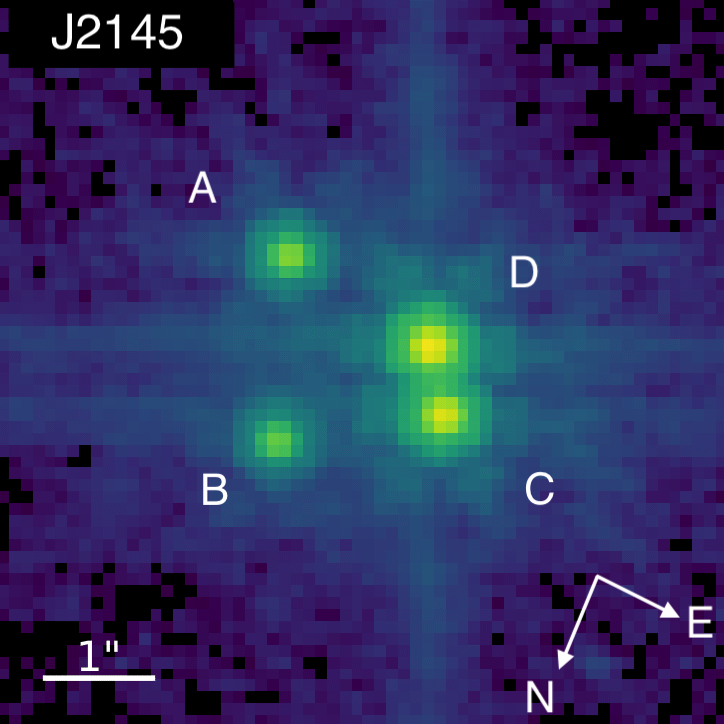}
  \end{subfigure}
  \begin{subfigure}[t]{0.15\textwidth}
    \centering
    \includegraphics[width=\textwidth]{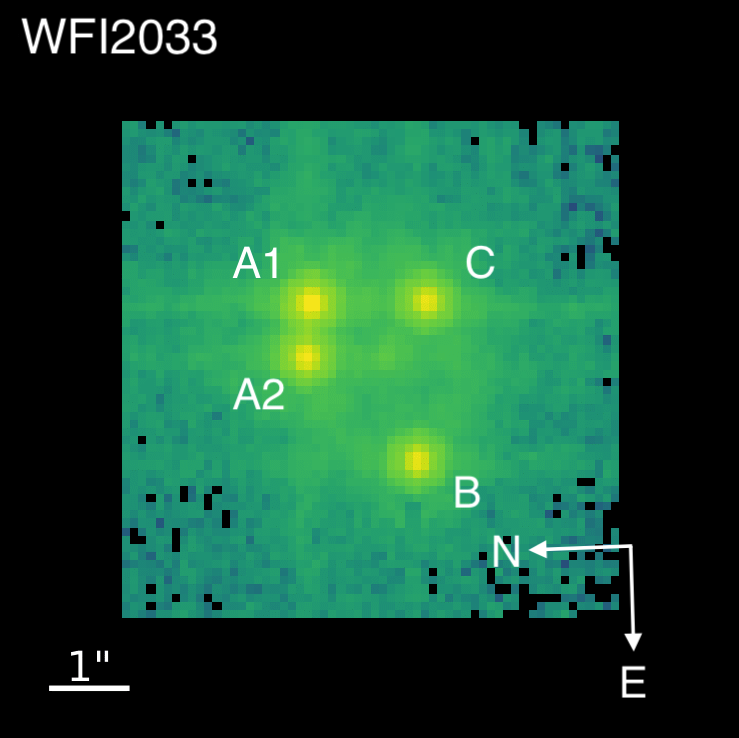}
  \end{subfigure}
  \begin{subfigure}[t]{0.15\textwidth}
    \centering
    \includegraphics[width=\textwidth]{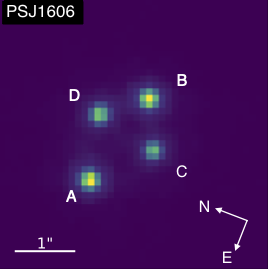}
  \end{subfigure}
  \begin{subfigure}[t]{0.15\textwidth}
    \centering
    \includegraphics[width=\textwidth]{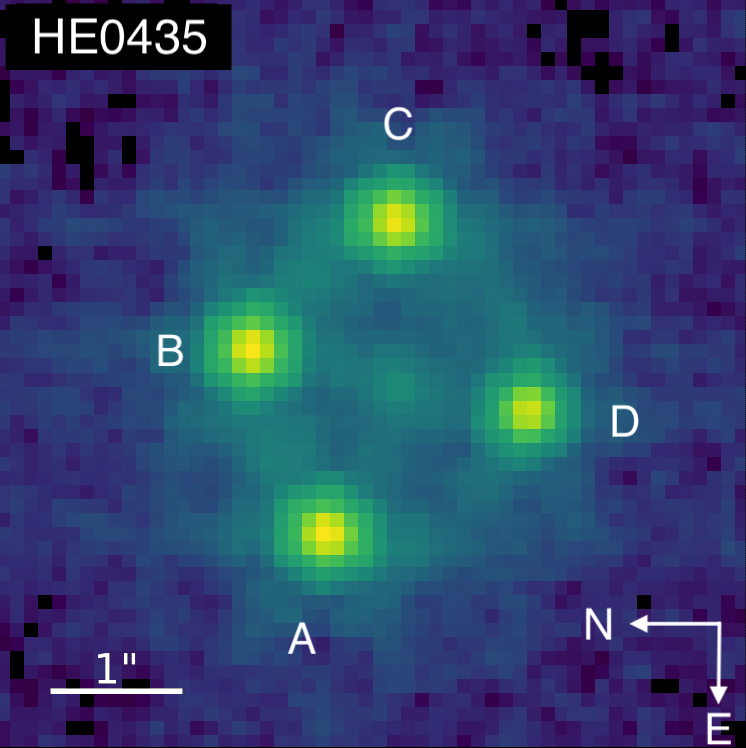}
  \end{subfigure}
  
  \begin{subfigure}[t]{0.15\textwidth}
    \centering
    \includegraphics[width=\textwidth]{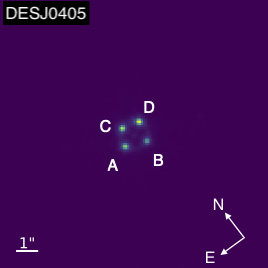}
  \end{subfigure}
  \begin{subfigure}[t]{0.15\textwidth}
    \centering
    \includegraphics[width=\textwidth]{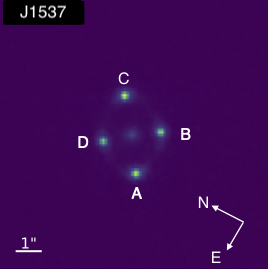}
  \end{subfigure}
  \begin{subfigure}[t]{0.15\textwidth}
    \centering
    \includegraphics[width=\textwidth]{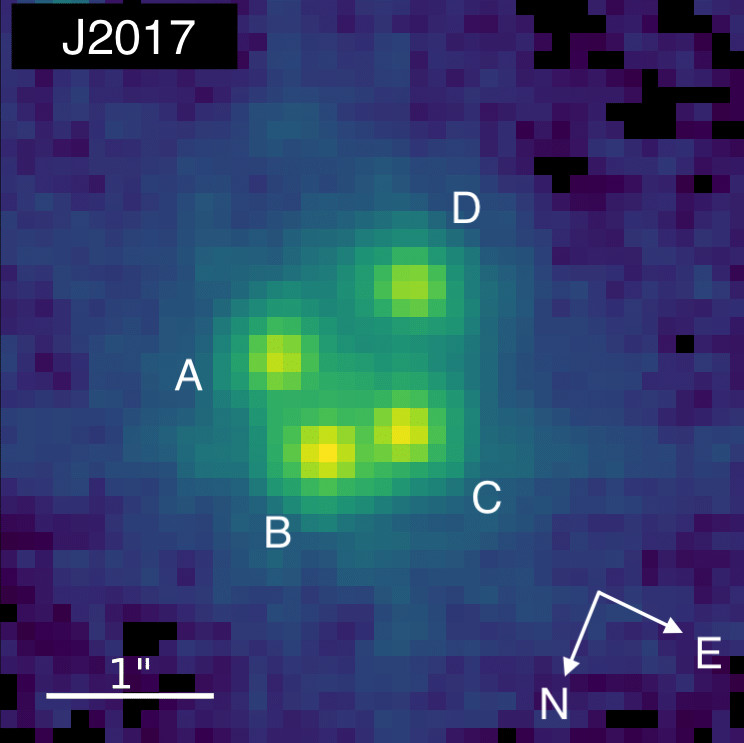}
  \end{subfigure}
  \begin{subfigure}[t]{0.15\textwidth}
    \centering
    \includegraphics[width=\textwidth]{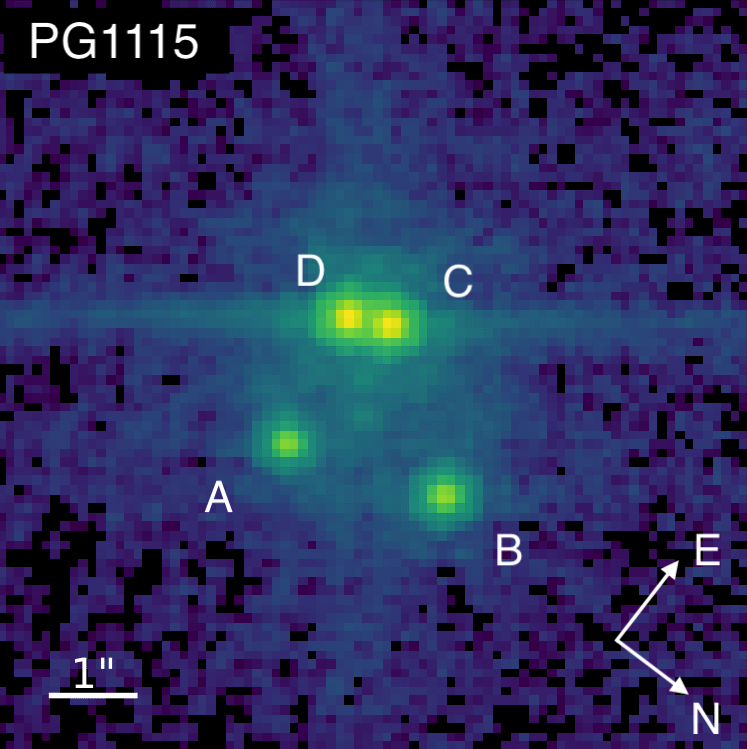}
  \end{subfigure}
  \begin{subfigure}[t]{0.15\textwidth}
    \centering
    \includegraphics[width=\textwidth]{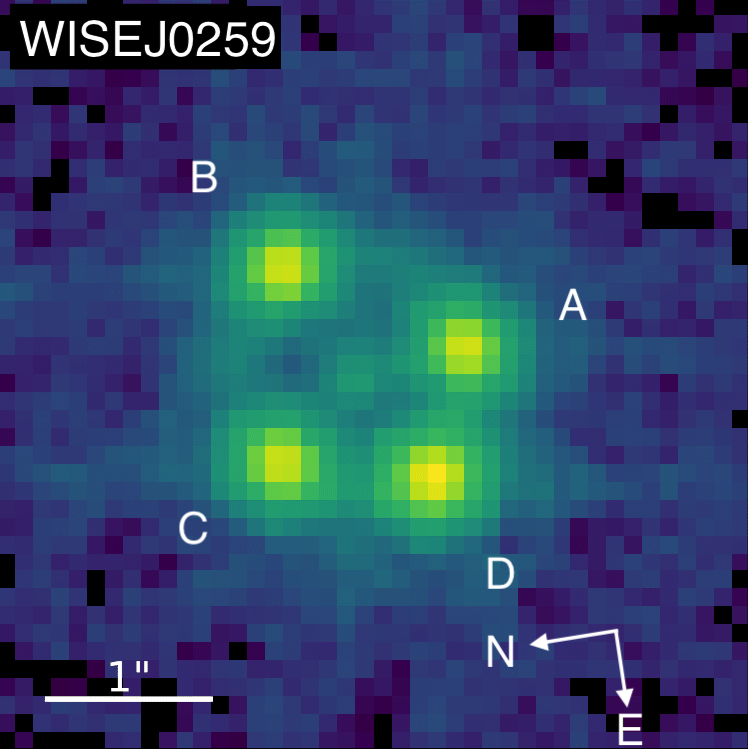}
  \end{subfigure}
  \begin{subfigure}[t]{0.15\textwidth}
    \centering
    \includegraphics[width=\textwidth]{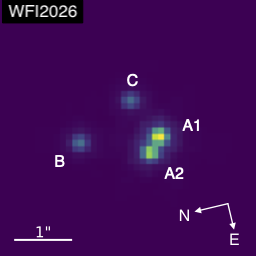}
  \end{subfigure}
  
  \begin{subfigure}[t]{0.15\textwidth}
    \centering
    \includegraphics[width=\textwidth]{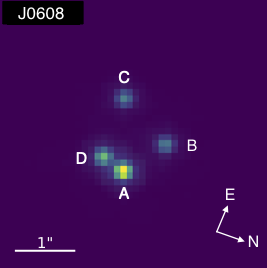}
  \end{subfigure}
  \begin{subfigure}[t]{0.15\textwidth}
    \centering
    \includegraphics[width=\textwidth]{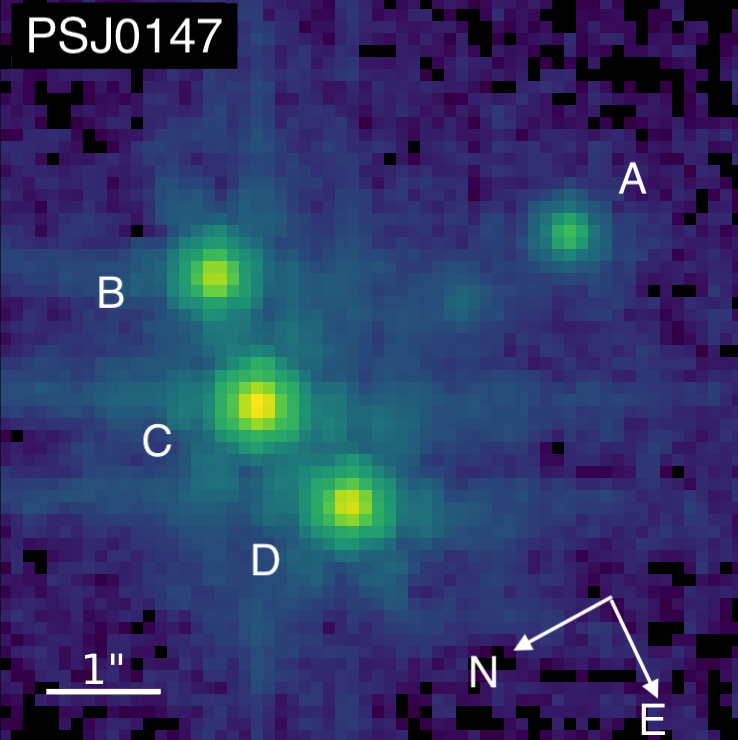}
  \end{subfigure}
  \begin{subfigure}[t]{0.15\textwidth}
    \centering
    \includegraphics[width=\textwidth]{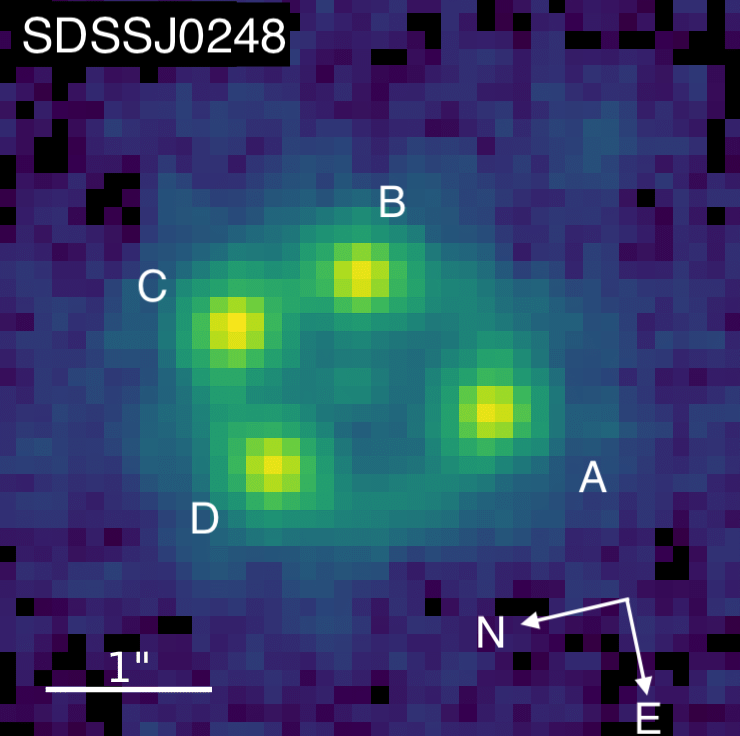}
  \end{subfigure}
  \begin{subfigure}[t]{0.15\textwidth}
    \centering
    \includegraphics[width=\textwidth]{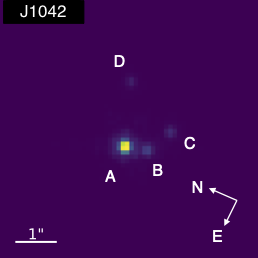}
  \end{subfigure}
  \begin{subfigure}[t]{0.15\textwidth}
    \centering
    \includegraphics[width=\textwidth]{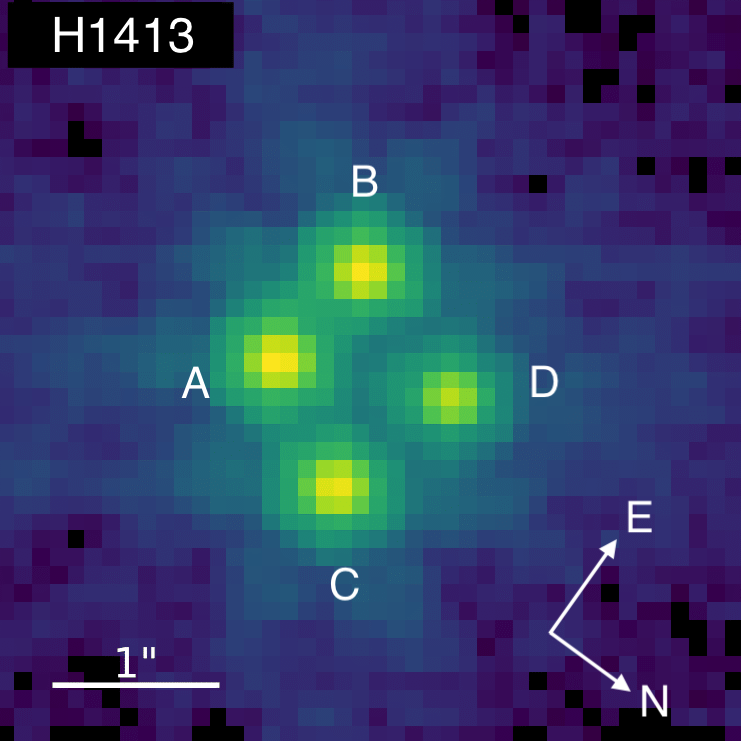}
  \end{subfigure}
  \begin{subfigure}[t]{0.15\textwidth}
    \centering
    \includegraphics[width=\textwidth]{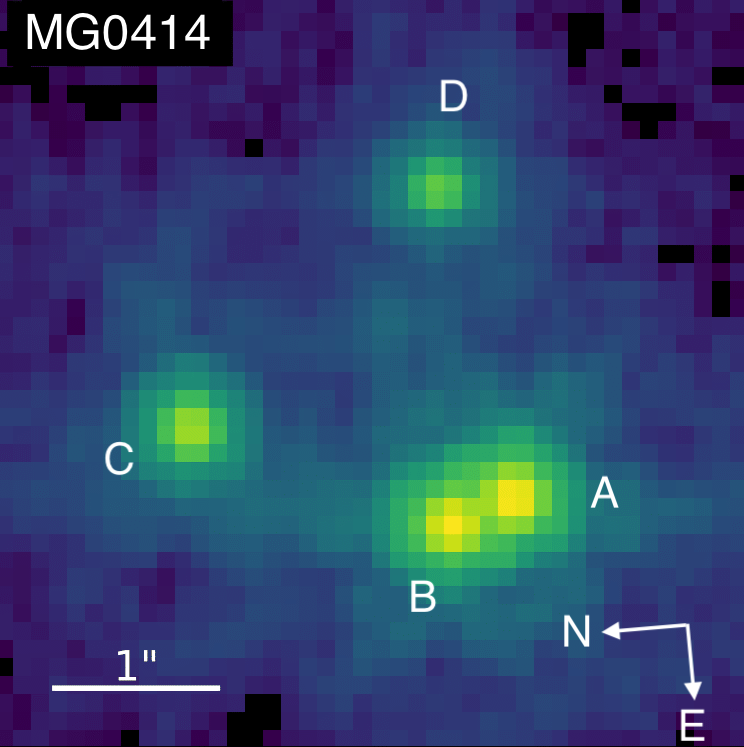}
  \end{subfigure}
  
  \begin{subfigure}[t]{0.15\textwidth}
    \centering
    \includegraphics[width=\textwidth]{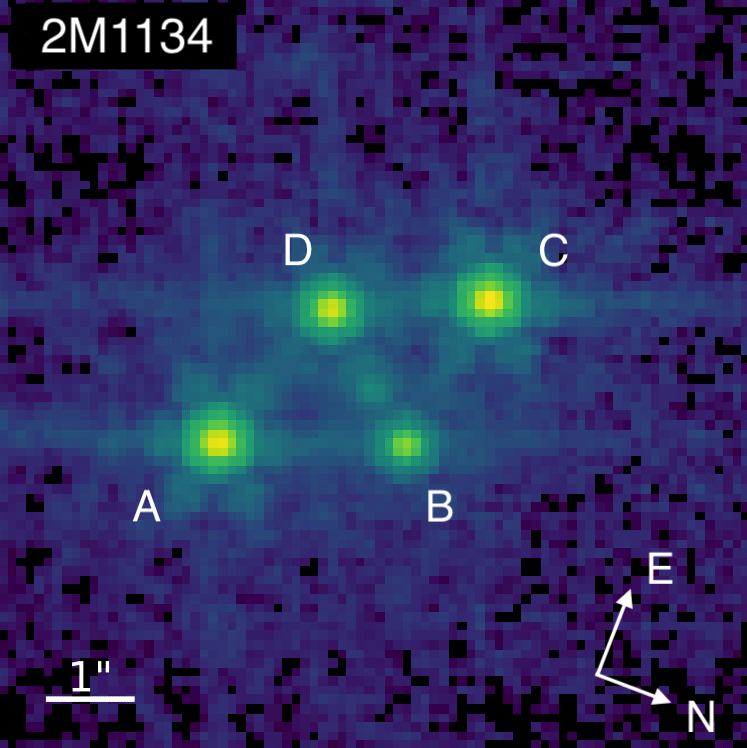}
  \end{subfigure}
  \begin{subfigure}[t]{0.15\textwidth}
    \centering
    \includegraphics[width=\textwidth]{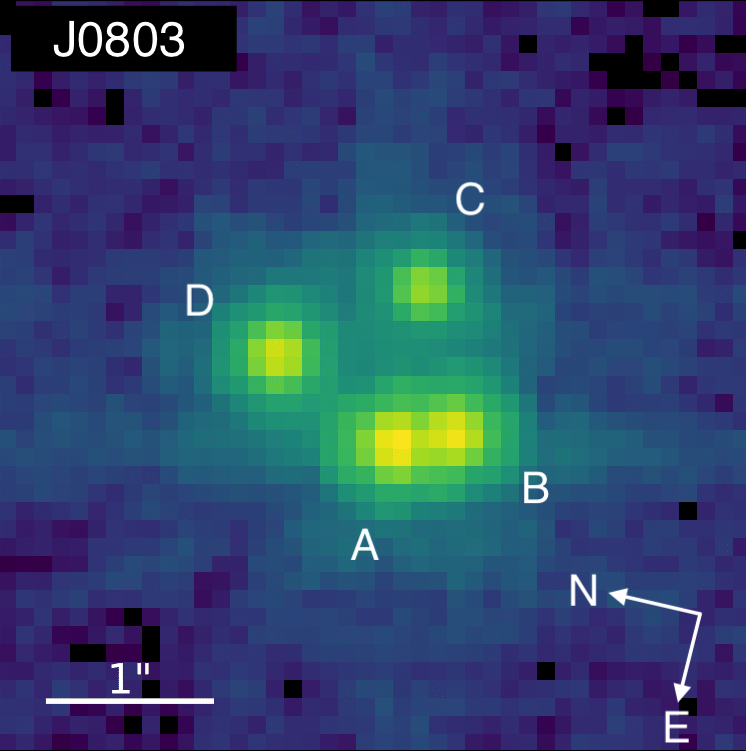}
  \end{subfigure}
  \begin{subfigure}[t]{0.15\textwidth}
    \centering
    \includegraphics[width=\textwidth]{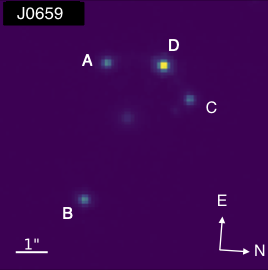}
  \end{subfigure}
   
\caption{Image labeling of the four lensed images in our lensed quasar systems. All images correspond to observations in the F560W band of JWST/MIRI (\citealp{2024MNRAS.530.2960N,2024MNRAS.535.1652K,2025arXiv251107765K}).}
\label{fig:lens_image}
\end{figure*}


\begin{table*}
\centering
\caption{Optical/UV photometry from HST images}
\begin{tabular}{cccc}
\hline
\hline
\specialrule{0.1em}{0em}{0.5em}
\vspace{1mm}
Object & Image & F475X & F814W\\
\hline\\
WISE2344 & A & $22.88\pm0.19$ & $21.69\pm0.24$\\
         & B & $21.71\pm0.18$ & $21.17\pm0.19$\\
         & C & $21.12\pm0.15$ & $20.82\pm0.29$\\
\vspace{2mm}
         & D & $21.64\pm0.16$ & $21.05\pm0.22$\\
WISEJ0259 & A & $21.01\pm0.23$ & $19.74\pm0.19$\\
          & B & $20.96\pm0.20$ & $19.45\pm0.16$\\
          & C & $20.50\pm0.20$ & $19.10\pm0.17$\\
\vspace{2mm}
          & D & $20.27\pm0.17$ & $18.79\pm0.16$\\
J1042 & A & $21.21\pm0.17$ & $20.36\pm0.24$\\
      & B & $22.22\pm0.14$ & $22.05\pm0.15$\\
      & C & $22.70\pm0.19$ & $22.47\pm0.31$\\
\vspace{2mm}
      & D & $23.25\pm0.24$ & $22.94\pm0.39$\\
J0659 & A & $20.17\pm0.06$ & $19.78\pm0.12$\\
      & B & $20.09\pm0.06$ & $19.46\pm0.10$\\
      & C & $20.30\pm0.06$ & $19.78\pm0.13$\\
\vspace{2mm}
      & D & $18.75\pm0.10$ & $18.21\pm0.09$\\
\hline
\end{tabular}
\label{tab:hst}
\end{table*}


\end{appendix}

\end{document}